%% file: main_arxiv.tex
\newif\ifmsanon      
\newif\ifjhr         
\newif\ifsplit       
\newif\ifsupp        
\newif\ifmsappendix  
\edef\msjobname{\jobname}
\edef\msmainjobname{\detokenize{main_ms}}
\edef\msecjobname{\detokenize{main_ms_ec}}
\edef\jhrmainjobname{\detokenize{main_jhr}}
\edef\jhrappjobname{\detokenize{main_jhr_appendix}}
\ifx\msjobname\msmainjobname\msanontrue\splittrue\fi
\ifx\msjobname\msecjobname\msanontrue\splittrue\supptrue\fi
\ifx\msjobname\jhrmainjobname\jhrtrue\splittrue\fi
\ifx\msjobname\jhrappjobname\jhrtrue\splittrue\supptrue\fi
\ifmsanon\PassOptionsToClass{11pt}{article}\else\PassOptionsToClass{12pt}{article}\fi
\documentclass{article}

\usepackage[utf8]{inputenc}
\usepackage[english]{babel}
\usepackage{amsmath}
\usepackage{amsthm}
\usepackage{amsfonts}
\ifjhr\usepackage{newtxtext,newtxmath}\fi 
\usepackage{bm}
\usepackage{bbm}
\usepackage{setspace}
\usepackage[margin=1in]{geometry}
\usepackage[shortlabels]{enumitem}
\usepackage{graphicx}
\usepackage[dvipsnames]{xcolor}
\usepackage{booktabs}
\usepackage{multirow}
\usepackage{threeparttable}
\usepackage{adjustbox}
\usepackage{caption}
\usepackage{subcaption}
\usepackage{float}
\usepackage{pdflscape}
\usepackage{placeins}
\usepackage[]{natbib}
\ifsplit\usepackage{xr-hyper}\fi
\usepackage{hyperref}

\hypersetup{colorlinks=true, linkcolor=BrickRed, citecolor=BrickRed, filecolor=BrickRed, urlcolor=BrickRed, hypertexnames=true}
\ifmsanon
  \usepackage{endnotes}
  \let\footnote\endnote
  
\fi
\ifsplit
  \ifmsanon
    \ifsupp\else\fi
  \else
    \ifsupp\else\fi
  \fi
\fi
\bibpunct[:]{(}{)}{,}{a}{}{,}
\newtheorem{proposition}{Proposition}

\theoremstyle{definition}

\theoremstyle{plain}

\title{Task-Specific Affinity and Team Assignment:\\
Evidence from Elite Bobsleigh}
\newcommand{\fundingnote}{ This work was supported by JST ERATO JPMJER2301, 25K16620, and JST SPRING JPMJSP2123.}
\ifjhr\renewcommand{\fundingnote}{}\fi
\ifmsanon
\author{}
\date{}
\else
\author{Masaya Nishihata\thanks{\href{mailto:nishihata-masaya@keio.jp}{nishihata-masaya@keio.jp}, Graduate School of Economics, Keio University and Mitsubishi UFJ Research and Consulting Co., Ltd.}
\and Suguru Otani\thanks{\href{mailto:suguru.otani@e.u-tokyo.ac.jp}{suguru.otani@e.u-tokyo.ac.jp}, Department of Economics, The University of Tokyo. An earlier version of this paper circulated under the title \emph{Team for Speed: Nonparametric Evidence on Heterogeneous Skill-Specific Affinity in Team Production}.\fundingnote}}
\date{\today}
\fi

\newcommand{\abstracttext}{Organizations and managers observe collaboration histories, but how these should inform team assignment is unclear when production involves multiple tasks. We ask whether pairs can be ranked with a task-agnostic affinity index or assignment requires task-specific affinity, and characterize when a common ranking suffices. We estimate task-specific production functions and pair affinities in women's bobsleigh, where Start and Riding are timed separately and monobob races measure individual skills. Pair affinity persists within each task, and its Start and Riding components are positively correlated yet differ by task. In counterfactual reassignments, using both tasks' affinities improves predicted performance over either alone.}
\newcommand{\keywordstext}{\textbf{Keywords}: team production, team assignment, team affinity, team-specific human capital, managerial decision making, nonparametric estimation}
\newcommand{\jeltext}{\textbf{JEL code}: J24, D24, D23, C14, M54}

\begin{document}

\ifsupp
\begin{center}
{\LARGE\bfseries \ifmsanon Electronic Companion\else Online Appendix\fi}\\[1ex]
\makeatletter{\large\@title}\makeatother
\ifjhr\\[1ex]{\large Masaya Nishihata and Suguru Otani}\fi
\end{center}
\else
\maketitle

\ifjhr
\begin{center}\bfseries Abstract\end{center}
\begin{doublespace}
\noindent\abstracttext
\end{doublespace}
\vspace{1ex}
\noindent\keywordstext\\
\jeltext
\newpage
\else
\begin{abstract}
\abstracttext
\quad \\
\keywordstext
\ifmsanon\else
\\
\jeltext
\fi
\end{abstract}
\fi

\section{Introduction}

A central question in the economics of team production is whether the productivity of a pair of workers beyond their individual skills is a single attribute of the pair or depends on the task they perform. The answer matters both for how organizations form teams and for how economists measure worker complementarities: theories of matching rank pairs by a scalar match output \citep{becker1973theory} or match quality \citep{jovanovic1979job}, and empirical work on team-specific human capital treats it as a single pair-level stock that accumulates with collaboration \citep{jaravel2018team}, so both treat pair productivity as task-independent. Managers often know who has worked with whom and how those teams performed, but such information is difficult to use when production has several tasks. A natural simplification is to rank each pair with one task-agnostic affinity index. That simplification is useful when the component that makes a pair productive generates a common ranking across tasks. If affinity is task-specific, the same pair can be a good assignment for one part of production and a poor assignment for another, so an index that discards task information can rank pairs differently from the task-specific production objective. Testing whether affinity is task-specific requires output observed separately by task, individual skills measured along the same dimensions, and variation in who works with whom, a combination that standard personnel data rarely provide.

This paper answers that question through the economics of team production. We ask whether unobserved pair affinity is task-specific. We define affinity as the pair-task component of production that remains after conditioning on observed task-specific skills, roles, task environments, and teammate skill. It captures how effectively two workers' skills are jointly converted into output in a given task and may reflect coordination, communication, timing, physical compatibility, psychological fit, or other teamwork inputs not directly observed in standard data. The empirical question is whether the same observable skill composition implies a common affinity ranking across tasks or instead requires task-specific affinity information for assignment.

The distinction matters because a large literature documents that experience working together predicts team performance \citep{reagans2005individual,huckman2009team}, and empirical specifications or assignment heuristics often use that relationship as if it were a task-independent team attribute. A common ranking is adequate, for example, when the two task-affinity matrices are positive affine transformations of one another. It can be inadequate when affinity is task-specific: a pair that coordinates well in an execution-intensive task may not be the best pair in a technical-control task, and high individual skill need not imply high joint productivity. The same issue arises in empirical production analysis. A production function that treats ability as a single index may attribute task-specific frictions to individual skill, while a model that ignores latent affinity may misread heterogeneous coordination as measurement noise.

We provide new evidence using a nonparametric framework for team production with multidimensional skills and asymmetric roles, following \citet{matzkin2003nonparametric} and \citet{lange2020beyond}.\footnote{Applications of related nonparametric identification strategies include labor markets \citep{otani2024nonparametric,otani2024onthejob,kanayama2024nonparametric}, transportation \citep{brancaccio2020geography}, and marriage markets \citep{otani2025nonparametric}.} The method is a tool for the assignment question: it recovers task-specific production functions and latent pair affinities without imposing a Cobb-Douglas, CES, or finite-type structure that would predetermine how skills and affinity interact. The central empirical object is a task-level production function in which one athlete's observed skill is multiplied by a latent affinity term before combining with the partner's skill. Under constant returns to scale and a conditional independence restriction, variation in teammate assignments identifies the distribution of latent affinity and the production technology up to normalization. Descriptive pairing evidence is consistent with the conditional-independence assumption, while Monte Carlo sensitivity results indicate that modest departures from it need not substantially impair affinity recovery.

The same assignment problem arises in production lines, warehouses, and project teams whose output combines distinct tasks. Our application uses elite women's bobsleigh as a high-resolution field setting for this broader problem. The setting has four advantages. First, the task structure is unusually transparent. A race consists of two separately timed phases, which we refer to as the Start and Riding tasks: the Start phase depends heavily on explosive power and synchronized pushing, and the Riding phase depends on steering, line choice, balance, and sled control. Second, roles are sharply asymmetric. In two-woman bobsleigh, Player~1 is the driver and Player~2 is the brakeman. Both athletes push at the start, but the driver controls the sled during the ride while the brakeman's contribution during descent is mainly balance and weight placement. Third, the sport provides separate measures of individual skill. In monobob, a single athlete performs all roles; monobob performances therefore provide a way to estimate individual start and riding skills outside the observed two-woman team. Fourth, the underlying official race-result records are publicly available. Their run-level crew identities and phase times make bobsleigh an illustrative application of granular task-level data to team production and assignment.

The empirical design combines official International Bobsleigh \& Skeleton Federation (IBSF) results from two-woman and monobob events in the 2018--19 through 2024--25 seasons, using partner rotation across teams as the assignment variation for estimation. We residualize two-woman performance using event fixed effects, estimate athlete fixed effects from monobob races with event and starting-order controls, and transform lower race times into positive performance measures so that larger values indicate better output.

Estimated affinity varies substantially within each task--attempt panel, and neither phase follows a simple diagonal ranking in solo skills. 
To organize the assignment question, we embed the production structure in a two-task assignment benchmark. It characterizes when a common pair ranking is sufficient and when a reversal makes the preferred pairing depend on the relative task weights.

Pair affinity is persistent across attempts, with correlations of $0.976$ for Start affinity and $0.902$ for Riding affinity. The persistent Start and Riding affinity components are positively correlated, at $0.321$, but their 95 percent confidence interval $[0.166,0.686]$ excludes perfect correlation. This indicates that task-specific affinities cannot be treated as the same underlying measure expressed on different linear scales. The fitted production surface displays role asymmetry: Player~2's riding elasticities have a smaller median absolute magnitude than Player~1's in both attempts, although the projection provides no uncertainty bands.

We then use counterfactual team reassignments to quantify the value of task-specific affinity information. Holding the performance objective and feasible national athlete pools fixed, we compare assignments based on both tasks' affinities with those using a single task, pooled affinity information, or individual skills alone. Using both tasks improves predicted combined performance by $0.073$ seconds per weighted two-attempt participation relative to assignments that use Start affinity only (Start-only), with a 95 percent confidence interval excluding zero. The corresponding gain over Riding-only is smaller, at $0.017$ seconds, and its confidence interval includes zero. Separate within-event reassignments also show improvements when pairings are changed among the athletes available at the same competition.

The implications extend beyond this setting, for organizations that form teams and for the measurement of worker complementarities. Collaboration histories contain persistent pair-level information that individual skill measures do not, so assignment rules that rely on skills alone forgo fitted performance gains. Because the persistent affinity components differ across tasks, an organization that records who worked with whom should also record how each pair performed in each task rather than keep a single team-level score: in our application, dropping the riding-task history from the assignment rule lowers fitted performance. The gains do not require hiring, since improvements arise even when pairings are rearranged among the workers already present at the same site and time. For the measurement of team-specific human capital and worker complementarities, the same result implies that a single match-quality or familiarity measure omits task-specific pair productivity, since the persistent Start and Riding components are not collinear. In production lines, warehouses, and project teams whose output decomposes into tasks, managers should therefore keep collaboration histories task-specific when assigning teams.

\subsection{Related Literature}

The paper contributes first to research on team production, team design, and the assignment value of pair-history information. Existing studies document learning spillovers, coworker effects, complementarities among workers, or productivity gains from familiarity \citep{falk2006clean, mas2009peers, bandiera2010social, reagans2005individual, huckman2009team, jarosch2021learning, jager2022substitutable, herkenhoff2024production}. Sports and sales settings have been especially useful for measuring individual and team output and for studying role hierarchies within teams \citep{candelon2015hierarchical, arcidiacono2017productivity, cohen2024effort, xu2024heterogeneous}. Recent structural work separates individual contributions to team output \citep{bonhomme2021teams, xu2024heterogeneous}, while experimental work directly measures teamwork ability \citep{weidmann2021team}. Studies of team-specific human capital, in which the premature death of a collaborator lowers the productivity of the coworkers who remain \citep{jaravel2018team}, likewise summarize pair productivity by a single pair-level stock. Relative to this literature, our contribution is to study whether the affinity object used for assignment varies by task even when it is persistent and economically meaningful.

Second, the paper contributes to work on multidimensional skills, skill segregation, task-based production, and assignment. Models of worker assignment increasingly emphasize that workers have multiple skills and that jobs require different task bundles \citep{kremer1996wage,lindenlaub2017sorting,lise2020multidimensional,guvenen2020multidimensional,woessmann2024skills,deming2023multidimensional}. Empirical implementation is often limited by coarse proxies for skills, such as education or occupation. Our setting provides direct, objective, task-specific performance measures. The start and riding phases map naturally into different skill dimensions, while monobob records allow us to measure those skills at the individual level. This makes it possible to ask whether team affinity differs by skill dimension rather than treating all teamwork ability as a single residual, and to illustrate the fitted assignment consequences of alternative information structures.

The rest of the paper proceeds as follows. Section~\ref{sec:framework} presents the empirical framework, the model, and its assignment implications. Section~\ref{sec:data} describes the data, pairing patterns, and descriptive evidence. Section~\ref{sec:estimation} explains the estimation procedure. Section~\ref{sec:results} reports the estimates. Section~\ref{sec:counterfactual} presents the counterfactual assignment exercises, and Section~\ref{sec:conclusion} concludes.

\section{Conceptual and Theoretical Framework}\label{sec:framework}

\subsection{Team Production with Task-Specific Affinity}

Consider a two-person team $\ell$ at time $t$ performing $K$ tasks indexed by $k=1,\ldots,K$. Let $X_{\ell kt}$ denote Player~1's task-$k$ skill input and let $Y_{\ell kt}$ denote Player~2's task-$k$ skill input. In the bobsleigh application, Player~1 is the driver and Player~2 is the brakeman. For each task, team output is
\begin{align}
H_{\ell kt}=m_k(A_{\ell kt}X_{\ell kt},Y_{\ell kt}),\label{eq:task_specific_team_production}
\end{align}
where $m_k(\cdot,\cdot)$ is a task-specific production function, strictly increasing in both inputs on the relevant support, and $A_{\ell kt}$ is a latent team-task-specific affinity term. Conceptually, $A_{\ell kt}$ captures the unobserved part of a pair that changes how effectively Player~1's task-specific skill is converted into team output conditional on Player~2's task-specific skill. It is the input-augmenting residual that remains after conditioning on observed skills, and it absorbs any pair-specific factor not captured by those skills.

The model is intentionally not a full equilibrium model of team formation. In many settings, search costs make it infeasible to evaluate every potential pair, while availability, mobility costs, institutional restrictions, managerial discretion, and historical pairings constrain assignment. We therefore avoid imposing equilibrium sorting or assuming that observed teams solve a planner's problem. We treat observed pairings as the assignment process through which the econometrician sees variation in $(X_{\ell kt},Y_{\ell kt},H_{\ell kt})$ and use that variation to recover production and affinity. The assignment problem below is a decision benchmark under these constraints, not a claim that the historical process was globally optimal.

Two assumptions organize the empirical strategy. First, each $m_k(\cdot,\cdot)$ satisfies constant returns to scale (CRS) on the transformed positive performance scale used in the estimation. Because the raw data are race times, the production objects below are not raw seconds: they are the output-like indices constructed in Section~\ref{sec:data}, where larger values mean better performance. CRS should therefore be read as a scale restriction on this transformed index, not as a claim that homogeneity is preserved by arbitrary affine transformations of raw times. Under CRS,
\begin{align}
\frac{H_{\ell kt}}{X_{\ell kt}}=m_k\left(A_{\ell kt},\frac{Y_{\ell kt}}{X_{\ell kt}}\right).
\end{align}
The performance transformation introduced in Section~\ref{sec:data} constructs positive performance measures, with larger values indicating better performance and $X_{\ell kt}>0$, $Y_{\ell kt}>0$, and $H_{\ell kt}>0$ throughout the estimation sample. We maintain positive latent affinity, $A_{\ell kt}>0$, so that the ratio representation and the multiplicative form $A_{\ell kt}X_{\ell kt}$ are well defined. CRS is weaker than imposing a particular parametric form and allows flexible complementarities between the effective Player~1 input and the Player~2 input, but all such statements are made on the transformed performance scale.

Second, latent affinity is conditionally independent of Player~2 skill given Player~1 skill:
\begin{align}
A_{\ell kt}\perp Y_{\ell kt}\mid X_{\ell kt}.\label{eq:conditional_independence}
\end{align}
In the bobsleigh application, this restriction requires that, conditional on the driver's observed task-specific skill, the brakeman's observed task-specific skill is not systematically related to unobserved pair-task affinity. Pair formation is constrained by nationality rules, roster depth, and event availability, while coaches may also observe aspects of pair compatibility that the econometrician does not. Conditional independence is therefore a substantive assumption rather than a mechanical consequence of the sport's rules. We assess its plausibility using descriptive pairing patterns and examine the sensitivity of affinity recovery to departures from it in Monte Carlo simulations.\footnote{Appendix~\ref{app:pairing_selection} provides descriptive evidence that pair continuation does not increase systematically with estimated affinity, and that first-observed pairs have modestly higher Start affinity but no statistically significant Riding difference. These patterns are consistent with the maintained identifying assumption. Separately, Appendix~\ref{app:monte_carlo} shows that modest departures from conditional independence need not substantially worsen affinity-recovery accuracy in the simulated designs.}

\subsection{Task Additivity and Interpretation}

Additive task structures arise whenever total time or output is assembled from separately measurable stages, as in production lines, warehouse operations, or project work divided into subtasks. Let $H_{\ell t}$ denote the aggregate performance index for run $t$, defined as the sum of the transformed task-level performance indices. We write
\begin{align}
H_{\ell t}=\sum_{k=1}^{K}H_{\ell kt}
=\sum_{k=1}^{K}m_k(A_{\ell kt}X_{\ell kt},Y_{\ell kt}).\label{eq:additive_tasks}
\end{align}
Additivity requires only that aggregate performance be representable as the sum of separately measured task indices. Equation~\eqref{eq:task_specific_team_production} does, however, require task-$k$ output to depend on task-$k$ inputs only; in the application, any carryover from the start phase into riding performance is therefore absorbed in the riding-phase affinity term. This decomposition allows us to estimate task-specific production relationships and compare latent affinity across tasks.

The interpretation of $A_{\ell kt}$ is task-specific. A high $A_{\ell kt}$ means that the pair produces more output in task $k$ than the model attributes to its observed skills, under the chosen normalization; it does not imply uniformly high affinity in every task. Under a common-ranking structure, a pair with high affinity in one task would also rank highly in the other. With task-specific affinity, the same pair can have high affinity in one task and low affinity in another.

The framework also clarifies the role of Player~2. The latent term multiplies Player~1's input for identification and normalization reasons, not because Player~2 is conceptually unimportant. Player~2 enters the production function directly through $Y_{\ell kt}$, and the marginal return to Player~2 skill can differ across tasks.

\subsection{Task-Specific Affinity and Team Assignment}\label{sec:theory_assignment}

The team-assignment question is whether task-level collaboration histories can be compressed into one task-agnostic pair ranking without changing assignment choices. We abstract from moral hazard and treat affinity as a technological pair component rather than unobserved effort. We then specialize to $K=2$ and index tasks by $k\in\{1,2\}$.\footnote{The benchmark is a linear assignment problem in the tradition of \citet{koopmans1957assignment}, \citet{becker1973theory}, and \citet{sattinger1993assignment}; see \citet{chade2017sorting} for a survey. The moral-hazard abstraction follows \citet{alchian1972production} and \citet{holmstrom1982moral}.}

Let $\mathcal I$ and $\mathcal J$ be two fixed role sets with $n$ members each, indexed by $i$ and $j$. A feasible perfect matching is a bijection $\sigma:\mathcal I\to\mathcal J$. Let $A_k(i,j)$ denote the persistent, assignment-relevant component of pair $(i,j)$'s affinity in task $k$. We define the first-order assignment benchmark around reference skill levels by
\begin{align}
v(i,j)=f(i)+g(j)+\omega_1 A_1(i,j)+\omega_2 A_2(i,j),\label{eq:pair_value}
\end{align}
where $f$ and $g$ collect additively separable individual contributions and $\omega_k>0$ is the task-$k$ marginal value of affinity at the reference skills, common across candidate pairs. The manager chooses $\sigma$ to maximize $\sum_{i\in\mathcal I}v(i,\sigma(i))$. Because $\sigma$ is a bijection, $\sum_{i\in\mathcal I}[f(i)+g(\sigma(i))]=\sum_{i\in\mathcal I}f(i)+\sum_{j\in\mathcal J}g(j)$ for every feasible matching. Assignment in this benchmark is therefore governed by the two affinity matrices.

For distinct $i,i'\in\mathcal I$ and $j,j'\in\mathcal J$, define the task-$k$ assortativity gap
\begin{align}
\Delta_k=A_k(i,j)+A_k(i',j')-A_k(i,j')-A_k(i',j),\qquad k\in\{1,2\}.\label{eq:assortativity_gap}
\end{align}
$\Delta_k>0$ means that task-$k$ affinity favors $\{(i,j),(i',j')\}$ over $\{(i,j'),(i',j)\}$; $\Delta_k<0$ means the reverse. A reversal, $\Delta_1\Delta_2<0$, is the simplest case in which the preferred pairing depends on the relative task weights.

\begin{proposition}[Task-Agnostic Affinity Rankings Can Be Insufficient]\label{prop:assignment}
(i) Suppose the two affinity matrices have the common-ranking structure $A_2(i,j)=\eta+bA_1(i,j)$ for every pair, where $\eta$ is constant across pairs and $b>0$. Any matching that maximizes $\sum_{i\in\mathcal I}A_1(i,\sigma(i))$ is then optimal for every $(\omega_1,\omega_2)>0$, and no reversal exists. (ii) If a two-by-two subroster exhibits a reversal, then, holding all other assignments fixed, there are two strictly positive technology-weight vectors under which the two sub-matchings are each strictly optimal in turn. Consequently, no technology-invariant choice between the two sub-matchings is optimal for every positive weight vector. At given weights, the loss from choosing the wrong sub-matching is $\lvert\omega_1\Delta_1+\omega_2\Delta_2\rvert$.
\end{proposition}

Part (ii) is a local exchange result; whether a reversal changes the roster-wide optimum also depends on the rest of the roster. For a fixed technology, the correctly weighted task-aware value $\omega_1A_1+\omega_2A_2$ is itself a scalar. The proposition rules out only a task-agnostic rule whose choice, including tie-breaking, is fixed across technology weights. The condition in part (i) can be weakened to $A_2(i,j)=\eta_i+\eta'_j+bA_1(i,j)$ with role-specific constants $\eta_i$ and $\eta'_j$: the separable terms are constant across feasible matchings, as in Equation~\eqref{eq:pair_value}, and cancel in every $\Delta_k$, so the same conclusions hold.

\subsection{Empirical Mapping}\label{sec:theory_to_measurement}

In the bobsleigh application, task 1 is start performance and task 2 is riding performance, while $\mathcal I$ contains drivers and $\mathcal J$ contains brakemen. We capitalize Start and Riding when they refer to the two tasks. From this point onward, $S$ and $R$ denote tasks 1 and 2: for example, $A_S\equiv A_1$, $\omega_S\equiv\omega_1$, and $\Delta_S\equiv\Delta_1$, with analogous definitions for $R$. The phase-level indices follow the additive accounting convention in Equation~\eqref{eq:additive_tasks}.

To connect assignment-relevant affinity to the two observed attempts, index a pair by $p=(i,j)$ and let $Z_{pka}$ be its pair-level estimated affinity in task $k$ from attempt $a\in\{1,2\}$, measured up to a positive affine transformation within each task. We use the repeated-measurement model
\begin{align}
Z_{pka}=P_{pk}+\varepsilon_{pka},\qquad P_{pk}\equiv A_k(i,j),\label{eq:repeated_measurement}
\end{align}
where $P_{pk}$ is the persistent component used in the assignment benchmark and $\varepsilon_{pka}$ is an attempt-specific deviation that may include transitory conditions and affinity-recovery error. Across pairs, assume finite second moments, $\operatorname{E}[\varepsilon_{pka}\mid P_{p1},P_{p2}]=0$, $\operatorname{Cov}(\varepsilon_{pk1},\varepsilon_{ph2})=0$ for all $k,h\in\{1,2\}$, and $\operatorname{Var}(\varepsilon_{pk1})=\operatorname{Var}(\varepsilon_{pk2})$ for each task $k$. These conditions allow the two task errors within the same attempt to be correlated.

Define
\begin{align}
r_k&=\operatorname{Corr}(Z_{pk1},Z_{pk2}),\qquad k\in\{1,2\},\nonumber\\
c&=\frac{1}{2}\left\{\operatorname{Corr}(Z_{p11},Z_{p22})+\operatorname{Corr}(Z_{p12},Z_{p21})\right\}.\label{eq:repeated_measurement_moments}
\end{align}
If $\operatorname{Var}(P_{pk})>0$ for both tasks, the repeated-measurement conditions imply
\begin{align}
\rho_P\equiv\operatorname{Corr}(P_{p1},P_{p2})=\frac{c}{\sqrt{r_1r_2}}.\label{eq:persistent_correlation}
\end{align}
Under the nondegenerate common-ranking restriction in Proposition~\ref{prop:assignment}(i), $P_{p2}=\eta+bP_{p1}$ with $b>0$ and $\operatorname{Var}(P_{p1})>0$, so
\begin{align}
\rho_P=1.\label{eq:common_ranking_correlation}
\end{align}
The estimated persistent-component correlation therefore evaluates this exact positive-affine restriction among observed pairs under the repeated-measurement conditions. A value below one does not identify a unique decomposition of affinity into common and task-specific components. Under the weaker pair-separable form of the common-ranking condition, $\rho_P$ is generally below one even though no reversal exists; the persistent-component correlation therefore evaluates the exact positive-affine restriction, whereas the two-by-two reversal exercise in Section~\ref{sec:counterfactual} is invariant to separable task differences.

The estimated production functions and affinities also enter the assignment exercises in Section~\ref{sec:counterfactual}. Those exercises evaluate the estimated nonseparable projection directly rather than Equation~\eqref{eq:pair_value}. For candidate pairs never observed in the data, they additionally require conditional exchangeability with observed pairs and sufficient common support, and they impute conditional mean affinity given driver skill. These assumptions are not implied by Equation~\eqref{eq:conditional_independence}.

\section{Data and Institutional Background}\label{sec:data}

We apply the framework of Section~\ref{sec:framework} to elite women's bobsleigh, which serves as a high-resolution field setting for a broader team-assignment problem. The setting is useful because output is measured precisely, the production process decomposes cleanly into two phases with distinct skill content, and the same athletes compete both alone in monobob and in specialized two-person roles, so individual skills and latent match affinity can be separated. The contribution is accordingly methodological and substantive rather than sport-specific: the lessons are meant to transfer to team settings in which output can be decomposed by task, individual skills can be measured along corresponding dimensions, and organizations or managers must decide whether pair-history information should be summarized by a single pair-level object or kept task-specific.

\subsection{Competition Design}

Figure~\ref{fg:bobsleigh_overview} summarizes the two competition phases and the athletes' roles. Bobsleigh competitions, in both the monobob and two-woman formats, divide each run into two distinct and separately timed phases: a start phase and a riding phase, each emphasizing a different skill set. In the start phase, the athletes push the sled from a standing position to generate initial momentum and then load into it. This phase rewards explosive pushing power, acceleration, and precise timing, and in two-woman events it additionally rewards coordination between the two athletes, because synchronized pushing and a clean, well-timed transition into the sled minimize the time lost at loading. Because start time is recorded separately from the subsequent descent, it provides a distinct measure of propulsion and loading performance, although the momentum generated at the start carries into the riding phase.

\begin{figure}[htbp]
  \begin{center}
    \includegraphics[width=0.90\textwidth]{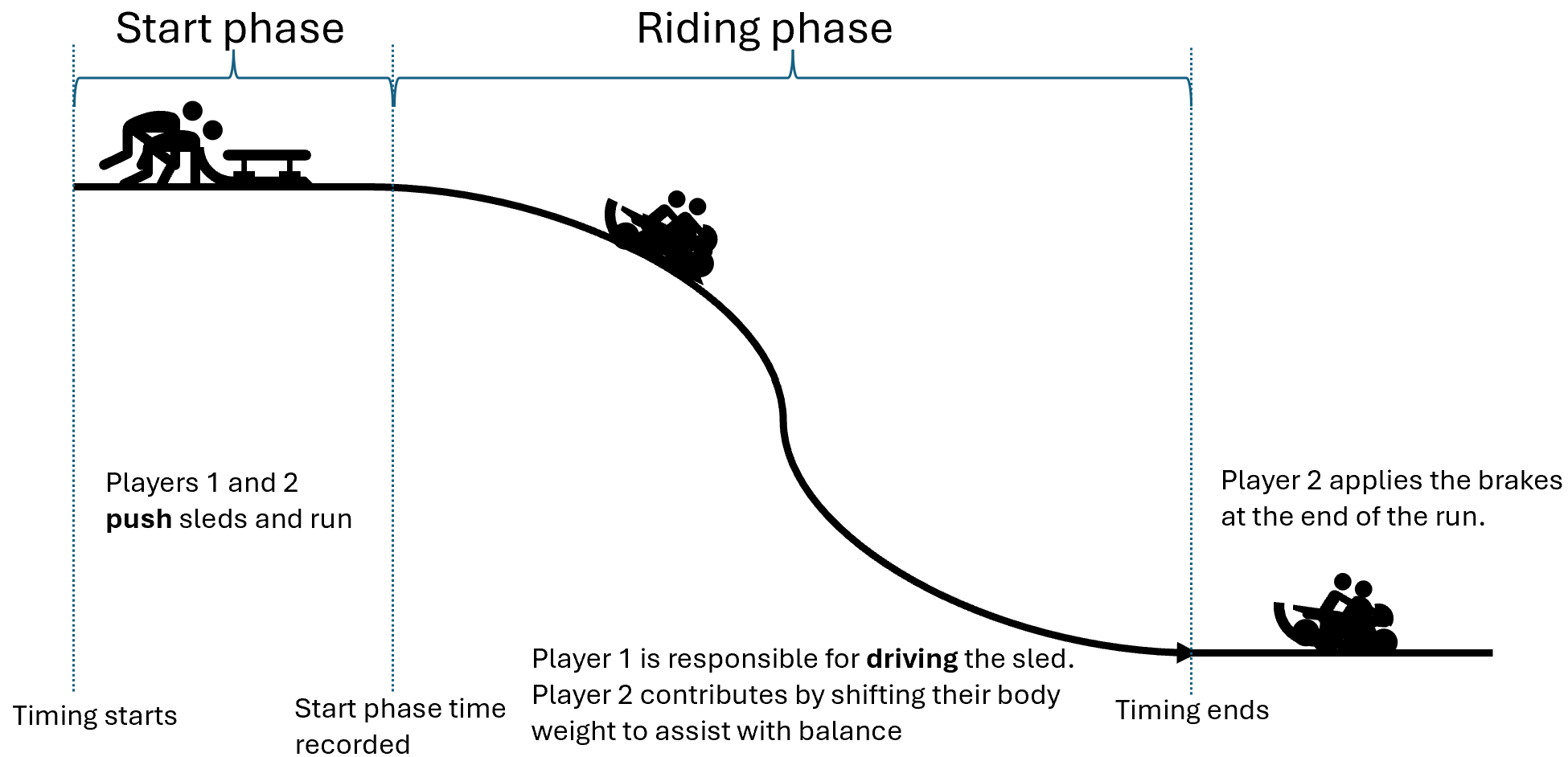}
    \caption{Bobsleigh Competition Design}
    \label{fg:bobsleigh_overview}
  \end{center}
  {\footnotesize \textit{Note}: The figure summarizes the two-woman competition design: both athletes push during the start phase, Player~1 drives during the riding phase, and Player~2 contributes to balance before braking after the timed run.}
\end{figure}

Once the sled completes the timed Start section, the Riding phase begins. Speed during the descent reflects both the momentum inherited from the start and the athletes' riding skill, which comprises steering precision, course navigation, balance, and the maintenance of an aerodynamic line. In monobob a single athlete manages every task, driving and braking while holding her form. In two-woman bobsleigh the driver steers the sled through the descent, while the brakeman contributes to balance and weight distribution during the ride and applies the brakes at the end. Both the inherited momentum and technical riding ability therefore shape performance, and in the team event coordination again matters through the management of weight and movement during the descent.

The two-woman format thus embodies a clear division of labor: the driver specializes in controlling the sled during the descent, while the brakeman specializes in initial propulsion and final braking. Crucially, the final braking occurs after the sled crosses the finish line and the official time has already been recorded, so the brakeman's measured contribution during the ride operates primarily through mass and balance rather than through steering or timing. Monobob, by contrast, requires the same athlete to combine the explosive strength of the start with the fine motor control of the ride.

Several institutional features help separate skill and coordination from other sources of variation. Outcomes are recorded in standardized time units, and strict regulations govern the total weight of the sled and crew as well as equipment and race procedures, limiting equipment- and body-mass heterogeneity so that measured performance is driven primarily by athlete skill and coordination. Final rankings are determined by cumulative time across attempts rather than by any single run. For comparability, the baseline construction retains the first two observed attempts from each crew--event record.\footnote{Appendix~\ref{app:robustness} reports a robustness check restricted to crew--event records with exactly two observed attempts.}

Competition across teams follows a fixed starting protocol. In the first attempt, teams start in ascending order of their assigned start numbers; in the second attempt, the order is reversed according to the first-attempt ranking, so that slower teams run earlier on fresher ice.

\subsection{Data Source and Sample Construction}

The primary data source is the publicly available official results archive of the International Bobsleigh \& Skeleton Federation (IBSF), accessed on May 5, 2025.\footnote{\url{https://www.ibsf.org/en/races-and-results}} We digitize race-level results from two-woman and monobob events, including World Cup and Olympic competitions. For each run we collect the start time, the finish time, intermediate splits, athlete identities, nationalities, and starting order. The standardized format of the IBSF records ensures consistency across events and supports a reliable decomposition of each run into its start and riding components. The archive thus provides the granular run-level crew and phase-timing information needed to construct task-specific team outcomes across events.

The empirical analysis combines two data components. The first is the monobob data, used to measure individual task-specific skill. Because a monobob athlete performs both the start and the ride by herself, her monobob results reveal individual start ability, riding ability, and overall performance uncontaminated by a partner. The second is the two-woman data, used to estimate team production; for these races we observe the driver, the brakeman, the attempt, and phase-level team performance. Merging the two components restricts the main team-production sample to athletes who also have monobob records, because the production-function inputs are individual skill measures. This requirement keeps the estimation sample modest in size, but it is what allows individual skill to be separated from latent match affinity.

We estimate individual skills from monobob races using athlete, event, and starting-order fixed effects. We adjust two-woman team outcomes for event effects. We then transform the adjusted times into positive performance measures so that larger values indicate better performance. To retain a scale comparable to observed times, we define
\begin{align}
P(\tau)=2\bar{\tau}-\tau,\label{eq:positive_transform}
\end{align}
where $\tau$ denotes the adjusted time and $\bar{\tau}$ its mean for the corresponding measure in the matched estimation sample. Means are defined separately by task and role for individual inputs and by task and attempt for team output. The resulting measures retain seconds as units and are strictly positive in our estimation sample. Shifting the worst performance arbitrarily close to zero can create near-zero inputs and make ratio-based affinity recovery numerically unstable. CRS is imposed on these performance indices rather than inherited from raw race times.

\subsection{Individual Skills from Monobob Fixed Effects}\label{sec:individual_fixed_effect}

A key input to the analysis is the vector of individual skill fixed effects. We estimate these fixed effects from solo monobob competition data for start and riding performance, which are the two task inputs used in the production-function estimates. We also estimate finish-performance fixed effects as a descriptive aggregate performance measure. For athlete $i$ in event $e$, starting order $s$, and run $t$, we estimate
\begin{align}
Y_{iest}=\alpha_i+\mu_e+\rho_s+\varepsilon_{iest},\label{eq:individual_fe}
\end{align}
where $Y_{iest}$ is the raw time outcome, $\alpha_i$ is an athlete fixed effect, $\mu_e$ is an event fixed effect, and $\rho_s$ is a starting-order fixed effect. Event fixed effects absorb track, weather, and competition-specific conditions. Starting-order fixed effects control for sequence-related variation in ice conditions and race procedure.

The athlete fixed effects provide phase-specific measures of individual skill. After restoring the common time level, we apply the transformation in Equation~\eqref{eq:positive_transform} to construct the skill inputs $X_{\ell kt}$ and $Y_{\ell kt}$. The same individual skills are used for both attempts. Finish performance combines the two phases, so it is reported only as a descriptive aggregate and is not added as a third input to the phase-specific production functions.

The monobob fixed-effect sample covers international IBSF women's monobob events from the 2018--2019 through 2024--2025 seasons. The current data construction retains the first recorded run for each athlete--event, yielding 1,729 athlete--event observations across 143 athletes. The fixed-effect sample includes all athletes with usable women's monobob records; 18 athletes appear in one event, and the 59 athletes matched to the two-woman sample appear in between 1 and 53 events.\footnote{Appendix Table~\ref{tb:monobob_events_runs_matched_athletes_summary} reports athlete--event support for the full and matched samples.} The matched two-woman team-production sample for the same seasons contains 59 athletes, 63 observed driver--brakeman pairs, and 260 team participations, observed for the first and second attempts. Table~\ref{tb:summary_statistics_player_level} reports its athlete- and participation-level summary statistics. Fixed effects for sparsely observed athletes are less precisely estimated, and the baseline production estimates treat these estimated inputs as fixed.\footnote{Appendix Table~\ref{tb:skill_pairing_start} reports the corresponding observed-pair support.}

\subsection{Summary Statistics and Skill Measures}\label{sec:summary_skills}

Table~\ref{tb:summary_statistics_player_level} reports player-, team-, and nationality-level summaries, and Figure~\ref{fg:cdf_residualized_player_riding_time} plots the two solo-skill distributions. These displays document the scale and support of the analysis. Table~\ref{tb:summary_statistics_player_level}, panel (c), shows that Canada and the United States contribute relatively many matched athletes and team participations, while several countries appear only a few times. Comparisons of dispersion across start, riding, and finish times are partly mechanical because the phases differ in duration. We therefore do not interpret cross-phase variance differences as evidence about skill heterogeneity.

\begin{table}[!htbp]
  \begin{center}
    \caption{Summary Statistics}
    \label{tb:summary_statistics_player_level}
    \begin{subtable}{\textwidth}
      \centering
      \caption{Player-level}
      \begin{adjustbox}{max width=\textwidth}
        \input{figuretable/summary_statistics_player_level}
      \end{adjustbox}
    \end{subtable}

    \vspace{0.75em}
    \begin{subtable}{\textwidth}
      \centering
      \caption{Team-level}
      \begin{adjustbox}{max width=\textwidth}
        \input{figuretable/summary_statistics_team_level}
      \end{adjustbox}
    \end{subtable}

    \vspace{0.75em}
    \begin{subtable}{\textwidth}
      \centering
      \caption{Number of Players and Team Participations by Nationality}
      \begin{adjustbox}{max width=\textwidth}
        \input{figuretable/num_of_players_team_plays}
      \end{adjustbox}
    \end{subtable}
  \end{center}
  {\footnotesize \textit{Note}: Players in the main sample are athletes with recorded monobob performances that can be matched to two-woman bobsleigh events. Performance measures are defined in Equation~\eqref{eq:positive_transform} and expressed in seconds, with larger values indicating better performance.}
\end{table}

\begin{figure}[htbp]
  \begin{center}
    \begin{subfigure}{0.48\textwidth}
      \centering
      \includegraphics[width=\textwidth]{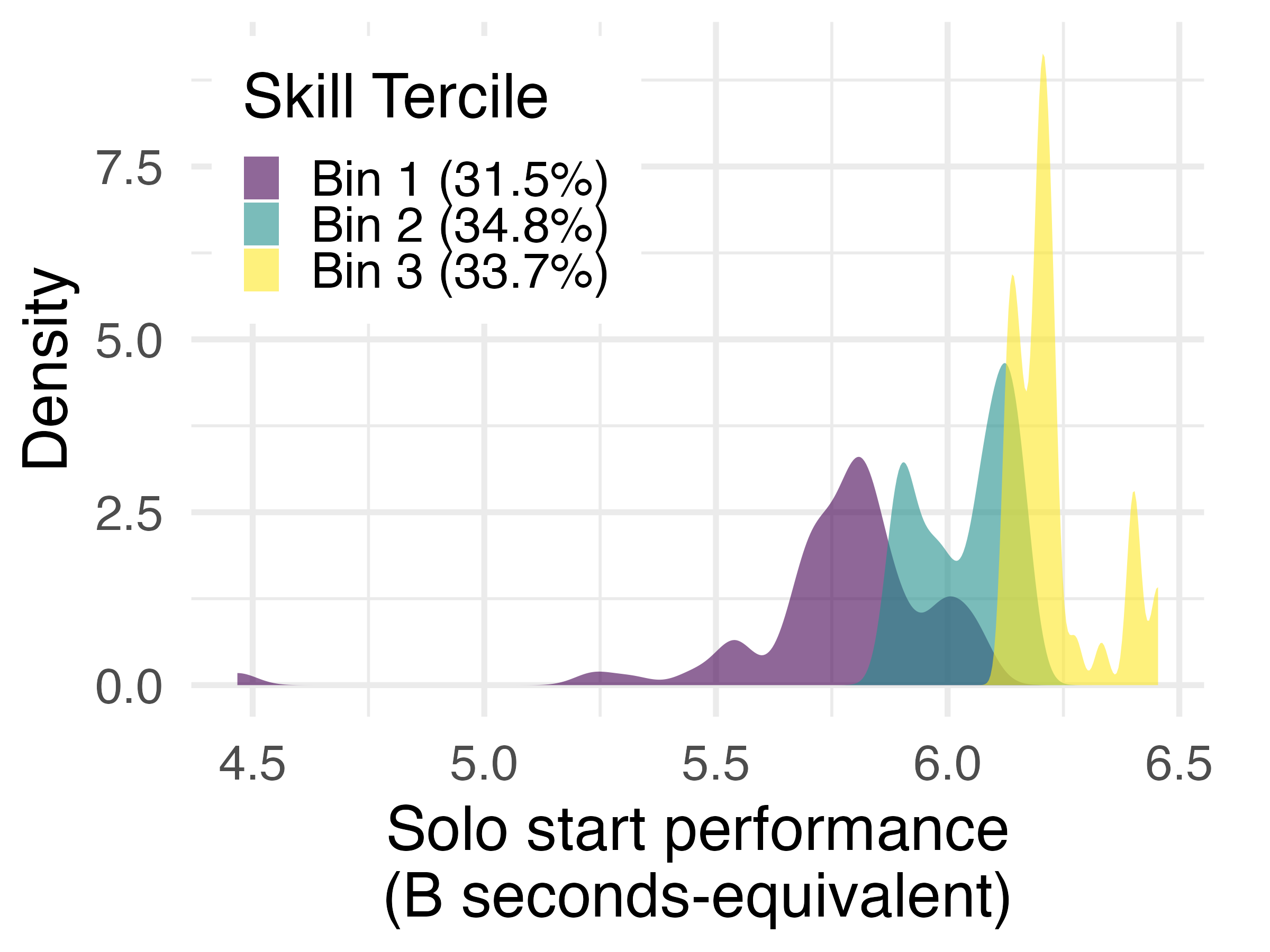}
      \caption{Start performance}
    \end{subfigure}
    \hfill
    \begin{subfigure}{0.48\textwidth}
      \centering
      \includegraphics[width=\textwidth]{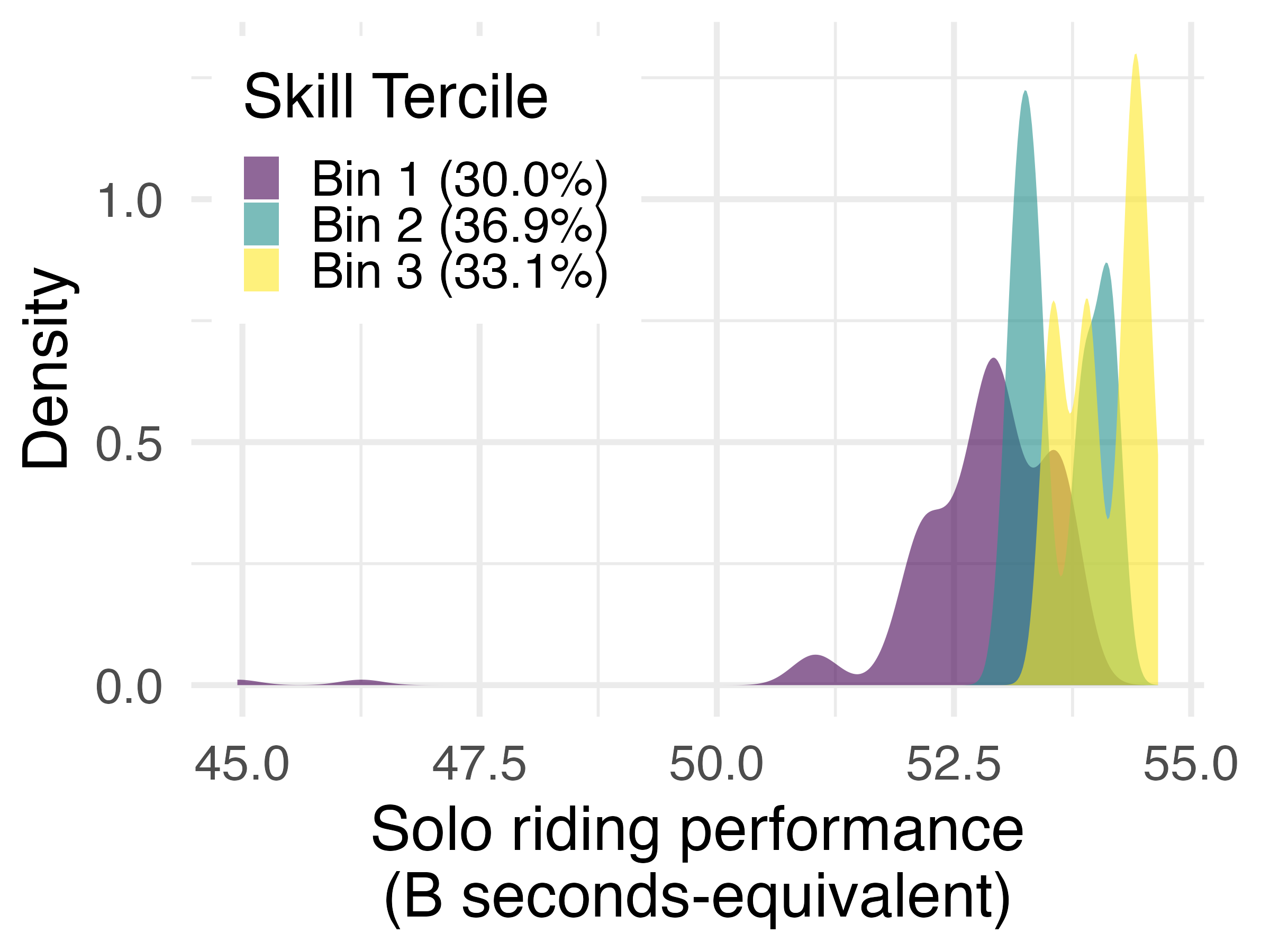}
      \caption{Riding performance}
    \end{subfigure}
    \caption{Residualized Solo Performance as Individual Skill}
    \label{fg:cdf_residualized_player_riding_time}
  \end{center}
  {\footnotesize \textit{Note}: Each panel displays kernel densities by participation-weighted skill tercile. The 260 team participations contribute two athlete--role observations each, so the densities use 520 athlete--role participation records and give more weight to athletes with more team appearances. Skills are performance measures constructed from monobob fixed effects using Equation~\eqref{eq:positive_transform}, expressed in seconds with larger values indicating faster performance.}
\end{figure}

\FloatBarrier

A natural question is whether start and riding skills are distinct or simply manifestations of general athletic ability. The start--riding skill correlation is 0.437 among the 143 athletes with women's monobob records and 0.334 in the matched two-woman sample.\footnote{Appendix Figure~\ref{fg:scatter_start_vs_riding_skill} and Appendix Table~\ref{tb:start_riding_skill_correlation_summary} report these correlations.} The weaker relationship in the matched sample may partly reflect restricted range, because the athletes who appear in two-woman events are a selected subset of the full monobob sample. The evidence does not establish that start and riding skills are fully separable. It shows only that the two fixed effects are not perfectly collinear in the analysis sample, which permits task-specific estimation while allowing both measures to contain a common component such as general athletic ability.

Identification benefits from variation in observed pairings. If the data contained only high-skill drivers paired with high-skill brakemen and low-skill drivers paired with low-skill brakemen, the estimator would have limited information to separate skill from affinity. Observed pairings span both diagonal and off-diagonal skill cells, providing variation in observed support, although this dispersion does not establish the conditional-independence restriction.\footnote{Appendix Tables~\ref{tb:skill_pairing_start}--\ref{tb:skill_pairing_run_counts_cross} report unique-pair and participation counts by skill cell. Participation-weighted skill terciles are assigned once within athlete, role, and task and then joined to the pair and participation data.}

\subsection{Descriptive Evidence} \label{sec:descriptive}

Before estimating the model, we examine descriptive residualized team performance across skill bins. Figure~\ref{fg:heatmap_start_time_2} reports participation-weighted averages by task and attempt using the fixed athlete--role--task terciles. 
%
These observed patterns cannot by themselves identify matching affinity. A bright off-diagonal cell could reflect true pair-specific coordination, nonlinear production in observed skills, sampling noise, or endogenous pairing by coaches. Similarly, a dark high-high cell could reflect poor latent affinity, a bad run, or unobserved assignment constraints. The descriptive evidence therefore motivates the estimator: we need a framework that conditions on observed skill inputs and recovers the residual match component as task-specific latent affinity, subject to explicit identification assumptions.

\begin{figure}[htbp]
  \begin{center}
    \begin{subfigure}{0.47\textwidth}
      \centering
      \includegraphics[width=\textwidth]{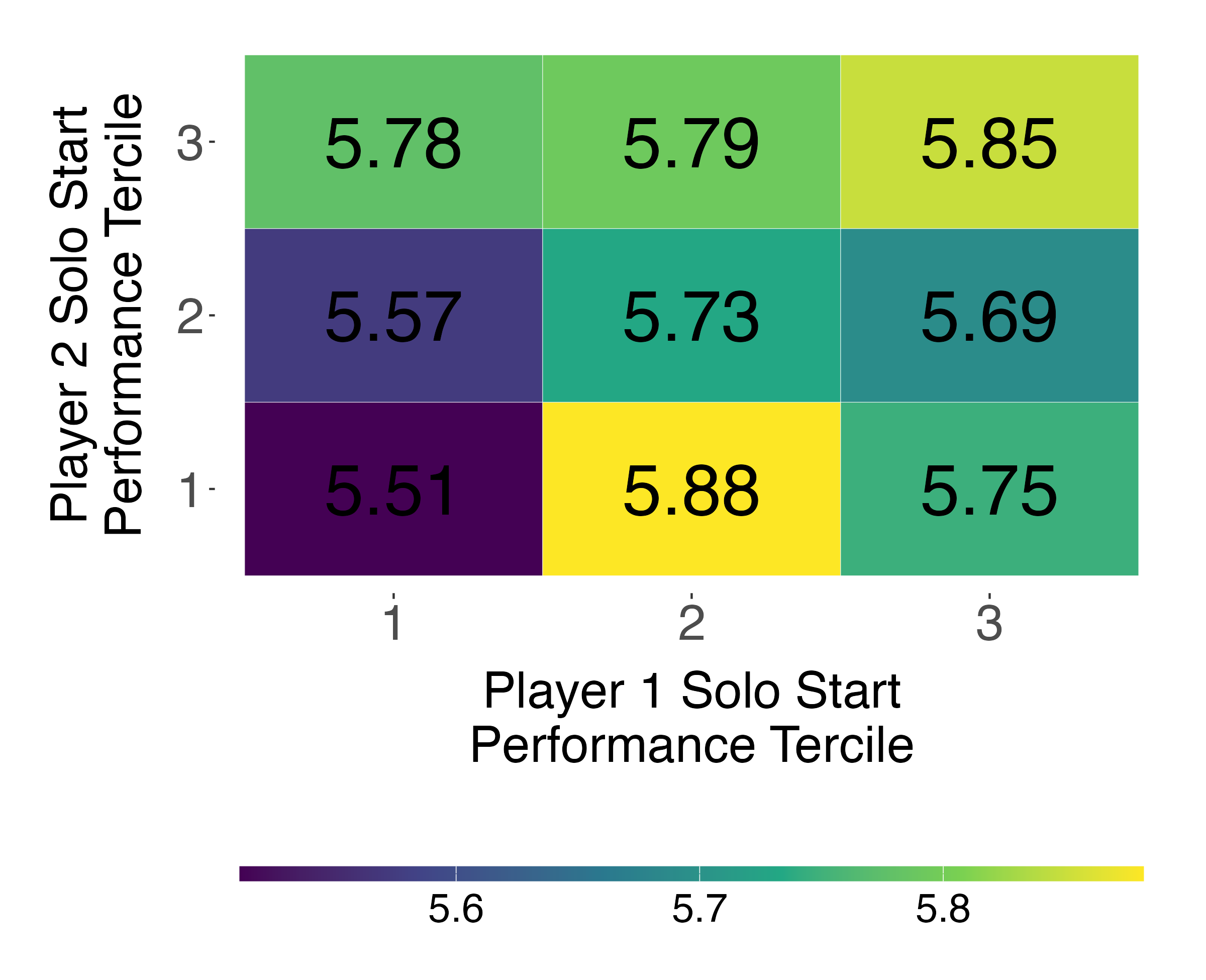}
      \caption{Start performance, first attempt}
    \end{subfigure}
    \hfill
    \begin{subfigure}{0.47\textwidth}
      \centering
      \includegraphics[width=\textwidth]{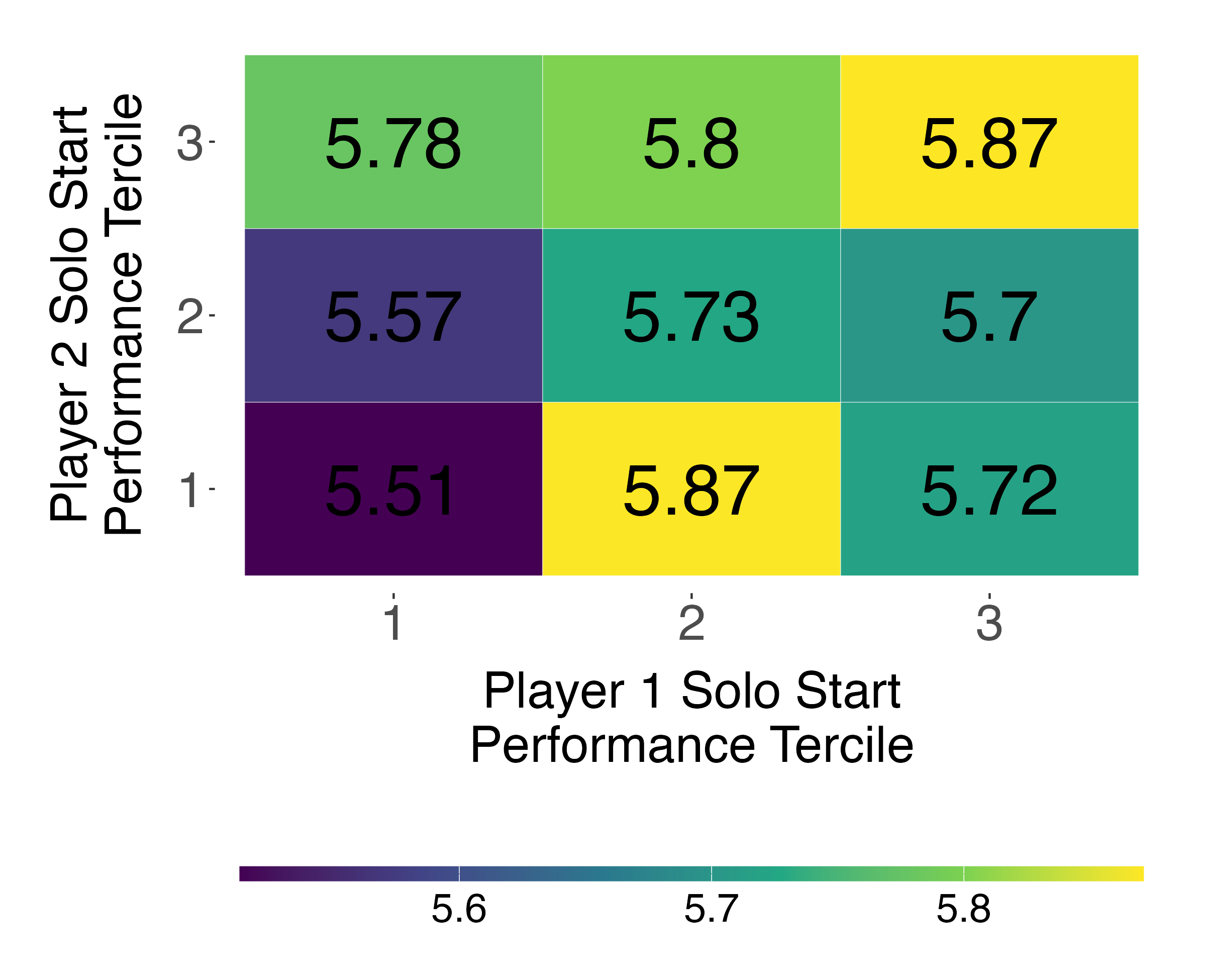}
      \caption{Start performance, second attempt}
    \end{subfigure}

    \vspace{0.75em}
    \begin{subfigure}{0.47\textwidth}
      \centering
      \includegraphics[width=\textwidth]{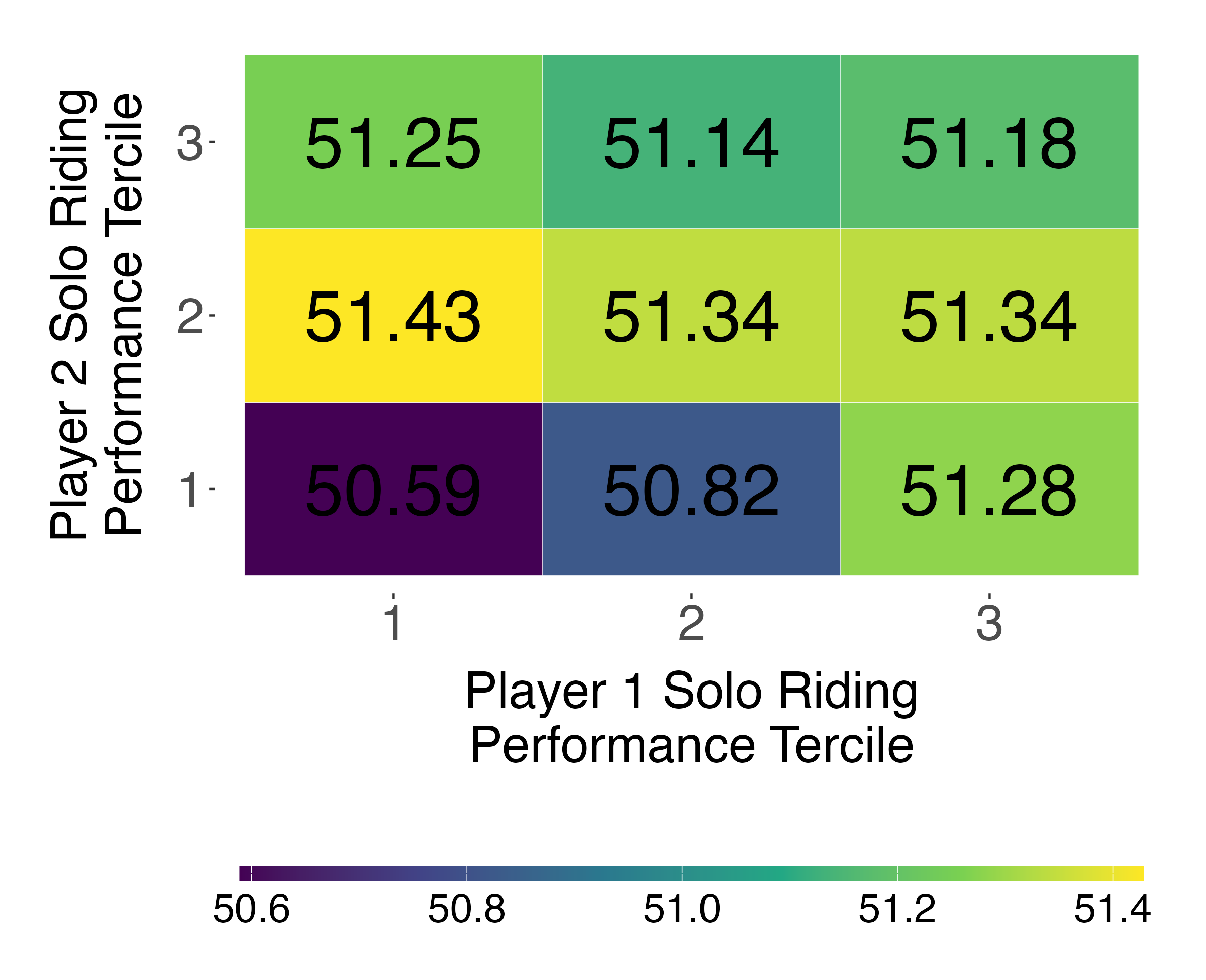}
      \caption{Riding performance, first attempt}
    \end{subfigure}
    \hfill
    \begin{subfigure}{0.47\textwidth}
      \centering
      \includegraphics[width=\textwidth]{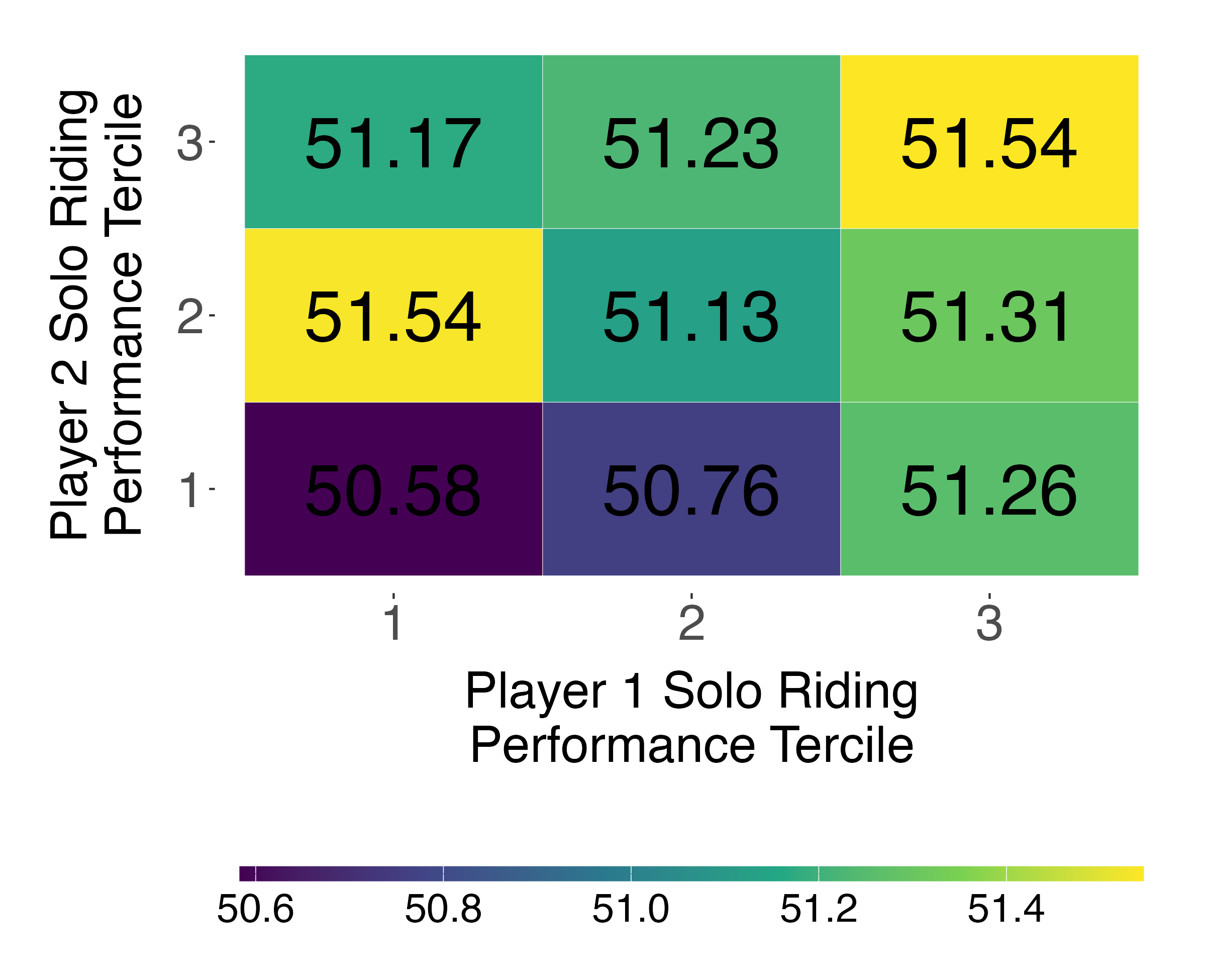}
      \caption{Riding performance, second attempt}
    \end{subfigure}
    \caption{Average Residualized Team Start and Riding Performance Conditional on Individual Skill-Bin Pairs}
    \label{fg:heatmap_start_time_2}
  \end{center}
  {\footnotesize \textit{Note}: 1 = lowest displayed performance tercile; 3 = highest displayed performance tercile. First and second attempts refer to the first and second runs within an event. Cell values are averages of the positive team performance index defined in Equation~\eqref{eq:positive_transform}, measured in seconds, with larger values indicating better performance; the color scale is set separately within each panel. Participation-weighted skill terciles are assigned once within athlete, role, and task, so repeated participations do not change bin membership.}
\end{figure}

\FloatBarrier

\section{Estimation}\label{sec:estimation}

\subsection{Nonparametric Recovery of Affinity}

Motivated by the irregular observed patterns in Section~\ref{sec:descriptive}, we recover task-specific affinity from conditional output ranks. For each task $k$, transformed output satisfies Equation~\eqref{eq:task_specific_team_production}. We assume that $m_k$ is strictly increasing, continuously differentiable, and CRS on the transformed performance scale.\footnote{Constant returns to scale give the ratio normalization $H_{\ell kt}/X_{\ell kt}=m_k(A_{\ell kt},\,Y_{\ell kt}/X_{\ell kt})$.} Let $G_k(H\mid X,Y)$ denote the conditional distribution of output for task $k$ given observed inputs. The recovery below is implemented through the conditional distribution $G_k(H\mid X,Y)$ and rank inversion, rather than by treating the skill ratio alone as the conditioning variable. This logic follows the nonparametric identification arguments in \citet{matzkin2003nonparametric} and \citet{lange2020beyond}.

We estimate conditional output distributions using kernel-weighted empirical distributions. For a reference point $(H_0,X_0,Y_0)$, the estimator takes the form\footnote{Here $K_h(\cdot,\cdot)$ is a bivariate Gaussian kernel. Within each task--attempt panel, $Y_0$ is the participation-weighted median of Player~2 skill, and $(H_0,X_0)$ is taken from an observation whose $Y$ is closest to that median. We set $A_0=100$ and write $\lambda=X/X_0$ and $\psi=A/A_0$. Bandwidths are 0.05 times the respective input ranges, with the Player~1 range bounded below by one.}
\begin{align}
\widehat F_{A,k}(\psi A_0\mid \lambda X_0)
&=\widehat G_k(\psi\lambda H_0\mid \lambda X_0,\psi\lambda Y_0) \notag\\
&=
\frac{
\sum_i \mathbbm{1}\{H_i\leq\psi\lambda H_0\}
K_h(X_i-\lambda X_0,Y_i-\psi\lambda Y_0)
}{
\sum_i K_h(X_i-\lambda X_0,Y_i-\psi\lambda Y_0)
}.\label{eq:kernel_estimator}
\end{align}
In finite samples, $\widehat F_{A,k}$ need not be monotone in affinity. We apply increasing rearrangement in affinity at fixed Player~1 skill, following \citet{chernozhukov2009improving}, and denote the result by $\widehat F^{\uparrow}_{A,k}$.\footnote{Rearrangement and inversion use the numerical interval $[1,1000]$; rearrangement is uniformly weighted in affinity. This interval is not an assumed economic support for latent affinity.} We then recover affinity through its generalized inverse:
\begin{align}
\hat{A}_{\ell kt}=c\,(\widehat F^{\uparrow}_{A,k})^{-1}\left(\widehat G_k(H_{\ell kt}\mid X_{\ell kt},Y_{\ell kt})\mid X_{\ell kt}\right).\label{eq:affinity_recovery}
\end{align}
Here $c>0$ is a panel-specific normalization factor chosen so that the selected reference observation has reported affinity 100. The normalization sets the scale of affinity, so the level of $\hat{A}_{\ell kt}$ should be interpreted relative to the chosen baseline rather than as an absolute physical quantity.

We repeat the procedure separately for the start and riding tasks and for each attempt. This yields estimates of task-specific affinity and production functions. Since task output is observed separately, estimation can proceed task by task. The resulting implied affinity heatmaps summarize average estimated affinity over the observed support of Player~1 and Player~2 skills.

\subsection{Elasticity Estimation}

To summarize marginal returns to individual skill inputs, we approximate the estimated production function using a second-order polynomial projection. For each task, we regress observed team output $H$ on terms in the effective Player~1 input $AX$, the Player~2 input $Y$, and their interactions:
\begin{align}
m(AX,Y)\approx \beta_0+\beta_1(AX)+\beta_2(AX)Y+\beta_3Y+\beta_4(AX)^2+\beta_5Y^2.\label{eq:polynomial}
\end{align}
This flexible approximation yields local derivatives of $m(AX,Y)$ with respect to each input. Let $\widehat m(AX,Y)$ denote the fitted value from Equation~\eqref{eq:polynomial}. The elasticity with respect to Player~1 skill is
\begin{align}
\frac{d\log \widehat m(AX,Y)}{d\log X}
&=\frac{\partial \widehat m}{\partial (AX)}\cdot A\cdot \frac{X}{\widehat m(AX,Y)} \notag\\
&=\left(\widehat\beta_1+\widehat\beta_2Y+2\widehat\beta_4AX\right)\frac{AX}{\widehat m(AX,Y)} \notag\\
&=\frac{d\log \widehat m(AX,Y)}{d\log(AX)},\label{eq:elasticity_x}
\end{align}
and the elasticity with respect to Player~2 skill is
\begin{align}
\frac{d\log \widehat m(AX,Y)}{d\log Y}
=\left(\widehat\beta_2(AX)+\widehat\beta_3+2\widehat\beta_5Y\right)\frac{Y}{\widehat m(AX,Y)}.\label{eq:elasticity_y}
\end{align}
These elasticities are descriptive summaries of the estimated production surface. They should not be interpreted as causal effects of training unless changes in skill can be exogenously induced. Their purpose here is to show whether the production function assigns larger marginal returns to one role or phase.

\section{Results}\label{sec:results}

\subsection{Task-Specific Affinity}

Figure~\ref{fg:efficiency_implied_riding_2} reports estimates of the task-specific latent affinity defined in Equation~\eqref{eq:task_specific_team_production}, recovered under the maintained CRS, monotonicity, and conditional-independence restrictions. The panels display implied affinity between Player~1 and Player~2 for the start and riding phases, separately for first and second attempts. Both axes run from lower to higher solo performance. White denotes the reported reference affinity of 100; red denotes higher implied affinity and blue denotes lower implied affinity.

\begin{figure}[htbp]
  \begin{center}
    \begin{subfigure}{0.47\textwidth}
      \centering
      \includegraphics[width=\textwidth]{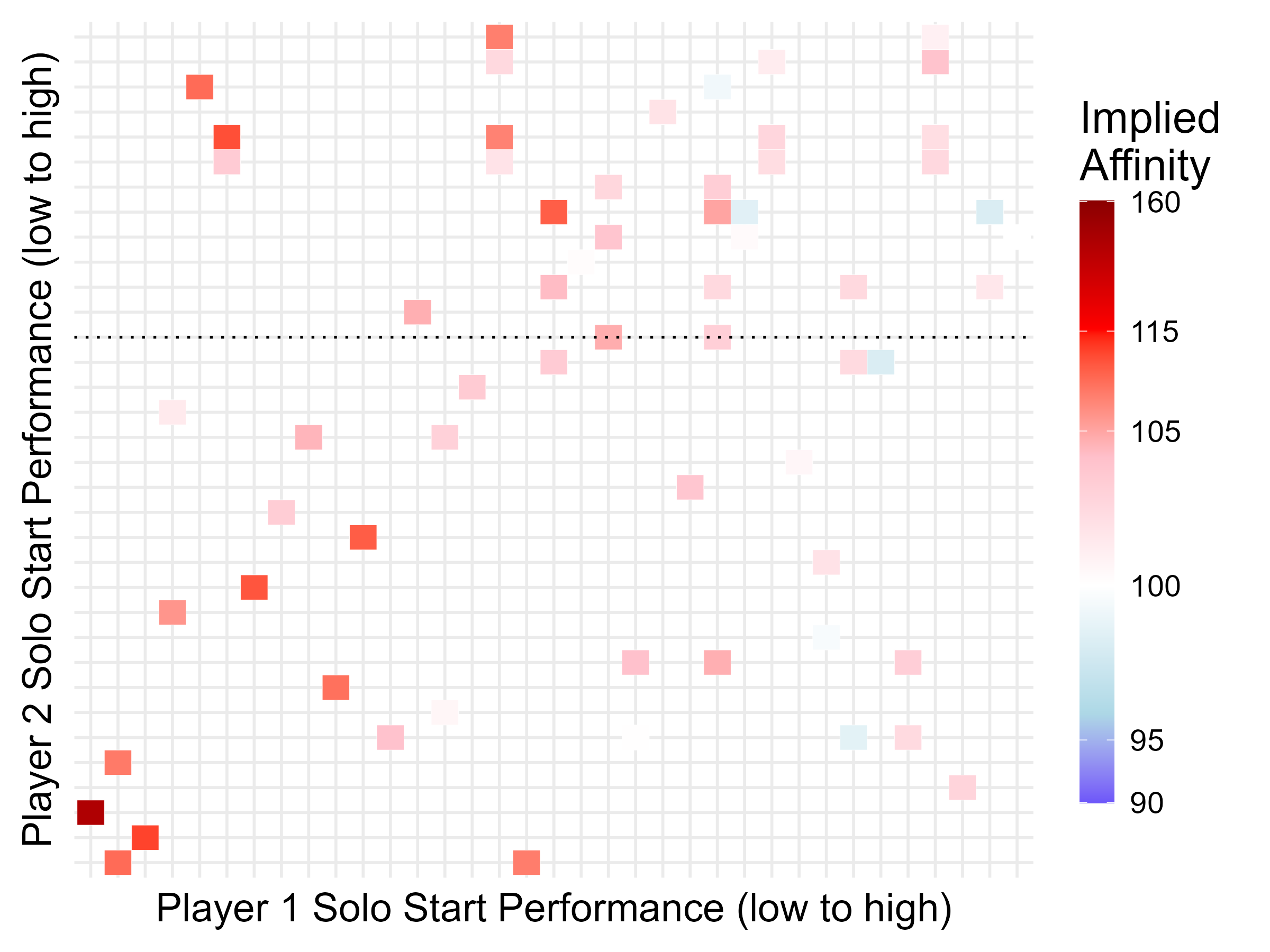}
      \caption{Start phase, first attempt}
    \end{subfigure}
    \hfill
    \begin{subfigure}{0.47\textwidth}
      \centering
      \includegraphics[width=\textwidth]{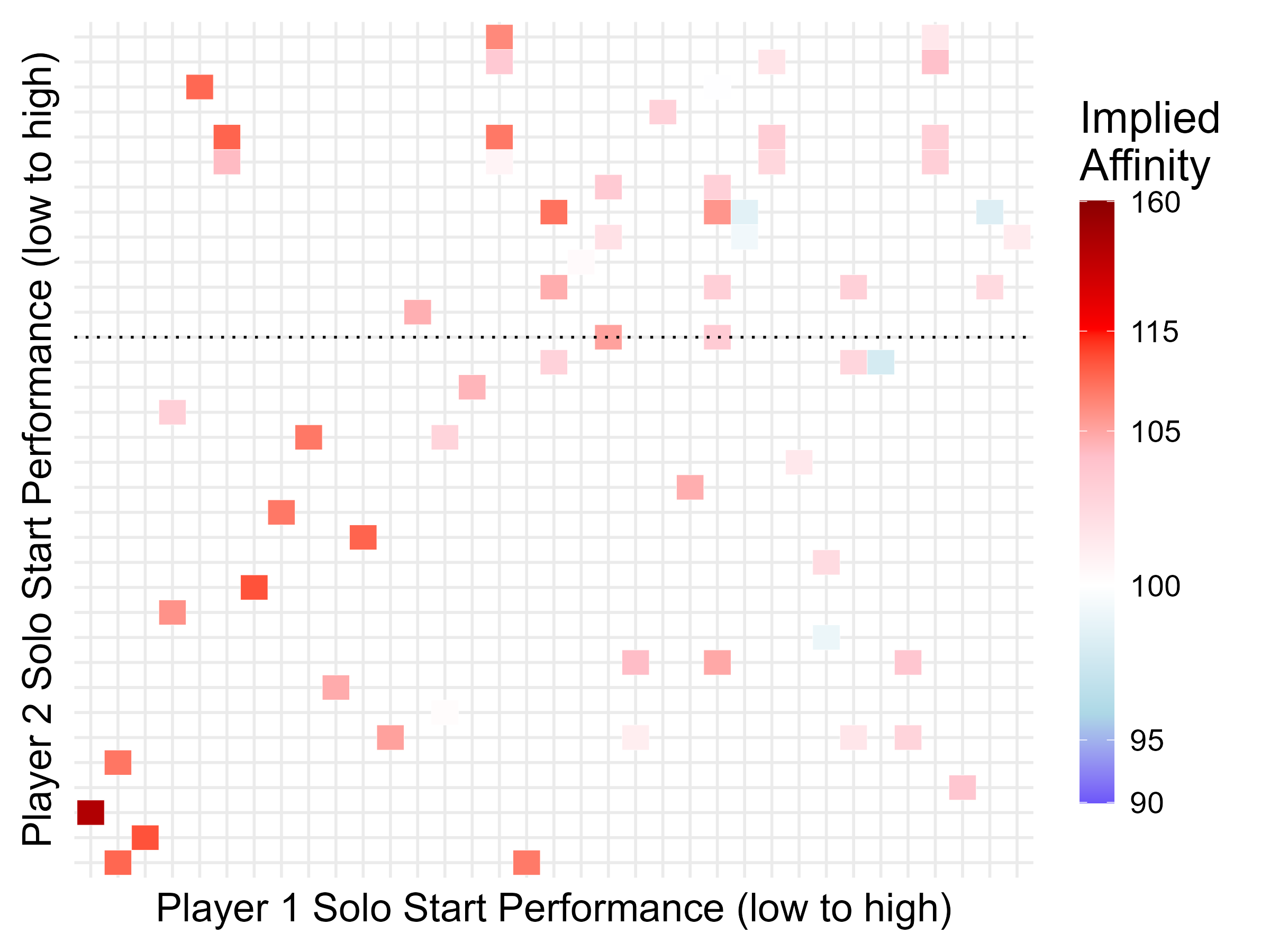}
      \caption{Start phase, second attempt}
    \end{subfigure}

    \vspace{0.75em}
    \begin{subfigure}{0.47\textwidth}
      \centering
      \includegraphics[width=\textwidth]{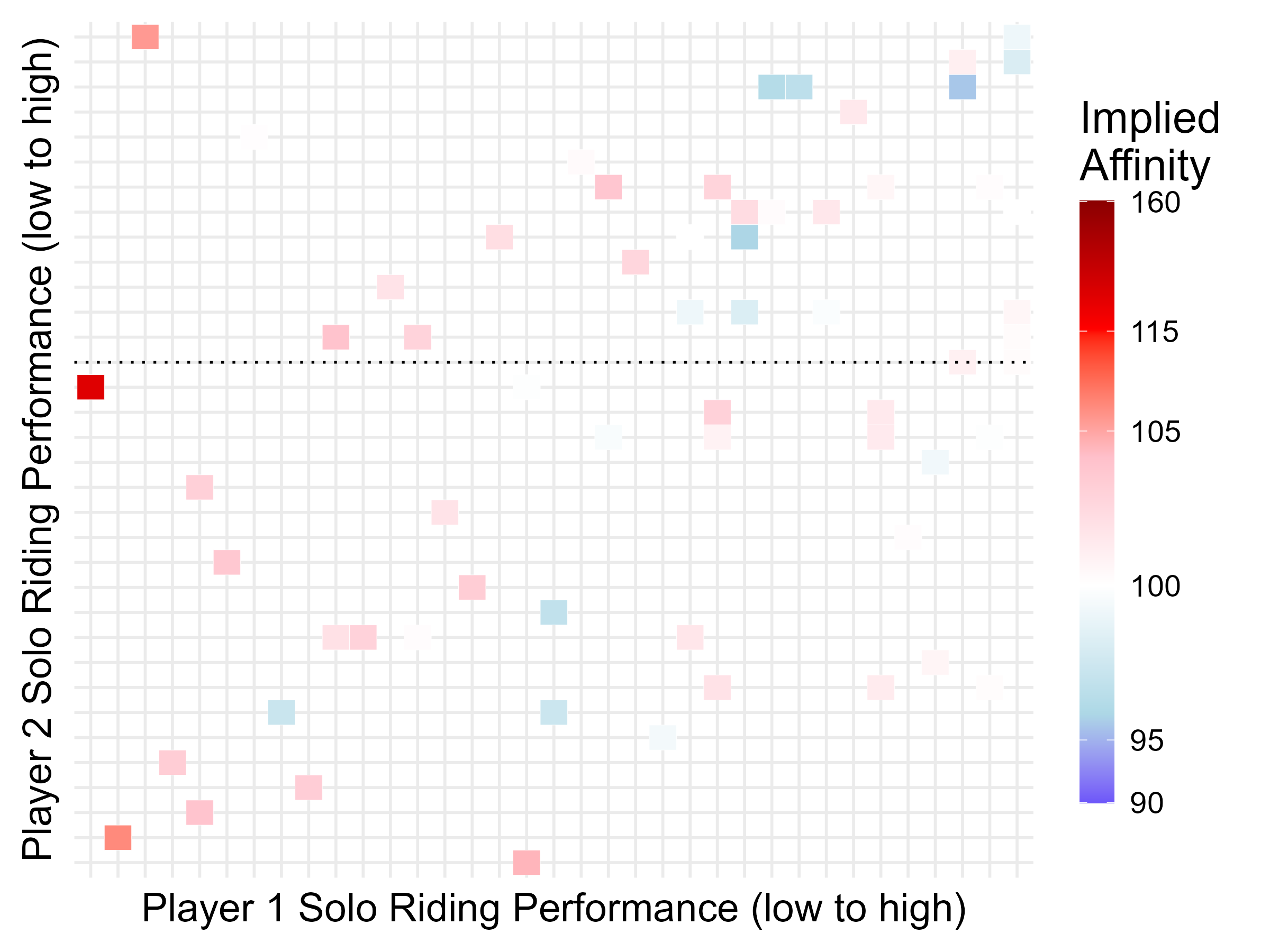}
      \caption{Riding phase, first attempt}
    \end{subfigure}
    \hfill
    \begin{subfigure}{0.47\textwidth}
      \centering
      \includegraphics[width=\textwidth]{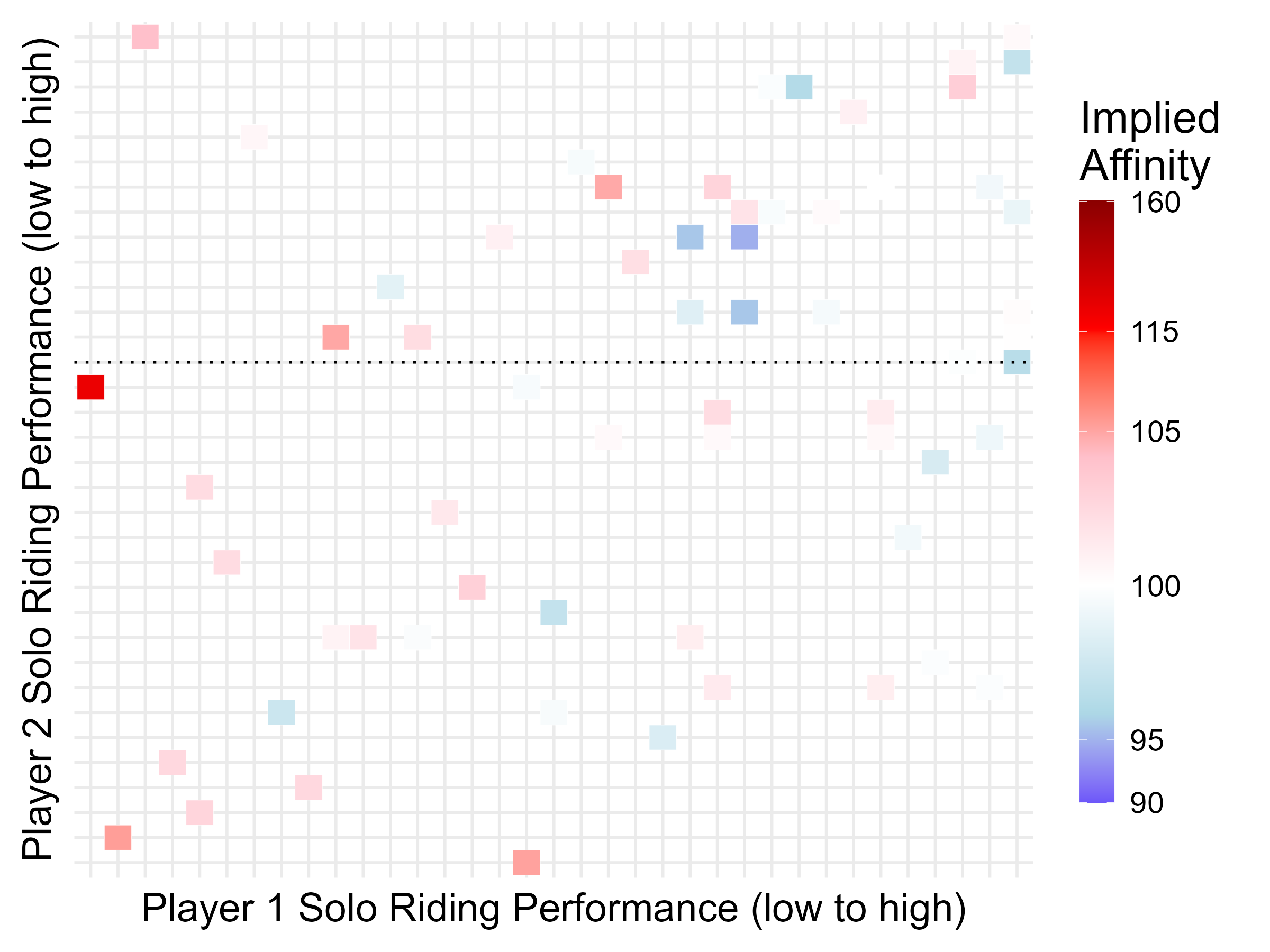}
      \caption{Riding phase, second attempt}
    \end{subfigure}
    \caption{Implied Affinity between Player 1 (Driver) and Player 2 (Brakeman)}
    \label{fg:efficiency_implied_riding_2}
  \end{center}
  {\footnotesize \textit{Note}: The dotted line marks the Player~2 skill of the normalization observation, whose reported affinity is set to 100 (white). Axes are ordered by fixed effects estimated from solo performance after event and starting-order controls. Red indicates higher implied affinity and blue indicates lower implied affinity. First and second attempts refer to the first and second runs within an event. Because each panel has its own normalization point, raw levels are not compared across panels or samples.}
\end{figure}

Figure~\ref{fg:efficiency_implied_riding_2} shows substantial variation within each task--attempt panel over the ordered individual-skill support. The pattern is nonmonotonic, and neither task follows a simple diagonal skill ranking. 
We next examine the common-ranking restriction implied by the positive-affine relationship in Proposition~\ref{prop:assignment}(i).\footnote{Appendix Tables~\ref{tb:summary_statistics_implied_efficiency} and \ref{tb:implied_affinity_structure_summary} report pair-level dispersion and skill-cell contrasts for the 63 observed pairs. The high-high, low-low, diagonal, and off-diagonal contrasts use the fixed within-role task-specific terciles.} Because implied affinity is normalized separately within each panel, raw means are not comparable across panels. Under Equation~\eqref{eq:task_specific_team_production}, the riding panels (c) and (d) display pair-specific affinity conditional on the two solo riding skills; they do not identify whether the underlying effort is symmetric.

\FloatBarrier

\subsection{Do Start and Riding Affinity Share a Common Ranking?}\label{sec:signal_correlation}

This subsection evaluates Proposition~\ref{prop:assignment}(i): under its nondegenerate positive-affine common-ranking restriction, Equation~\eqref{eq:common_ranking_correlation} requires the persistent-component correlation to equal one. Figure~\ref{fg:efficiency_implied_correlation_2} plots run-level start affinity against run-level riding affinity by attempt. Run-level correlations are suggestive but do not isolate persistent affinity, because start and riding measurements from the same run share run-specific conditions that can generate comovement even when the persistent pair components are unrelated.

\begin{figure}[htbp]
  \begin{center}
    \begin{subfigure}{0.47\textwidth}
      \centering
      \includegraphics[width=\textwidth]{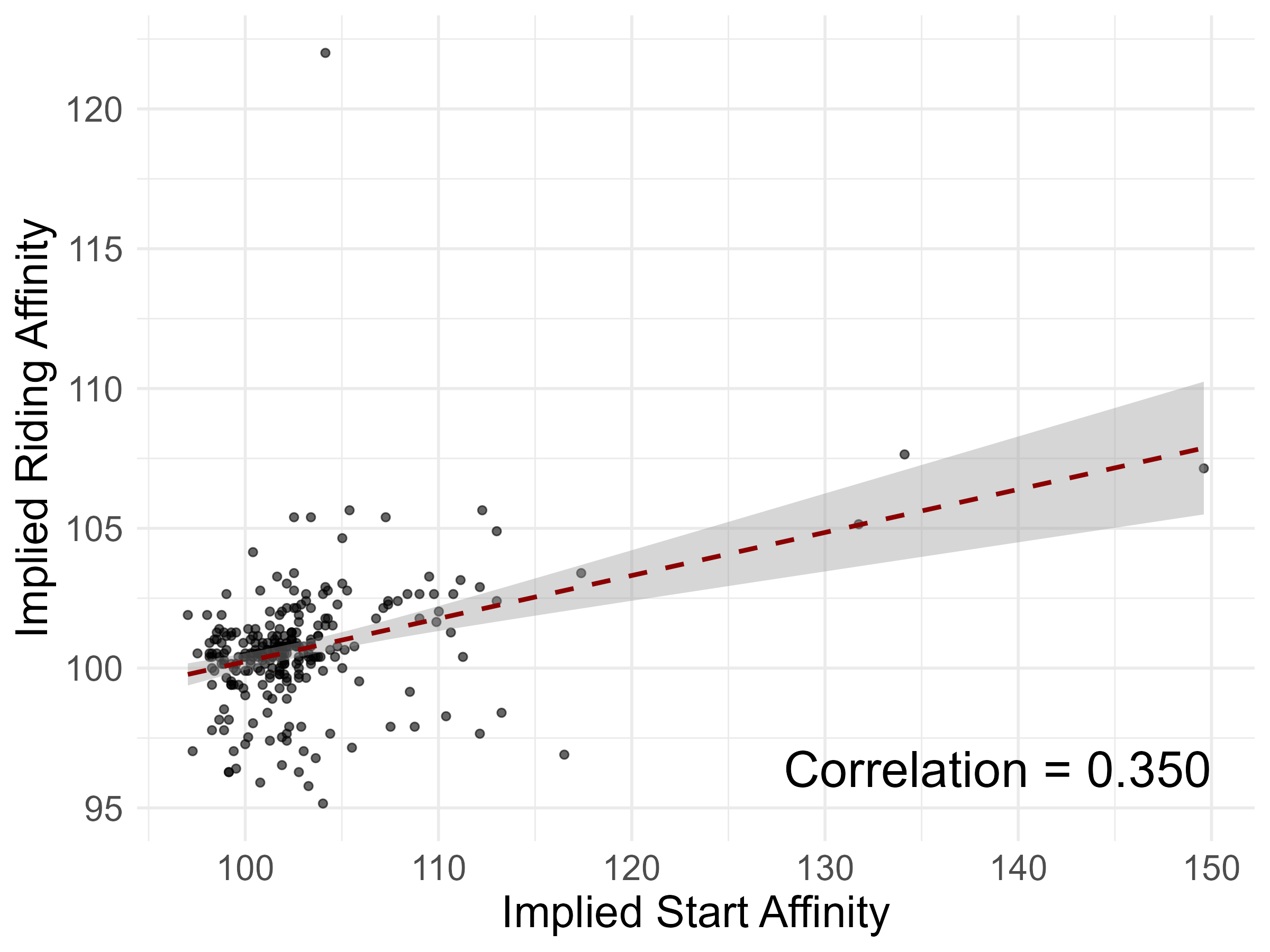}
      \caption{First attempt}
    \end{subfigure}
    \hfill
    \begin{subfigure}{0.47\textwidth}
      \centering
      \includegraphics[width=\textwidth]{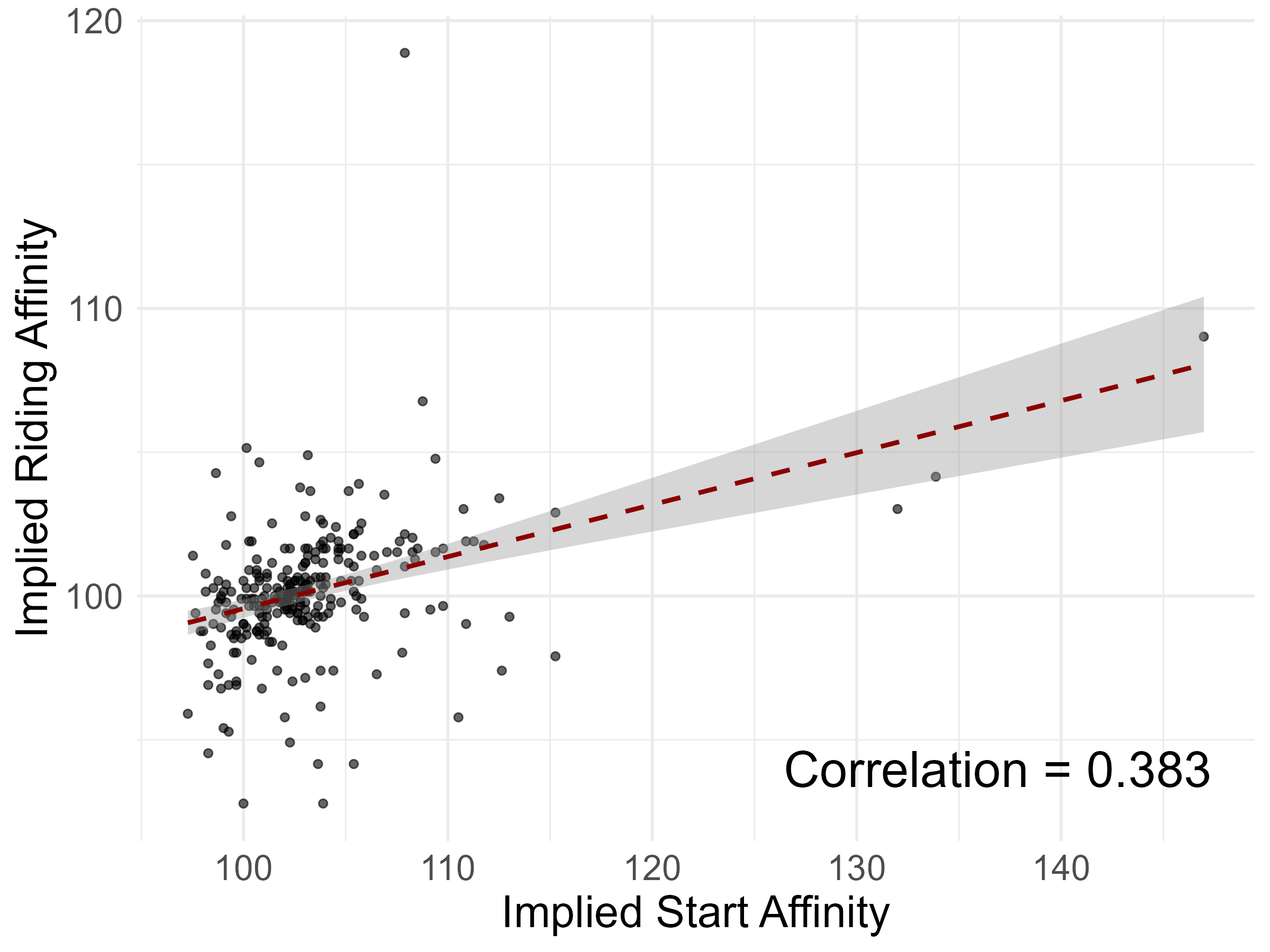}
      \caption{Second attempt}
    \end{subfigure}
    \caption{Run-Level Correlation between Start and Riding Affinity}
    \label{fg:efficiency_implied_correlation_2}
  \end{center}
  {\footnotesize \textit{Note}: Each point matches start and riding affinity for the same run using event, nation, start number, driver, and brakeman; both axes show the full observed range. Affinity is normalized relative to the median Player~2 skill baseline. Run-level comovement may partly reflect run-specific conditions common to both phases; the persistent-component correlations are in Table~\ref{tb:affinity_signal_correlation}}.
\end{figure}

Table~\ref{tb:affinity_signal_correlation} reports attempt-to-attempt correlations of $0.976$ for Start affinity and $0.902$ for Riding affinity. The average cross-task correlation is $0.289$ within attempts and $0.301$ across attempts. Adjusting the latter for within-task persistence yields a correlation of $0.321$ between the persistent Start and Riding affinity components.

\begin{table}[!htbp]
  \begin{center}
    \caption{Repeated-Attempt and Cross-Task Correlations of Pair-Level Affinity}
    \label{tb:affinity_signal_correlation}
    \begin{adjustbox}{max width=\textwidth}
      \input{figuretable/affinity_signal_correlation}
    \end{adjustbox}
  \end{center}
  {\footnotesize \textit{Note}: All statistics are computed on pair-mean implied affinities for the 63 observed directed pairs, using the two attempts as repeated measurements of the same pair-task component. The cross-attempt cross-task statistic is the average of $\operatorname{Corr}(S_1,R_2)$ and $\operatorname{Corr}(S_2,R_1)$. The persistent-component correlation divides this average by the geometric mean of the two within-task correlations under the repeated-measurement conditions in Section~\ref{sec:theory_to_measurement}. Correlations are invariant to positive affine panel rescaling; the persistent-component interpretation additionally assumes that the task--attempt normalizations preserve this correlation scale. Confidence intervals are based on 2,000 pair-cluster bootstrap replications that re-estimate affinity recovery conditional on the estimated individual skills and constructed performance measures.}
\end{table}

The 95 percent conditional affinity-recovery bootstrap confidence interval for the persistent correlation is $[0.166,0.686]$. Under the maintained repeated-measurement assumptions, this interval excludes both zero and the value of one implied by the positive-affine common-affinity restriction in Proposition~\ref{prop:assignment}(i).

\FloatBarrier

\subsection{Performance Elasticities by Role}\label{sec:role_asymmetry}

This subsection reports how the fitted performance elasticities differ between Player~1 and Player~2. Figure~\ref{fg:bobsleigh_overview} summarizes the institutional role distinction. During the start phase both athletes push, so affinity may capture coordination in synchronized joint effort. During the riding phase the driver directly controls the sled, while the brakeman contributes through weight placement, balance, aerodynamic posture, and compatibility with the driver's line choice.

Figure~\ref{fg:elasticity_riding_time_player_1} reports performance elasticities with respect to individual skill inputs, defined in Equations~\eqref{eq:elasticity_x} and \eqref{eq:elasticity_y}. An elasticity is the percentage change in fitted team performance in a task, measured on the transformed performance scale, associated with a one percent increase in one athlete's own solo skill in that task, holding the partner's skill and the pair's affinity fixed. A positive elasticity means that increasing an athlete's own start or riding skill is associated with higher team performance, holding the estimated production surface fixed. A larger elasticity means that team performance in that task is more sensitive to that athlete's skill, so that the fitted technology assigns a larger marginal return to that input. Comparing the elasticities of Player~1 and Player~2 within a task therefore shows whose solo skill carries the larger marginal return in the fitted technology.

\begin{figure}[htbp]
  \begin{center}
    \begin{subfigure}{0.47\textwidth}
      \centering
      \includegraphics[width=\textwidth]{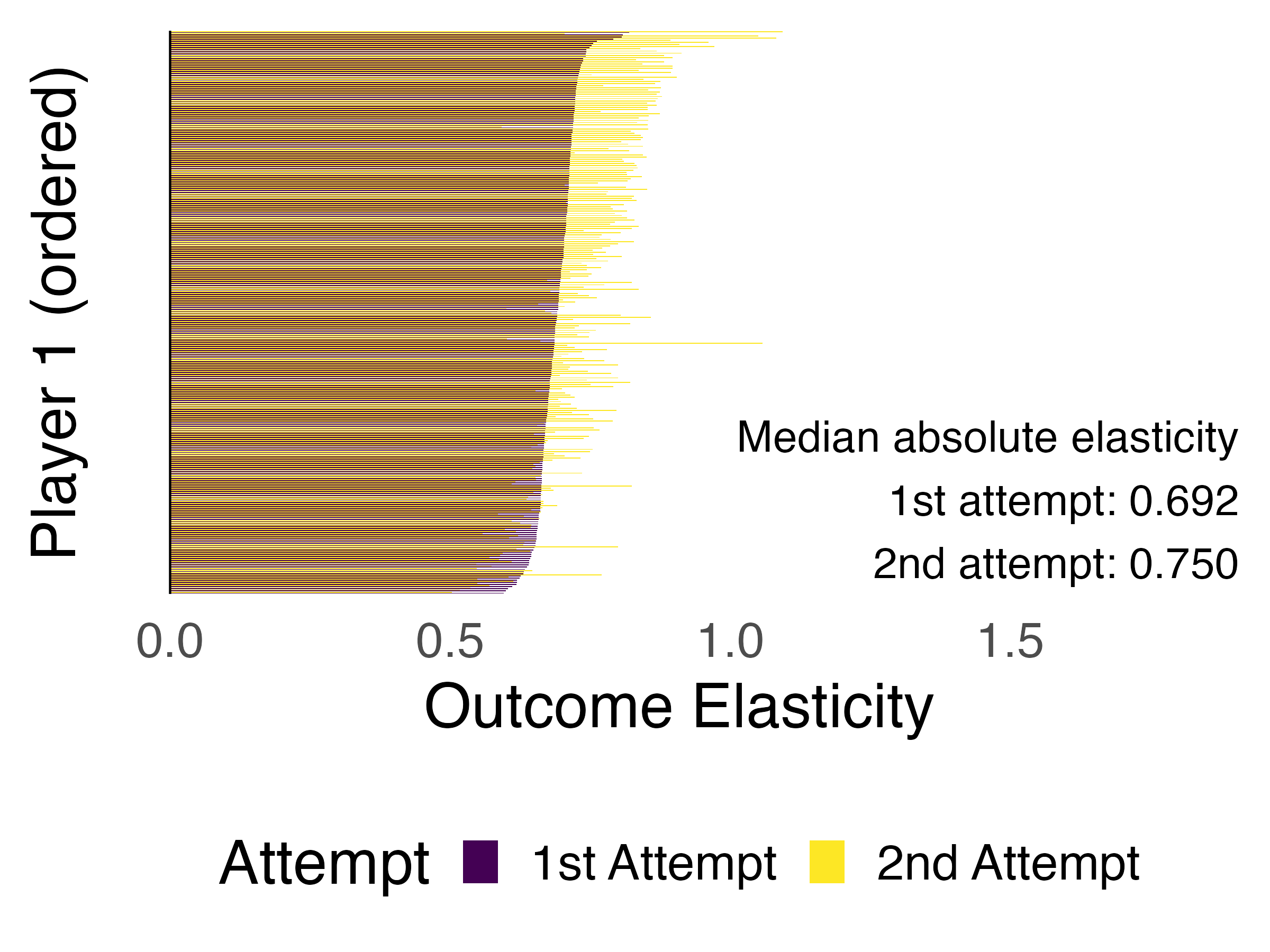}
      \caption{Player~1 start performance}
    \end{subfigure}
    \hfill
    \begin{subfigure}{0.47\textwidth}
      \centering
      \includegraphics[width=\textwidth]{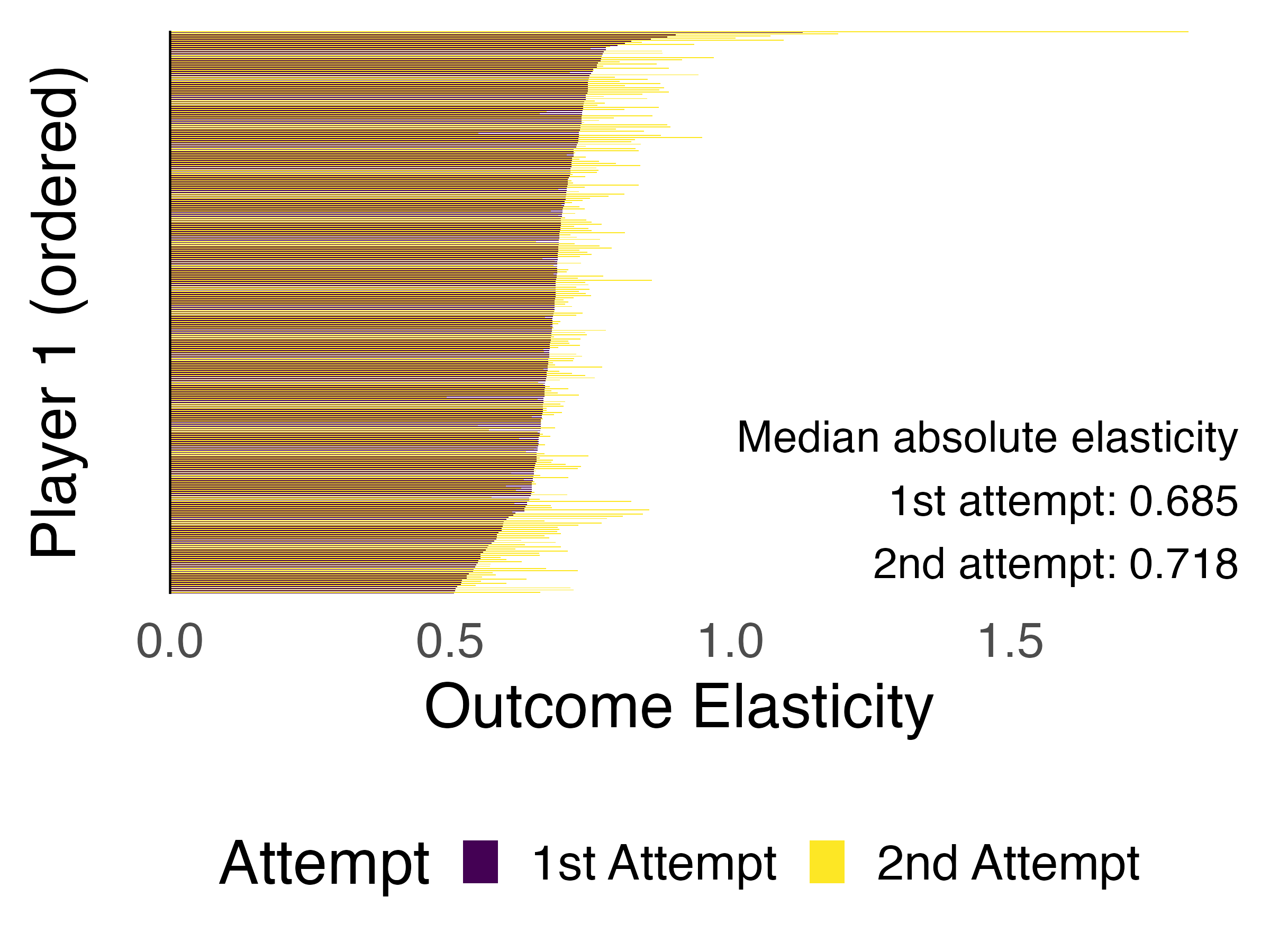}
      \caption{Player~1 riding performance}
    \end{subfigure}

    \vspace{0.75em}
    \begin{subfigure}{0.47\textwidth}
      \centering
      \includegraphics[width=\textwidth]{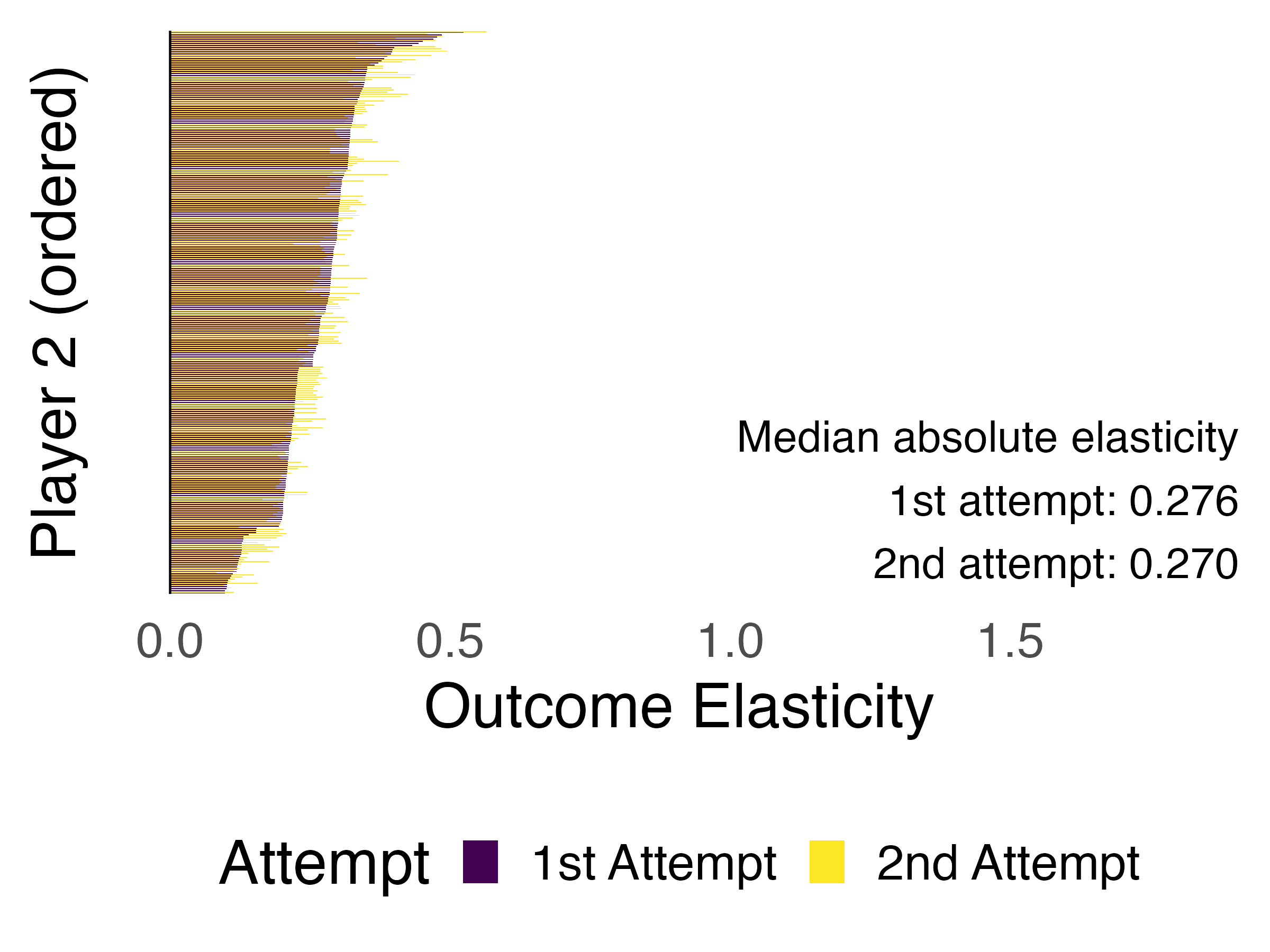}
      \caption{Player~2 start performance}
    \end{subfigure}
    \hfill
    \begin{subfigure}{0.47\textwidth}
      \centering
      \includegraphics[width=\textwidth]{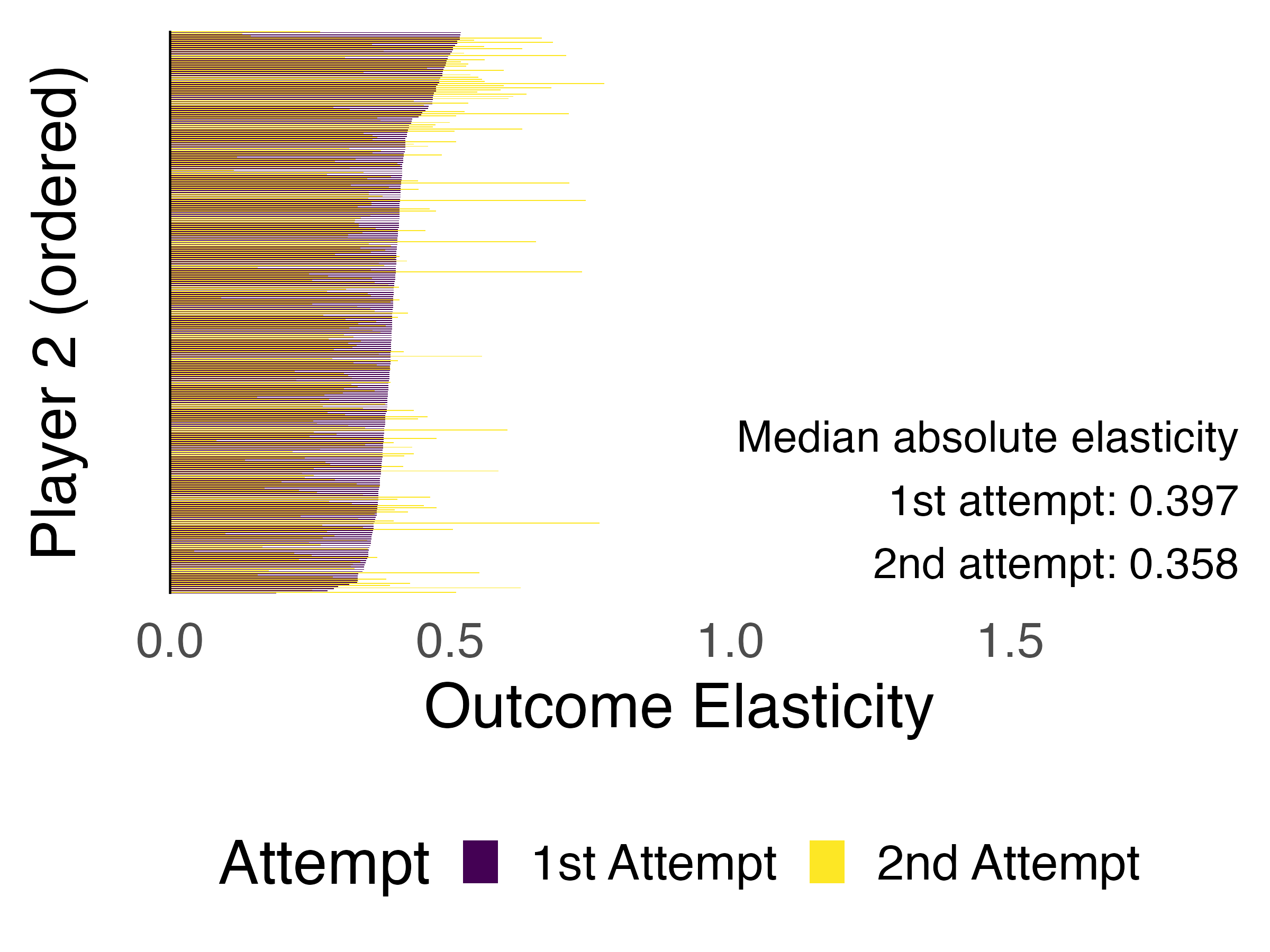}
      \caption{Player~2 riding performance}
    \end{subfigure}
    \caption{Performance Elasticity with Respect to Individual Skill Inputs}
    \label{fg:elasticity_riding_time_player_1}
  \end{center}
  {\footnotesize \textit{Note}: Elasticities are computed from the fitted unconstrained second-order polynomial using Equations~\eqref{eq:elasticity_x} and \eqref{eq:elasticity_y}. The approximation does not impose monotonicity, so negative local slopes can appear. Panels report first and second attempts for each role and task. The text in each panel reports the median absolute elasticity across observations for each attempt.}
\end{figure}

Panels (a) and (c) of Figure~\ref{fg:elasticity_riding_time_player_1} show that Player~2's start elasticities have a smaller median absolute magnitude than Player~1's in both attempts, so that team start performance responds less to the brakeman's solo start skill than to the driver's. Panels (b) and (d) show that Player~2's riding elasticities have a smaller median absolute magnitude than Player~1's in both attempts, so that team riding performance responds less to the brakeman's solo riding skill than to the driver's. The driver's solo skill therefore carries the larger marginal return in both tasks. 
Because the projection provides no uncertainty bands, these comparisons should be interpreted as descriptive features of the fitted production surface.

\FloatBarrier

\section{Counterfactual Team Assignment}\label{sec:counterfactual}

This section presents fitted-model benchmarks for how affinity information affects team assignment. We compare five hypothetical managers, who differ only in the affinity information they use when choosing pairs: SR, which uses separate Start and Riding affinities for both attempts; Start-only; Riding-only; Pooled, which uses their unweighted average; and Skills-only, which uses conditional mean affinities. All five optimize combined Start and Riding performance over both attempts and face the same feasible national rosters.

Start-only uses pair-specific Start affinity and conditional mean Riding affinity; Riding-only reverses these information choices. Each manager chooses a one-to-one assignment of brakemen to drivers within its roster to maximize the sum of predicted combined performance, weighted by each driver's observed participations in the national-pool exercise. Predicted performance for a candidate pair is the sum over the four task--attempt panels of the fitted second-order projection in Equation~\eqref{eq:polynomial}, evaluated at the two athletes' solo skills and at the affinity value available to that manager. The problem is a linear assignment problem, which we solve exactly. Every assignment is then evaluated using SR's fitted task--attempt values, so the managers differ only in the information used to choose, not in how outcomes are scored.\footnote{These benchmarks follow the reallocation logic of \citet{graham2014complementarity} and \citet{bhattacharya2009inferring}, with an affinity component that is here explicit and task-specific. Observed pairs enter with their realized pair-mean affinity for each task and attempt, while never-observed pairs enter with a kernel estimate of conditional mean affinity given the driver's skill, consistent with Equation~\eqref{eq:conditional_independence} but adding conditional-exchangeability and common-support assumptions. Evaluating the polynomial at conditional mean affinity generally differs from integrating expected output over the conditional affinity distribution; Appendix~\ref{app:counterfactual_details} gives implementation and bootstrap details.}

We first document descriptive reversals between the two tasks' affinity rankings of two-by-two assignments. We then reassign brakemen among the sleds fielded by a nation at the same event, before considering rematching across its sample-wide athlete pool under all five information sets. The reversal exercise illustrates Proposition~\ref{prop:assignment}(ii); the reassignment exercises evaluate the estimated nonlinear production functions rather than the linear benchmark in Equation~\eqref{eq:pair_value}.

\subsection{Reversals in Realized Affinities}

Proposition~\ref{prop:assignment} makes reversals the elementary unit of the assignment problem, so we first document their frequency in the observed support. In organizational terms, each block is a team-formation decision that pairs two workers in one role with two in another, and a reversal means that the collaboration records rank the two possible pairings differently across tasks. We restrict attention to two-driver, two-brakeman blocks from the same national athlete pool for which all four possible pairs are observed somewhere in the sample, so no imputation is needed. Table~\ref{tb:counterfactual_reversal_share} reports 21 such fully observed blocks: six in the Canadian sample and 15 in the U.S. sample. For each, we compute the assortativity gaps $\Delta_S$ and $\Delta_R$ in Equation~\eqref{eq:assortativity_gap} from attempt-averaged realized pair-mean affinities under the current task--attempt panel normalizations. This restriction makes the reversal comparison independent of the imputation used for never-observed pairs.

Four of the 21 blocks exhibit $\Delta_S\Delta_R<0$: two Canadian and two U.S. blocks (Table~\ref{tb:counterfactual_reversal_share}). These estimated affinity reversals illustrate the local assignment conflict in Proposition~\ref{prop:assignment}(ii); the reported share is descriptive. In these blocks, whichever pairing a task-agnostic rule selects is strictly suboptimal under some positive task weights, so the assignment rule needs the Start and Riding affinities separately to choose between them.

\begin{table}[!htbp]
  \begin{center}
    \caption{Task-Affinity Reversals in Fully Observed Two-by-Two Blocks}
    \label{tb:counterfactual_reversal_share}
    \begin{adjustbox}{max width=0.60\textwidth}
      \input{figuretable/counterfactual_reversal_share}
    \end{adjustbox}
  \end{center}
  {\footnotesize \textit{Note}: Blocks contain two drivers and two brakemen from the same national athlete pool, with all four pairs observed somewhere in the sample; no imputation is used. A reversal means the start and riding assortativity gaps have opposite signs.}
\end{table}

\subsection{Feasible Within-Event Reassignment}

Within each event, we reassign brakemen only among the athletes a nation fielded, holding its drivers and roster fixed. In organizational terms, this is re-pairing the workers already on site, without hiring, transferring, or changing who is on shift. The 32 eligible event--country cells comprise 28 two-sled and four three-sled cells (Table~\ref{tb:counterfactual_event_swap_summary}).

The observed assignment is fitted-optimal in nine cells; reassignment improves combined performance in the other 23. The mean gain is $0.744$ seconds per cell (95 percent conditional interval $[0.608,0.993]$), aggregating all sleds and both attempts. The mean among beneficial cells is $1.036$ seconds, and the maximum is $4.581$ seconds (Tables~\ref{tb:counterfactual_event_swap_summary} and \ref{tb:counterfactual_bootstrap_summary}). Because drivers, rosters, and the event schedule are held fixed, a nation could have realized these fitted gains by swapping brakemen among the sleds it had already entered, without changing which athletes competed.

\begin{table}[!htbp]
  \begin{center}
    \caption{Gains from Feasible Within-Event Brakeman Reassignment}
    \label{tb:counterfactual_event_swap_summary}
    \begin{adjustbox}{max width=\textwidth}
      \input{figuretable/counterfactual_event_swap_summary}
    \end{adjustbox}
  \end{center}
  {\footnotesize \textit{Note}: Cells are event--country combinations with two or three sleds. The counterfactual selects the best feasible brakeman reassignment within each cell. Gains aggregate predicted seconds across all sleds and both attempts.}
\end{table}

\subsection{National-Pool Rematching and the Value of Information}

The national-pool exercise allows rematching across each nation's sample-wide athlete pool. In organizational terms, this is forming pairings from the organization's full roster, as in seasonal staffing or project allocation, and the five information sets correspond to how finely the collaboration records are kept: by task, for one task only, pooled into a single score, or not at all. The feasible pools cover eight countries, 21 driver positions, and 178 weighted participations. Each included driver retains her observed participation weight, and one brakeman is assigned to each driver under common single-use athlete restrictions. National pools and participation weights are fixed, but same-event availability is not imposed.\footnote{Appendix Table~\ref{tb:counterfactual_nation_rematching} reports the eligible national pools and detailed fitted-sample gains.}

SR improves fitted combined performance by $0.073$ seconds per weighted two-attempt participation relative to Start-only (95\% CI: $[0.049,0.246]$). Its advantages over Riding-only and Pooled are smaller, at $0.017$ and $0.027$ seconds, respectively, with confidence intervals that include zero. Riding accounts for a larger share of total race time, which may partly explain why Riding-only assignments perform relatively well. The assignment-relevant information is therefore the Riding affinity: an assignment rule that ignores pair-specific Riding affinity forgoes $0.073$ seconds per participation, whereas ignoring pair-specific Start affinity shows no significant loss.

Relative to Skills-only, SR yields a gain of $0.444$ seconds. Its gain over observed historical assignments is $0.460$ seconds, close to the Skills-only comparison (Table~\ref{tb:counterfactual_bootstrap_summary}; Figure~\ref{fg:counterfactual_nation_rematching}). This similarity is consistent with observed pairings placing relatively little weight on the pair-specific affinities estimated here, in line with the descriptive pairing evidence in Appendix~\ref{app:pairing_selection}. Most of the fitted gain from rematching therefore comes from using pair-specific affinity at all rather than from how it is split across tasks: moving from Skills-only to SR is worth $0.444$ seconds, whereas the largest contrast among the affinity-using managers is $0.073$ seconds.

\begin{table}[!htbp]
  \begin{center}
    \caption{Bootstrap Summary of Counterfactual Gains}
    \label{tb:counterfactual_bootstrap_summary}
    \begin{adjustbox}{max width=\textwidth}
      \input{figuretable/counterfactual_bootstrap_summary}
    \end{adjustbox}
  \end{center}
  {\footnotesize \textit{Note}: The 95 percent confidence intervals are percentile intervals from 2,000 within-pair bootstrap resamples. Participations are resampled jointly across tasks and attempts within each directed pair, and assignments are reoptimized; production coefficients, individual skills, national rosters, and participation weights are held fixed. The within-event gain aggregates seconds across all sleds and both attempts in each cell. National gains and information contrasts are seconds per weighted two-attempt participation.}
\end{table}

\FloatBarrier

\begin{figure}[!htbp]
  \begin{center}
    \includegraphics[width=0.72\textwidth]{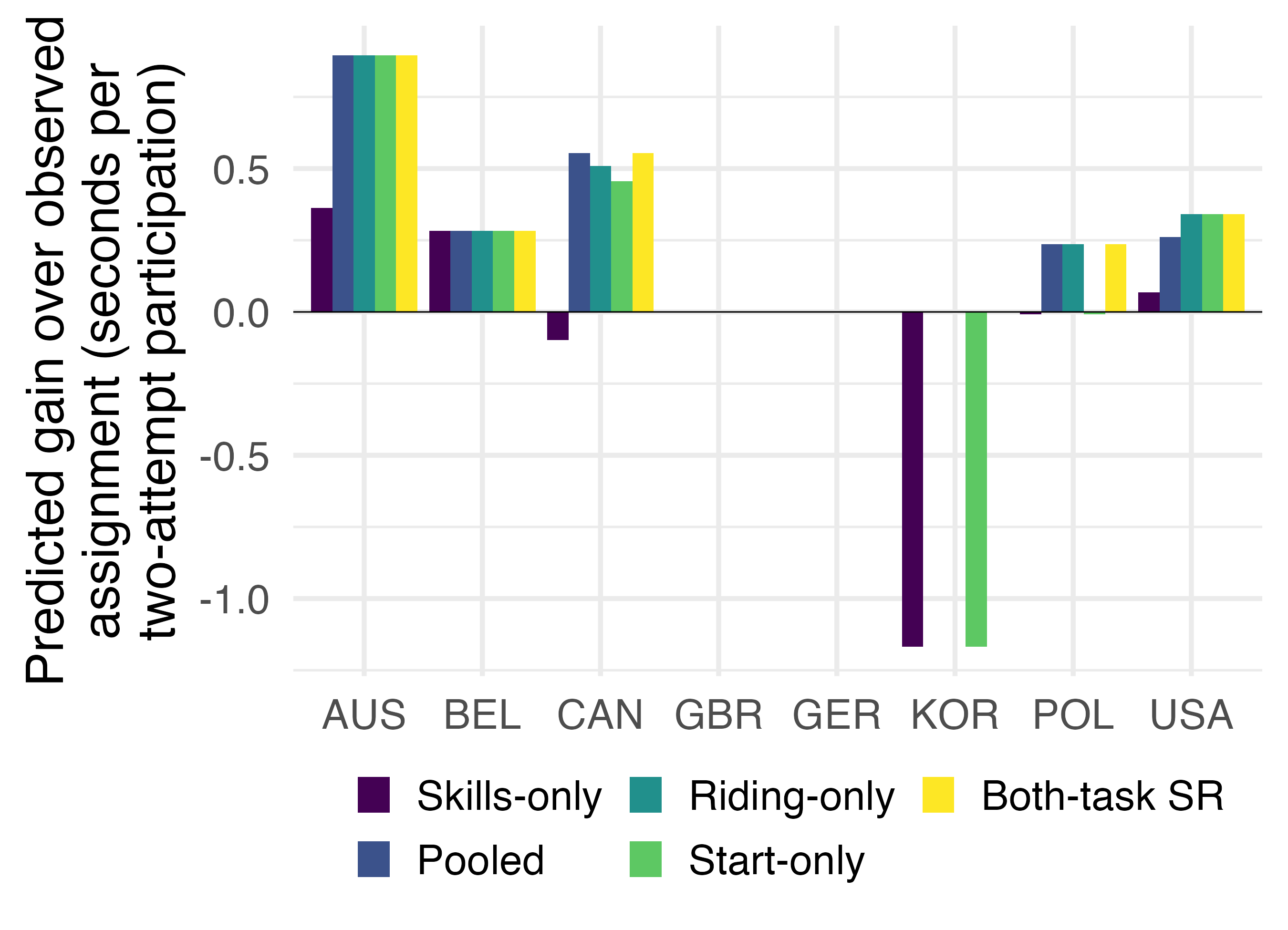}
    \caption{Rematching Gains by Manager Information Set}
    \label{fg:counterfactual_nation_rematching}
  \end{center}
  {\footnotesize \textit{Note}: Bars show predicted seconds gained over the observed assignment for each eligible nation and manager information set (SR denotes the manager using both tasks' affinities). Negative values mean the preferred assignment is predicted to perform worse than the observed one. A gain of exactly zero means that the manager selects the observed pairing; this holds for every manager in GBR and GER and for the Pooled, Riding-only, and SR managers in KOR.}
\end{figure}

\FloatBarrier

\section{Conclusion}\label{sec:conclusion}

This paper studies whether pair-history information supplies a common affinity ranking when teams perform multiple tasks. We develop a two-task assignment framework and apply a nonparametric team-production model to women's bobsleigh, separating individual start and riding skills measured in monobob from latent driver--brakeman affinity. Estimated affinity varies after conditioning on solo skill and persists across attempts within each task. The persistent start and riding components are positively correlated, but conditional affinity-recovery inference rejects the perfect correlation implied by the positive-affine common-affinity restriction under the maintained repeated-measurement assumptions. The fitted production surface also reflects the asymmetric riding roles. Task-specific affinities can reverse the locally preferred pairing across tasks, and feasible within-event reassignment improves fitted performance in some event--country cells. In national-pool reassignments, using both tasks' affinities improves fitted combined performance by $0.073$ seconds per weighted two-attempt participation relative to Start-only; the smaller advantages over Riding-only and Pooled have confidence intervals that include zero.

The broader implication is that organizations forming teams, and the managers who make the assignments, should examine whether pair histories carry task-specific information before compressing them into a single score, and that empirical measures of team-specific human capital should allow pair productivity to differ by task. Where output can be decomposed and individual skills measured along matching task dimensions, the framework provides a way to do so. Women's bobsleigh offers an illustrative application because publicly available official records report granular phase-level outcomes and crew identities. The same assignment question is relevant to production lines, warehouses, and project teams, and to the economics of team formation more generally, when distinguishable tasks, measurable individual skills, and collaboration histories inform the allocation of workers or partners.

\ifmsanon\theendnotes\fi
\bibliographystyle{ecca}
\bibliography{team_production}
\fi

\ifsplit\ifsupp\msappendixtrue\fi\else\msappendixtrue\fi
\ifmsappendix
\ifsupp\else\newpage\fi
\appendix
\ifsupp
  \ifmsanon
    \renewcommand{\theequation}{EC.\arabic{equation}}
  \else
    \makeatletter\@addtoreset{equation}{section}\makeatother
    \renewcommand{\theequation}{\thesection.\arabic{equation}}
  \fi
\fi
\setcounter{table}{0}
\setcounter{figure}{0}
\renewcommand{\thetable}{\thesection\arabic{table}}
\renewcommand{\thefigure}{\thesection\arabic{figure}}

\section{Proof of Proposition~\ref{prop:assignment}}\label{app:proofs}

\begin{proof}[Proof of Proposition~\ref{prop:assignment}]
For any bijection $\sigma$,
\begin{align*}
\sum_{i\in\mathcal I}\left[f(i)+g(\sigma(i))\right]=\sum_{i\in\mathcal I}f(i)+\sum_{j\in\mathcal J}g(j),
\end{align*}
so the separable terms are constant across feasible matchings.

(i) If $A_2=\eta+bA_1$, the objective is
\begin{align*}
\sum_{i\in\mathcal I}v(i,\sigma(i))=\mathrm{const}+(\omega_1+b\omega_2)\sum_{i\in\mathcal I}A_1(i,\sigma(i)).
\end{align*}
Because $\omega_1+b\omega_2>0$, any matching that maximizes the sum of task-1 affinity maximizes the objective. On every two-by-two subroster, $\Delta_2=b\Delta_1$, so $\Delta_1\Delta_2=b\Delta_1^2\geq0$ and no reversal exists.

(ii) Take a subroster $\{i,i'\}\times\{j,j'\}$ with $\Delta_1\Delta_2<0$; without loss of generality, let $\Delta_1>0>\Delta_2$. Hold all outside matches fixed and compare $\sigma_1=\{(i,j),(i',j')\}$ with $\sigma_2=\{(i,j'),(i',j)\}$. The separable terms cancel, and their value difference is
\begin{align*}
V(\sigma_1)-V(\sigma_2)=\omega_1\Delta_1+\omega_2\Delta_2.
\end{align*}
Choosing $\omega_2/\omega_1<\Delta_1/(-\Delta_2)$ makes $\sigma_1$ strictly optimal, while choosing $\omega_2/\omega_1>\Delta_1/(-\Delta_2)$ makes $\sigma_2$ strictly optimal; both choices permit strictly positive weights. Any rule that chooses the same sub-matching at both weight vectors is therefore strictly suboptimal at one of them. At given weights, the absolute value of the displayed difference is the loss from choosing the wrong sub-matching.
\end{proof}

\section{Supplementary Analyses}\label{app:supplementary}

\subsection{Additional Data Description}\label{app:data_description}\label{app:monobob_support}

This subsection collects the remaining data-description figures and tables referenced in Section~\ref{sec:data}: the distribution of observed monobob events per athlete and the relationship between the two solo skill measures.

\begin{figure}[htbp]
  \begin{center}
    \begin{subfigure}{0.47\textwidth}
      \centering
      \includegraphics[width=\textwidth]{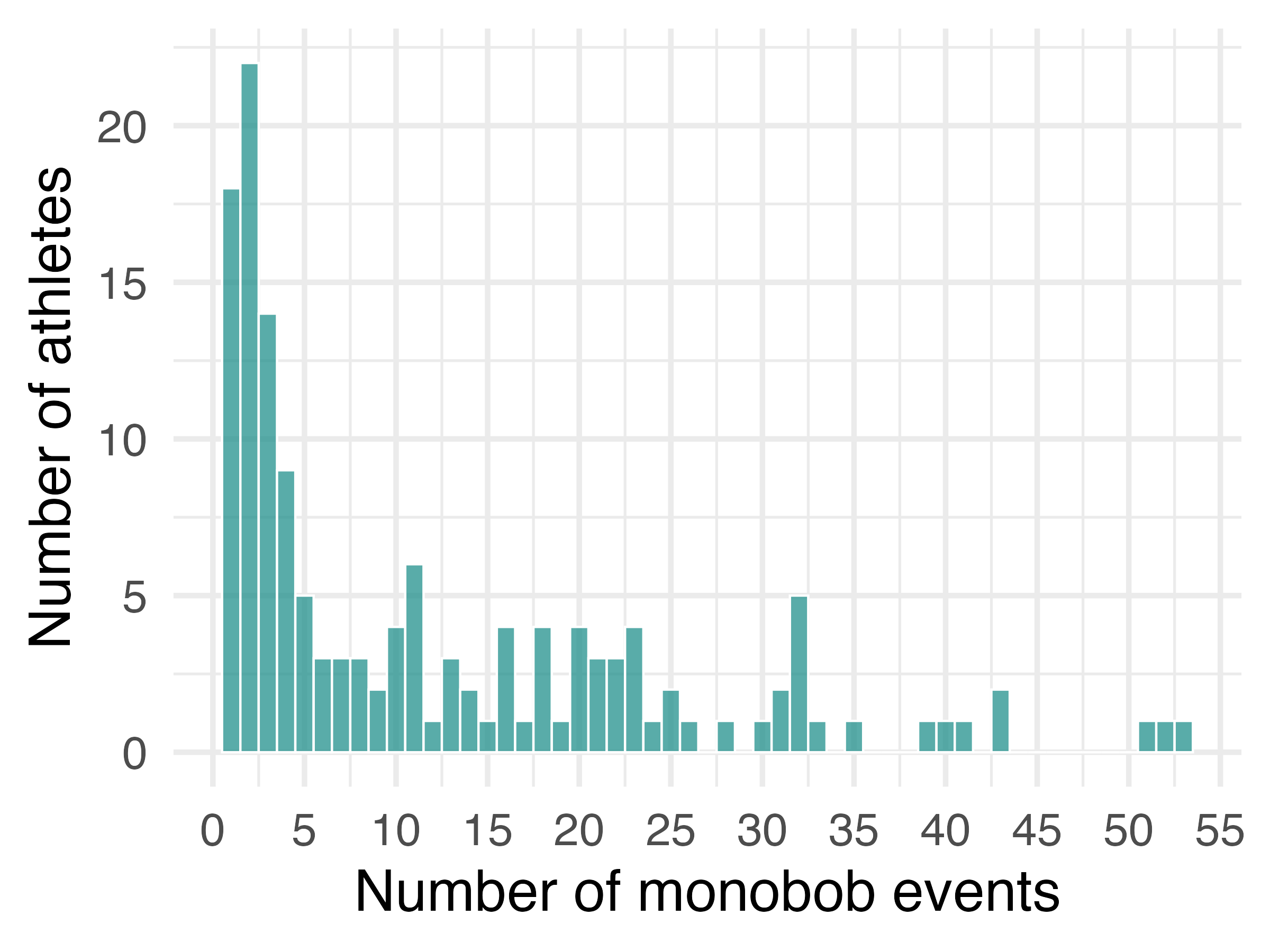}
      \caption{All monobob athletes}
    \end{subfigure}
    \hfill
    \begin{subfigure}{0.47\textwidth}
      \centering
      \includegraphics[width=\textwidth]{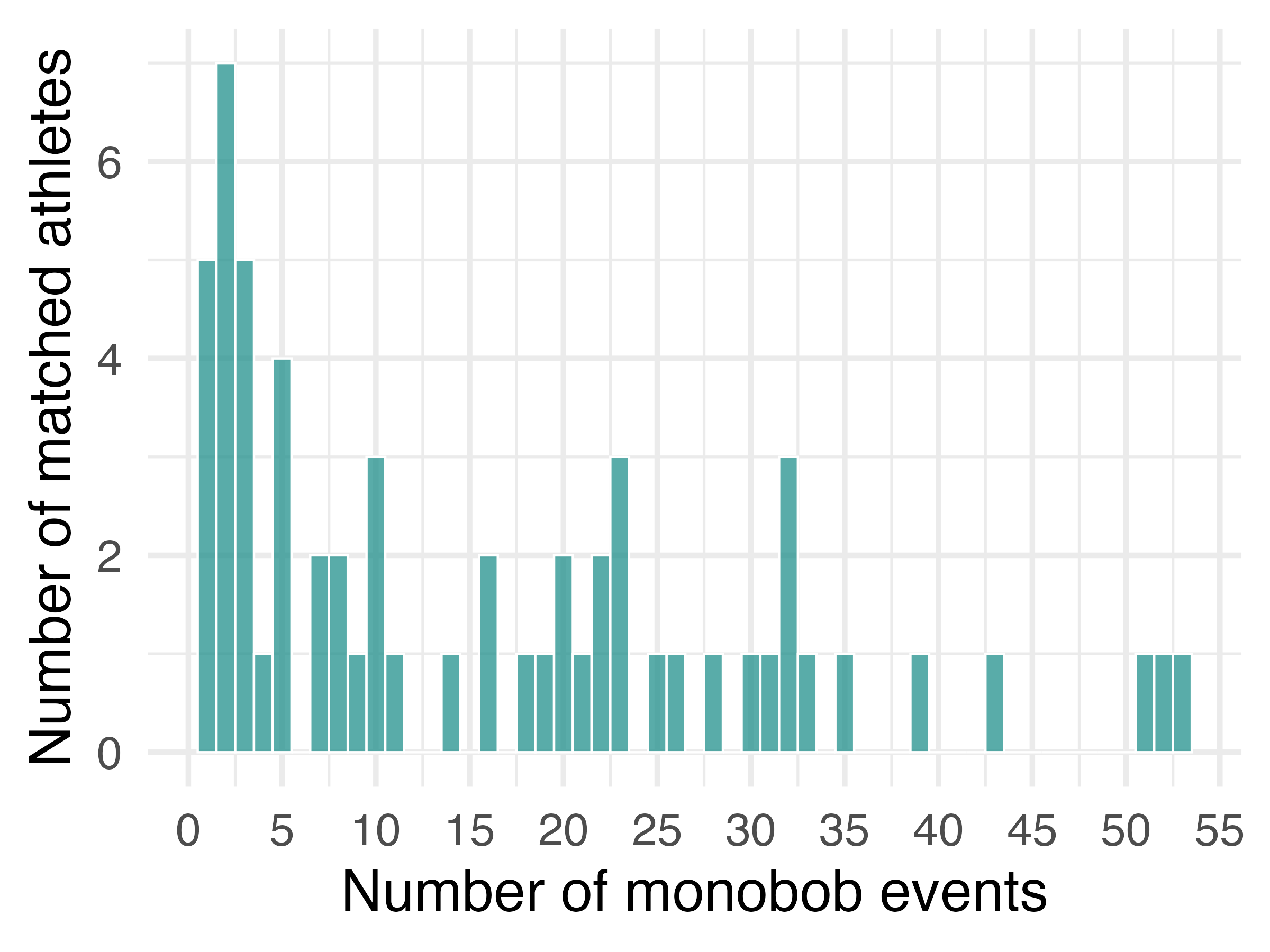}
      \caption{Matched two-woman athletes}
    \end{subfigure}
    \caption{Distribution of Observed Monobob Events per Athlete}
    \label{fg:histogram_monobob_runs_appendix}
  \end{center}
  {\footnotesize \textit{Note}: The current construction retains the first recorded run for each athlete--event. Panel~(a) reports the number of observed events for all athletes used to estimate individual fixed effects. Panel~(b) restricts the sample to athletes matched to the two-woman production sample.}
\end{figure}

\begin{table}[!htbp]
  \begin{center}
    \caption{Monobob Observation Support}
    \label{tb:monobob_events_runs_matched_athletes_summary}
    \begin{adjustbox}{max width=\textwidth}
      \input{figuretable/monobob_events_runs_matched_athletes_summary}
    \end{adjustbox}
  \end{center}
  {\footnotesize \textit{Note}: The table summarizes first-recorded-run athlete--event observations from the 2018--2019 through 2024--2025 seasons. All = all women's monobob athletes; Matched = the 59 athletes matched to two-woman events. Obs.\ = athlete--event observations; Athletes = number of athletes; One event = athletes observed in exactly one event. The remaining columns summarize the number of events per athlete.}
\end{table}

\begin{figure}[htbp]
  \begin{center}
    \begin{subfigure}{0.52\textwidth}
      \centering
      \includegraphics[width=\textwidth]{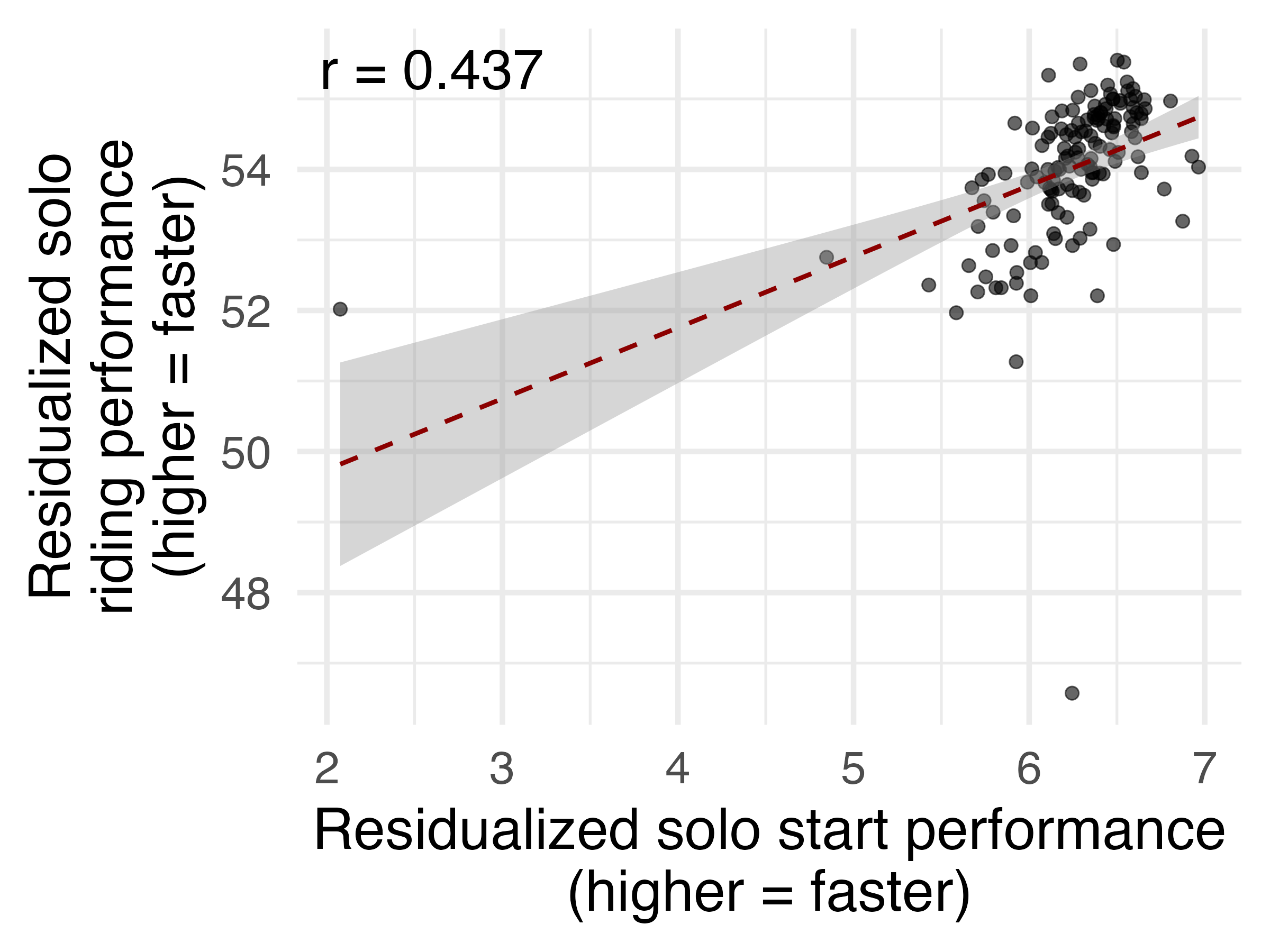}
      \caption{Players with monobob records}
    \end{subfigure}

    \vspace{0.75em}
    \begin{subfigure}{0.52\textwidth}
      \centering
      \includegraphics[width=\textwidth]{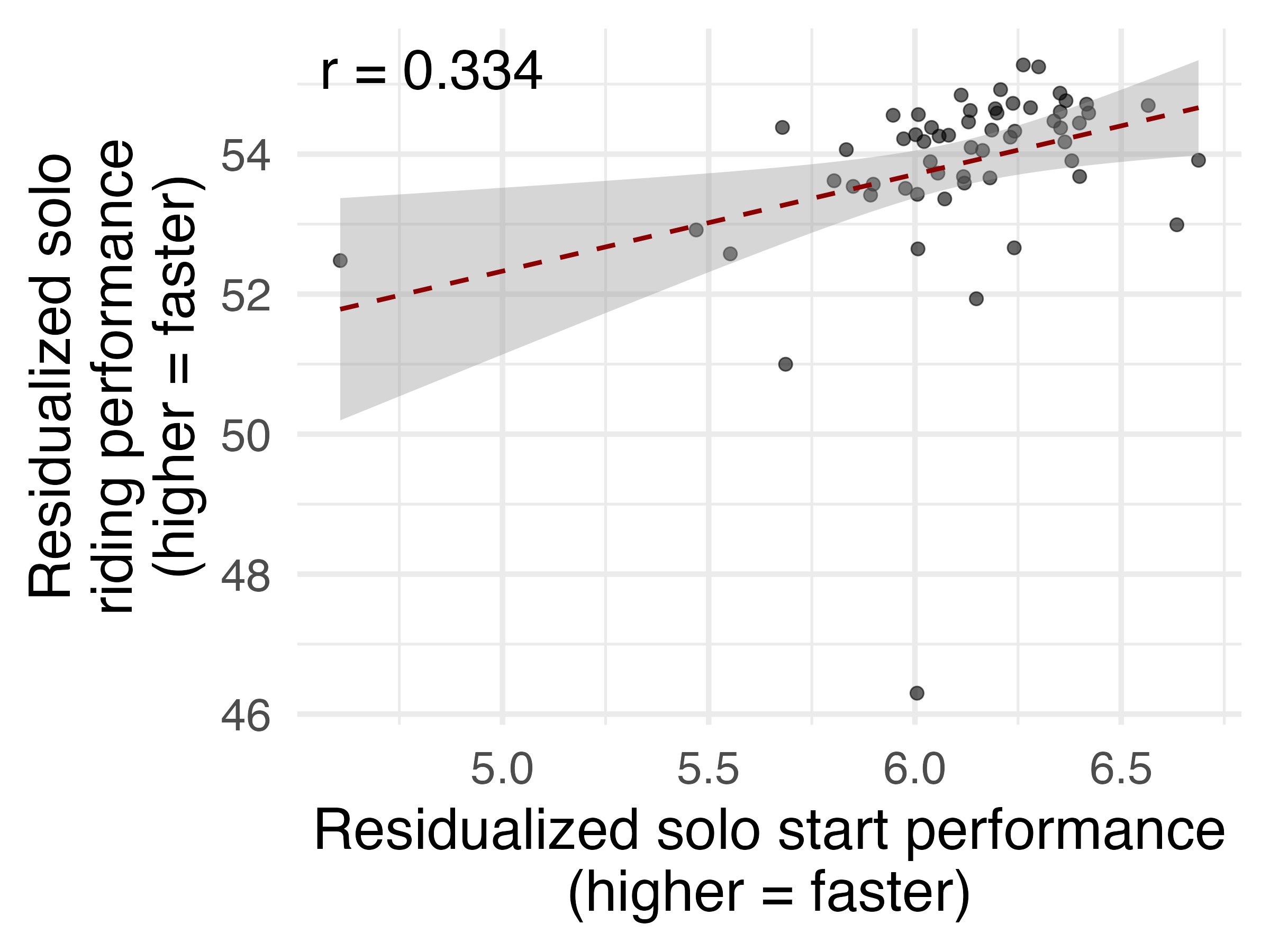}
      \caption{Players matched to two-woman races}
    \end{subfigure}
    \caption{Correlation between Start-Phase and Riding-Phase Individual Skills}
    \label{fg:scatter_start_vs_riding_skill}
  \end{center}
  {\footnotesize \textit{Note}: Each point represents an athlete. Skills are residualized monobob fixed effects, controlling for event and starting-order fixed effects, and transformed so that higher values indicate better performance. The dashed line shows the linear fit.}
\end{figure}

\begin{table}[!htbp]
  \begin{center}
    \caption{Correlation between Solo Start and Riding Skills}
    \label{tb:start_riding_skill_correlation_summary}
    \begin{adjustbox}{max width=\textwidth}
      \input{figuretable/start_riding_skill_correlation_summary}
    \end{adjustbox}
  \end{center}
  {\footnotesize \textit{Note}: Correlations are computed at the athlete level using residualized monobob fixed effects. Confidence intervals are from the standard Pearson correlation test.}
\end{table}

\FloatBarrier

\subsection{Pairing Patterns}\label{app:run_pairing_counts}

The pairing tables use participation-weighted skill terciles and assign one bin to each athlete within role and task before joining the bins to the 260 team-participation rows. Each of the 63 directed pairs therefore enters exactly one cell in each unique-pair table, while the participation tables retain repeated team appearances. These tables describe observed support but do not establish the identifying restriction.

\begin{table}[!htbp]
  \begin{center}
    \caption{Unique-Pair Within-Phase Pairing Counts}
    \label{tb:skill_pairing_start}
    \begin{subtable}{0.48\textwidth}
      \centering
      \caption{Start skill}
      \begin{adjustbox}{max width=\textwidth}
        \input{figuretable/skill_pairing_summary_statistics_start}
      \end{adjustbox}
    \end{subtable}
    \hfill
    \begin{subtable}{0.48\textwidth}
      \centering
      \caption{Riding skill}
      \begin{adjustbox}{max width=\textwidth}
        \input{figuretable/skill_pairing_summary_statistics_riding}
      \end{adjustbox}
    \end{subtable}
  \end{center}
  {\footnotesize \textit{Note}: Each of the 63 directed driver--brakeman pairs enters one cell. Terciles are defined within player role and task; 1 = lowest displayed performance and 3 = highest displayed performance.}
\end{table}

\begin{table}[!htbp]
  \begin{center}
    \caption{Unique-Pair Cross-Phase Pairing Counts}
    \label{tb:skill_pairing_p1start_p2riding}
    \begin{subtable}{0.48\textwidth}
      \centering
      \caption{Player~1 start skill vs. Player~2 riding skill}
      \begin{adjustbox}{max width=\textwidth}
        \input{figuretable/skill_pairing_summary_statistics_p1start_p2riding}
      \end{adjustbox}
    \end{subtable}
    \hfill
    \begin{subtable}{0.48\textwidth}
      \centering
      \caption{Player~1 riding skill vs. Player~2 start skill}
      \begin{adjustbox}{max width=\textwidth}
        \input{figuretable/skill_pairing_summary_statistics_p1riding_p2start}
      \end{adjustbox}
    \end{subtable}
  \end{center}
  {\footnotesize \textit{Note}: Each of the 63 directed driver--brakeman pairs enters one cell. The same fixed within-role task-specific terciles are used in both panels.}
\end{table}

\begin{table}[!htbp]
  \begin{center}
    \caption{Team-Participation Counts by Within-Phase Skill Pairing}
    \label{tb:skill_pairing_run_counts_start}
    \begin{subtable}{0.48\textwidth}
      \centering
      \caption{Start skill}
      \begin{adjustbox}{max width=\textwidth}
        \input{figuretable/skill_pairing_run_counts_start}
      \end{adjustbox}
    \end{subtable}
    \hfill
    \begin{subtable}{0.48\textwidth}
      \centering
      \caption{Riding skill}
      \begin{adjustbox}{max width=\textwidth}
        \input{figuretable/skill_pairing_run_counts_riding}
      \end{adjustbox}
    \end{subtable}
  \end{center}
  {\footnotesize \textit{Note}: Each cell reports team-participation rows using the fixed athlete--role--task terciles; the cells sum to 260. Tercile 1 = lowest displayed performance and tercile 3 = highest displayed performance.}
\end{table}

\begin{table}[!htbp]
  \begin{center}
    \caption{Team-Participation Counts by Cross-Phase Skill Pairing}
    \label{tb:skill_pairing_run_counts_cross}
    \begin{subtable}{0.48\textwidth}
      \centering
      \caption{Player~1 start skill vs. Player~2 riding skill}
      \begin{adjustbox}{max width=\textwidth}
        \input{figuretable/skill_pairing_run_counts_p1start_p2riding}
      \end{adjustbox}
    \end{subtable}
    \hfill
    \begin{subtable}{0.48\textwidth}
      \centering
      \caption{Player~1 riding skill vs. Player~2 start skill}
      \begin{adjustbox}{max width=\textwidth}
        \input{figuretable/skill_pairing_run_counts_p1riding_p2start}
      \end{adjustbox}
    \end{subtable}
  \end{center}
  {\footnotesize \textit{Note}: Each cell reports team-participation rows using the fixed athlete--role--task terciles; the cells sum to 260.}
\end{table}

\FloatBarrier

\subsubsection{Pairing Decisions and Estimated Affinity}\label{app:pairing_selection}

We examine pairing decisions in the 2018--19 through 2024--25 matched sample, using two-woman records from 2015--16 onward to identify previously observed partnerships. Continuation rates do not increase systematically with estimated affinity in either task (Table~\ref{tb:pairing_continuation_by_affinity}). First-observed pairs have modestly higher Start affinity (difference: $1.728$; 95 percent confidence interval: $[0.503,3.008]$), while the corresponding Riding difference is not statistically significant (Table~\ref{tb:first_observed_affinity_comparison}).

\begin{table}[!htbp]
  \centering
  \caption{Pair Continuation by Previous-Event Affinity}
  \label{tb:pairing_continuation_by_affinity}
  \input{figuretable/pairing_continuation_by_affinity}
\end{table}

\begin{table}[!htbp]
  \centering
  \caption{Affinity of First-Observed and Previously Observed Pairs}
  \label{tb:first_observed_affinity_comparison}
  \input{figuretable/first_observed_affinity_comparison}
\end{table}

\FloatBarrier

\subsection{Monte Carlo Evidence on Affinity Recovery}\label{app:monte_carlo}

This subsection examines how accurately the nonparametric (NP) estimator recovers latent affinity in small samples and under departures from conditional independence. Building on the framework and simulation analysis of \citet{matzkin2003nonparametric}, we adapt the simulation design of \citet{otani2024nonparametric} to the team-production notation $H=m(AX,Y)$. We retain the latter study's input transformations and two production technologies, while replacing the serially correlated primitive draws with independent draws. The experiment evaluates recovery of $A$ on its normalized level scale.

\subsubsection{Simulation Design}

For each replication, we draw $Z_{Xi}$, $Z_{Yi}$, and $Z_{Ai}$ independently across observations and series from $N(0,s_Z^2)$, where $s_Z^2=1/(1-0.2^2)$ preserves the stationary marginal variance in the reference design. Positive skill inputs are constructed using a positive shift of 10 before rescaling by 100, giving an unscaled sample minimum of 10 in every simulated sample:
\begin{align*}
X_i&=100\left(Z_{Xi}+\left|\min_j Z_{Xj}\right|+10\right), &
Y_i&=100\left(Z_{Yi}+\left|\min_j Z_{Yj}\right|+10\right).
\end{align*}
To vary the dependence between affinity and Player~2 skill, we set
\begin{align*}
\widetilde A_i
&=\overline Z_A+s_{Z_A}\left[
\rho\frac{Y_i-\overline Y}{s_Y}
+\sqrt{1-\rho^2}\frac{Z_{Ai}-\overline Z_A}{s_{Z_A}}\right],\\
Q_i&=\widetilde A_i+\left|\min_j\widetilde A_j\right|+10,
\qquad A_i=100\frac{Q_i}{Q_1},
\end{align*}
where bars and $s_{Z_A},s_Y$ denote sample means and standard deviations. Thus $A_1=100$ in every replication, and this normalized $A_i$ enters production. At $\rho=0$, $\widetilde A_i=Z_{Ai}$ and $A\perp Y\mid X$ holds. Positive values of $\rho$ introduce dependence through the same Cholesky mixing used in the reference design; $\rho$ is a mixing parameter, not the realized sample correlation. Although the primitive draws are independent, the shared sample transformations and normalization induce dependence among the transformed observations.

We consider two increasing, constant-returns technologies:
\begin{align*}
\text{Cobb--Douglas:}\quad
H_i&=\frac{(A_iX_i)^{0.3}Y_i^{0.7}}{10},\\
\text{Perfect substitutes:}\quad
H_i&=\frac{0.3A_iX_i+0.3Y_i}{10}.
\end{align*}
Output has no added noise or event effects. For each technology, we use 2,000 replications per design cell at sample sizes $n=50$, 100, and 200 and mixing parameters $\rho=0$, 0.1, 0.2, and 0.5.

\subsubsection{Estimators and Evaluation}

For NP, we apply Equations~\eqref{eq:kernel_estimator}--\eqref{eq:affinity_recovery} with Gaussian kernel weights. Bandwidths are 0.1 times each input's sample range, with the range of $X$ bounded below by one. We use $(H_1,X_1,Y_1)$ as the reference observation. Before generalized inversion, we apply increasing rearrangement in affinity at fixed Player~1 skill, using uniform weighting in the affinity value over the numerical interval $[1,1000]$. Recovered affinities are positively rescaled so that the reference observation has reported affinity exactly 100. The first-observation reference and bandwidth fraction 0.1 differ from the median-based reference and bandwidth fraction 0.05 in the empirical application.

The parametric benchmarks are Cobb--Douglas with multiplicative affinity (CD-M) and Cobb--Douglas with input-augmenting affinity (CD-I). Both estimate the same unrestricted OLS regression separately in each replication $r$:
\begin{align*}
\log(10H_{ri})=c_r+\alpha_r\log X_{ri}+\beta_r\log Y_{ri}+e_{ri}.
\end{align*}
Neither the restriction $\alpha_r+\beta_r=1$ nor the true technology coefficients are imposed. CD-M interprets affinity as a production multiplier, $10H=A^M X^\alpha Y^\beta\exp(e)$, and recovers $\widehat A_r^M=\exp(\widehat c_r)$. CD-I instead specifies $10H=(A^I X)^\alpha Y^\beta\exp(e)$ and recovers $\widehat A_r^I=\exp(\widehat c_r/\widehat\alpha_r)$, following the reference study's intercept conversion. Each benchmark returns one affinity level per replication, which is assigned to every observation.\footnote{Within replication $r$, one intercept-derived affinity parameter represents the $n$ distinct latent values $A_{r1},\ldots,A_{rn}$. This is the common-productivity restriction in the commonly used pooled Cobb--Douglas production function estimated by log-linear OLS. Observation-specific affinity variation is absorbed into the regression error, rather than recovered as separate affinity parameters. The input coefficients $\alpha_r$ and $\beta_r$ are also estimated from the same $n$ observations.} CD-M deliberately also misspecifies the location of $A$ in production: its reported level is compared with the input-augmenting $A_i$, rather than with $A_i^{0.3}$. Even under the Cobb--Douglas technology, the restriction to a common affinity omits the observation-specific variation in $A_i$. Thus this comparison evaluates two specified intercept-based measures, rather than all parametric estimators of heterogeneous affinity.

A residual-based Cobb--Douglas estimator (CD-R) instead permits observation-specific affinity. It estimates the CRS relationship $\log(H/Y)=c+\alpha\log(X/Y)+u$ and recovers $\widehat A_i=100\exp\{(\widehat u_i-\widehat u_1)/\widehat\alpha\}$. Positive input elasticities require $0<\alpha<1$. CD-R is examined separately from the three methods in Table~\ref{tb:monte_carlo_affinity}.

All $n$ observations are used for estimation. We randomly select the same 30 observations for all estimators within each replication and report bias and root mean squared error (RMSE), each computed within a replication and then averaged across replications. Parentheses report Monte Carlo standard errors of these averages. The intercept-based estimates receive no further rescaling, residual adjustment, or smearing correction. All three published estimators produce finite evaluated errors in 2,000 of 2,000 replications in every reported setting.

\subsubsection{Results}

Table~\ref{tb:monte_carlo_affinity} shows that, under conditional independence, NP's RMSE falls from $7.90$ at $n=50$ to $6.22$ at $n=100$ and $4.48$ at $n=200$ under perfect substitutes. Under Cobb--Douglas, the corresponding RMSEs are $9.76$, $8.37$, and $7.71$.

Under perfect substitutes, the deterioration in NP recovery accuracy is relatively limited at $\rho=0.1$ and $0.2$, but becomes more pronounced at $\rho=0.5$. Under Cobb--Douglas, NP's RMSE remains below its conditional-independence benchmark at all three positive dependence settings. Thus, modest departures from conditional independence need not substantially impair affinity recovery in the designs considered.

NP has lower RMSE than CD-M and CD-I in all 24 reported settings. Large errors for the intercept-based benchmarks reflect both their common-affinity restriction and the sensitivity of exponentiated intercepts to estimation error; CD-I can additionally magnify errors through division by $\widehat\alpha$. Its large Monte Carlo standard errors show that some averages are strongly affected by extreme estimates.

CD-R permits observation-specific affinity and yields lower numerical RMSE than NP in these simulations. However, under perfect substitutes with $\rho=0$, more than 40 percent of replications at each examined sample size produce elasticity estimates that violate the maintained positive-input restrictions. These estimates are numerically computable, but a small affinity-recovery error does not always correspond to an economically admissible fitted production function.

Together, the simulations illustrate NP affinity recovery in relatively small samples, with limited deterioration under modest departures from conditional independence in the designs considered. The NP approach accommodates observation-specific affinity without imposing a Cobb--Douglas production function.

\begin{landscape}
\begin{table}[!htbp]
  \begin{center}
    \caption{Monte Carlo Performance of Affinity Recovery}
    \label{tb:monte_carlo_affinity}
    \begin{adjustbox}{max width=\linewidth}
      \input{figuretable/monte_carlo_affinity_sensitivity}
    \end{adjustbox}
  \end{center}
  {\footnotesize \textit{Note}: The target is the affinity $A_i$ used in production, with $A_1=100$. NP is nonparametric rank inversion. CD-M and CD-I use the same unrestricted regression of $\log(10H)$ on a constant, $\log X$, and $\log Y$, recovering $\exp(\widehat c)$ and $\exp(\widehat c/\widehat\alpha)$, respectively. $N$ is sample size; Rho is the dependence-mixing parameter $\rho$. Bias and RMSE are averages across replications of statistics computed from the same 30 sampled observations for all methods. Parentheses give Monte Carlo standard errors. $R$ reports common valid replications divided by attempted replications. NP is normalized so that the estimated reference affinity is 100; the intercept-based estimates receive no additional normalization.}
\end{table}
\end{landscape}

\subsection{Implied-Affinity Summaries and Diagnostics}\label{app:affinity_structure}

Panel-specific normalization makes raw panel means incomparable across attempts or samples. Table~\ref{tb:summary_statistics_implied_efficiency} is therefore only a within-panel dispersion summary. The bin-based contrasts below use the fixed within-role task-specific terciles.

\begin{table}[!htbp]
  \begin{center}
    \caption{Summary Statistics of Mean Implied Affinity for 63 Observed Pairs}
    \label{tb:summary_statistics_implied_efficiency}
    \begin{adjustbox}{max width=\textwidth}
      \input{figuretable/summary_statistics_implied_efficiency}
    \end{adjustbox}
  \end{center}
  {\footnotesize \textit{Note}: Implied affinity corresponds to the panels in Figure~\ref{fg:efficiency_implied_riding_2}. Values are normalized relative to the median Player~2 skill baseline. The table should be read as a descriptive summary of within-panel dispersion rather than as a direct comparison of mean levels across panels.}
\end{table}

\begin{table}[!htbp]
  \begin{center}
    \caption{Structure-Based Summary of Implied Affinity}
    \label{tb:implied_affinity_structure_summary}
    \begin{subtable}{\textwidth}
      \centering
      \caption{Dispersion and high-low contrast}
      \begin{adjustbox}{max width=\textwidth}
        \input{figuretable/implied_affinity_structure_dispersion}
      \end{adjustbox}
    \end{subtable}

    \vspace{0.75em}
    \begin{subtable}{\textwidth}
      \centering
      \caption{Diagonal contrast}
      \begin{adjustbox}{max width=\textwidth}
        \input{figuretable/implied_affinity_structure_diagonal}
      \end{adjustbox}
    \end{subtable}
  \end{center}
  {\footnotesize \textit{Note}: N is the number of observed pairs, and Mean and SD are the within-panel mean and standard deviation of implied affinity. HH = mean implied affinity in high-high cells, where both players are in the highest skill tercile; LL = mean in low-low cells, where both players are in the lowest skill tercile; Diag.\ = mean in diagonal cells, where the two players are in the same skill tercile; Off-diag.\ = mean in all other cells. Terciles are assigned once within athlete, role, and task.}
\end{table}

\begin{figure}[htbp]
  \begin{center}
    \includegraphics[width=0.85\textwidth]{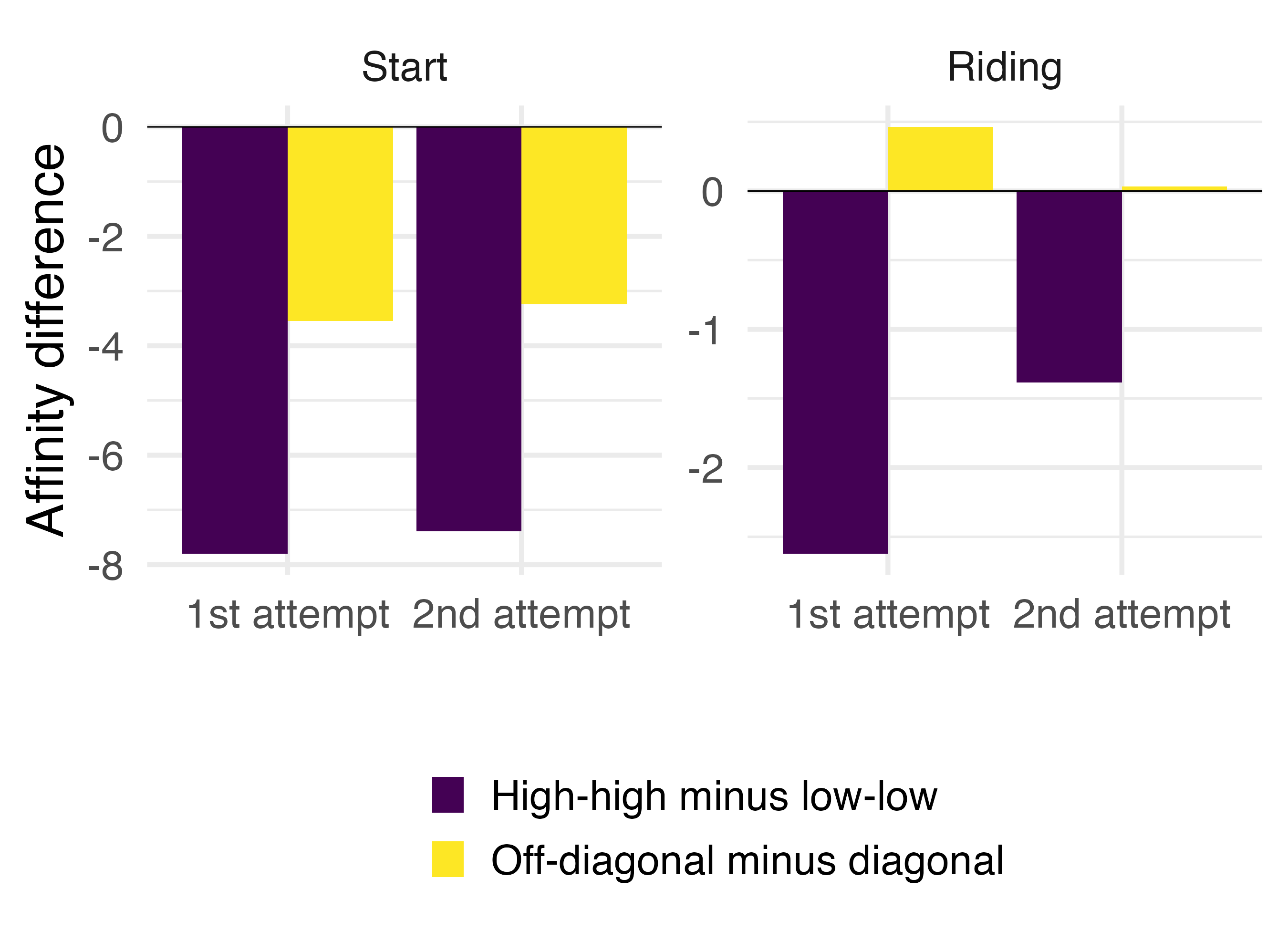}
    \caption{Structure-Based Summaries of Implied Affinity}
    \label{fg:implied_affinity_structure_summary}
  \end{center}
  {\footnotesize \textit{Note}: The figure plots within-panel contrasts from Table~\ref{tb:implied_affinity_structure_summary}. These contrasts summarize heatmap structure and are not direct level comparisons across panels.}
\end{figure}

\FloatBarrier

\subsection{Affinity Recovery in the Two-Attempt Sample}
\label{app:restricted_sample}\label{app:robustness}

The baseline sample retains the first two attempts even for crew--event records with more than two observed attempts. This choice maximizes comparability because rankings are based on cumulative times and because the first two attempts are observed across the relevant records. A concern is that records with additional attempts may come from competitions with different strategic considerations, pacing, or risk-taking.

Figure~\ref{fg:efficiency_implied_riding_2_only_two_attempts} repeats the main implied-affinity analysis using only crew--event records with exactly two observed attempts. The first-attempt start panel changes noticeably; the restricted sample also shifts the normalization point at the median Player~2 skill. Given panel-specific normalizations, we do not interpret cross-panel differences in raw affinity levels. In this restricted sample, the persistent Start--Riding affinity correlation is $0.504$ (95 percent conditional affinity-recovery bootstrap interval $[0.196,0.808]$). This interval also excludes the perfect-correlation benchmark implied by the positive-affine common-affinity restriction.

\begin{figure}[htbp]
  \begin{center}
    \begin{subfigure}{0.47\textwidth}
      \centering
      \includegraphics[width=\textwidth]{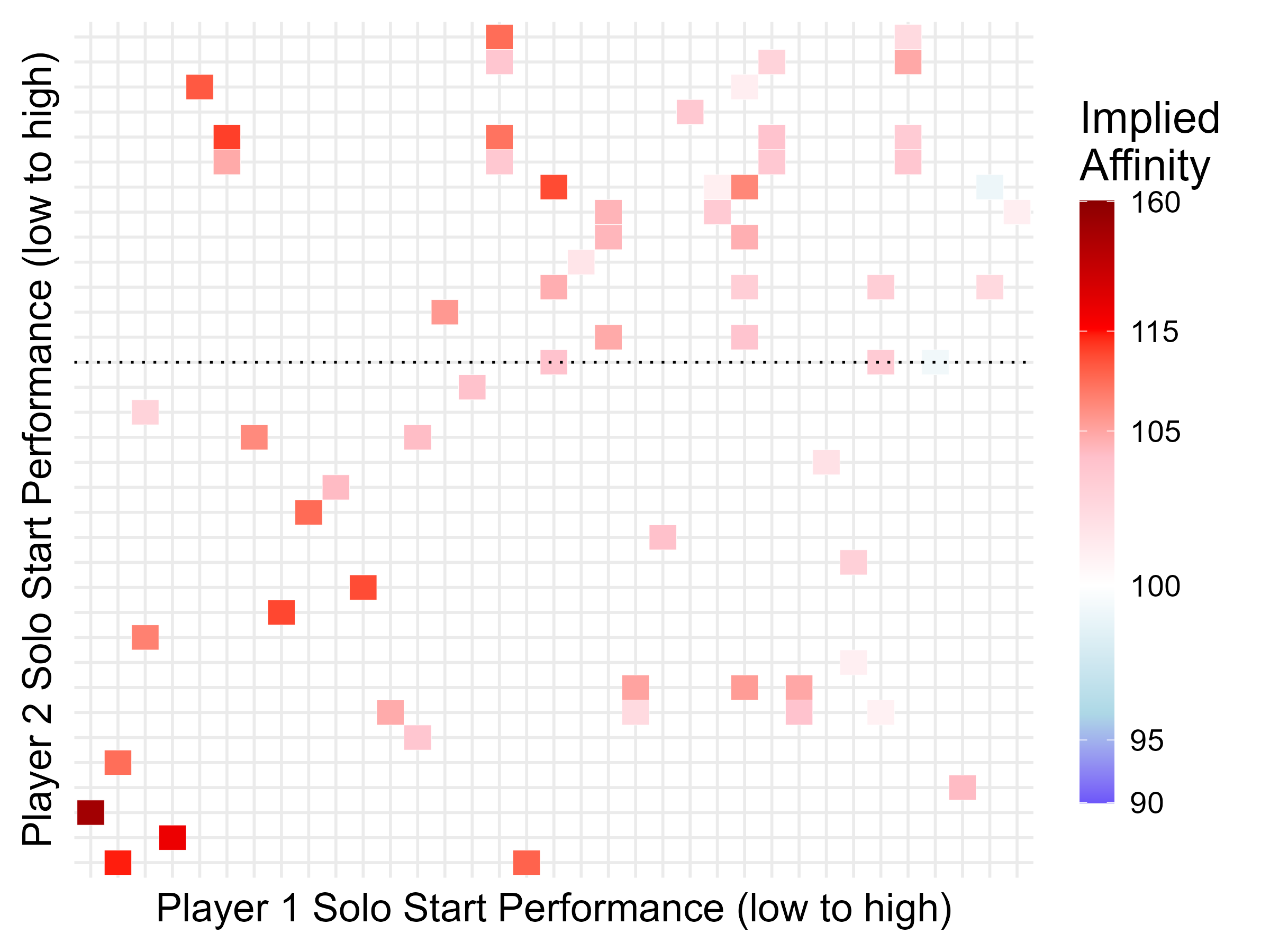}
      \caption{Start phase, first attempt}
    \end{subfigure}
    \hfill
    \begin{subfigure}{0.47\textwidth}
      \centering
      \includegraphics[width=\textwidth]{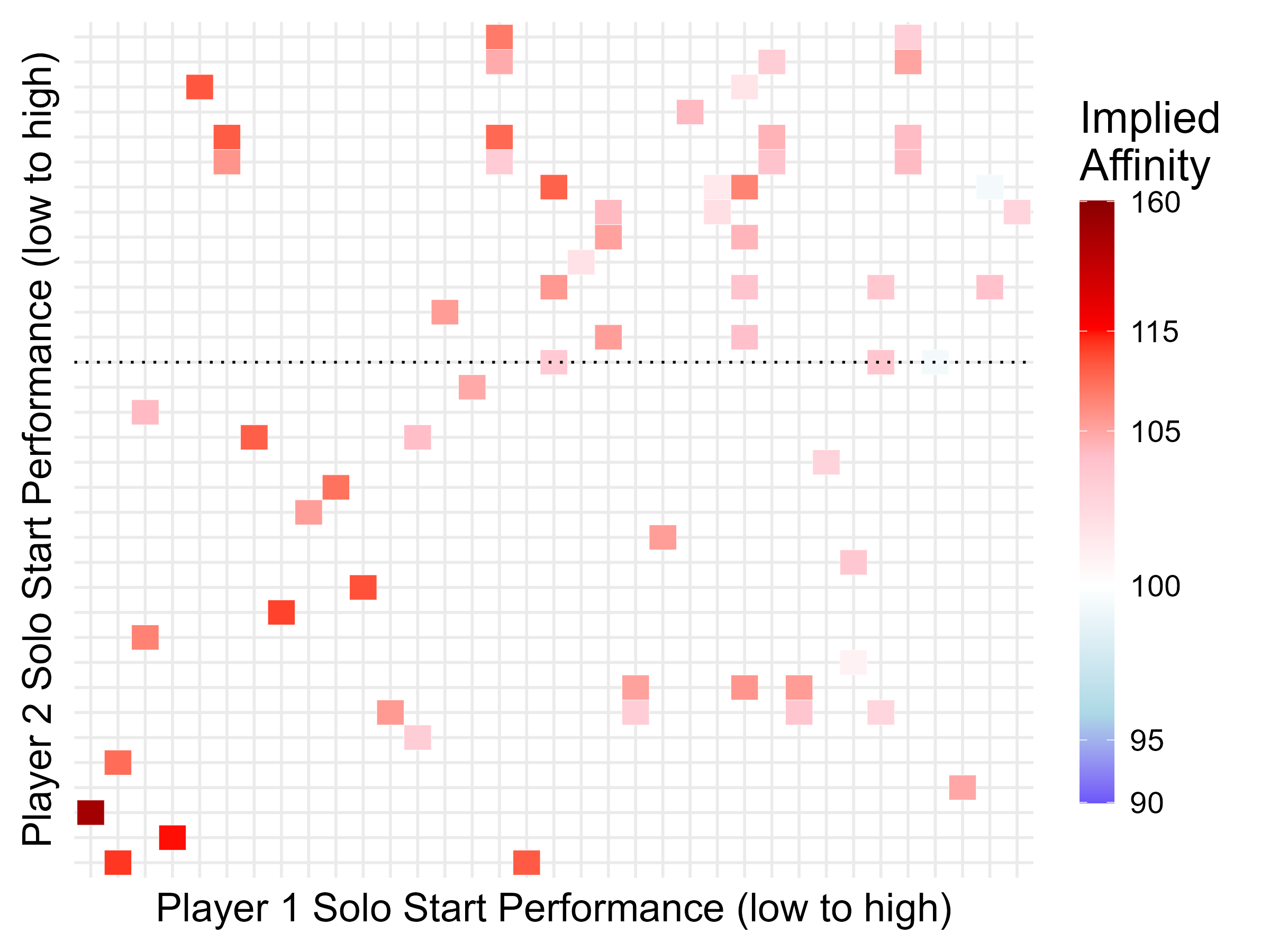}
      \caption{Start phase, second attempt}
    \end{subfigure}

    \vspace{0.75em}
    \begin{subfigure}{0.47\textwidth}
      \centering
      \includegraphics[width=\textwidth]{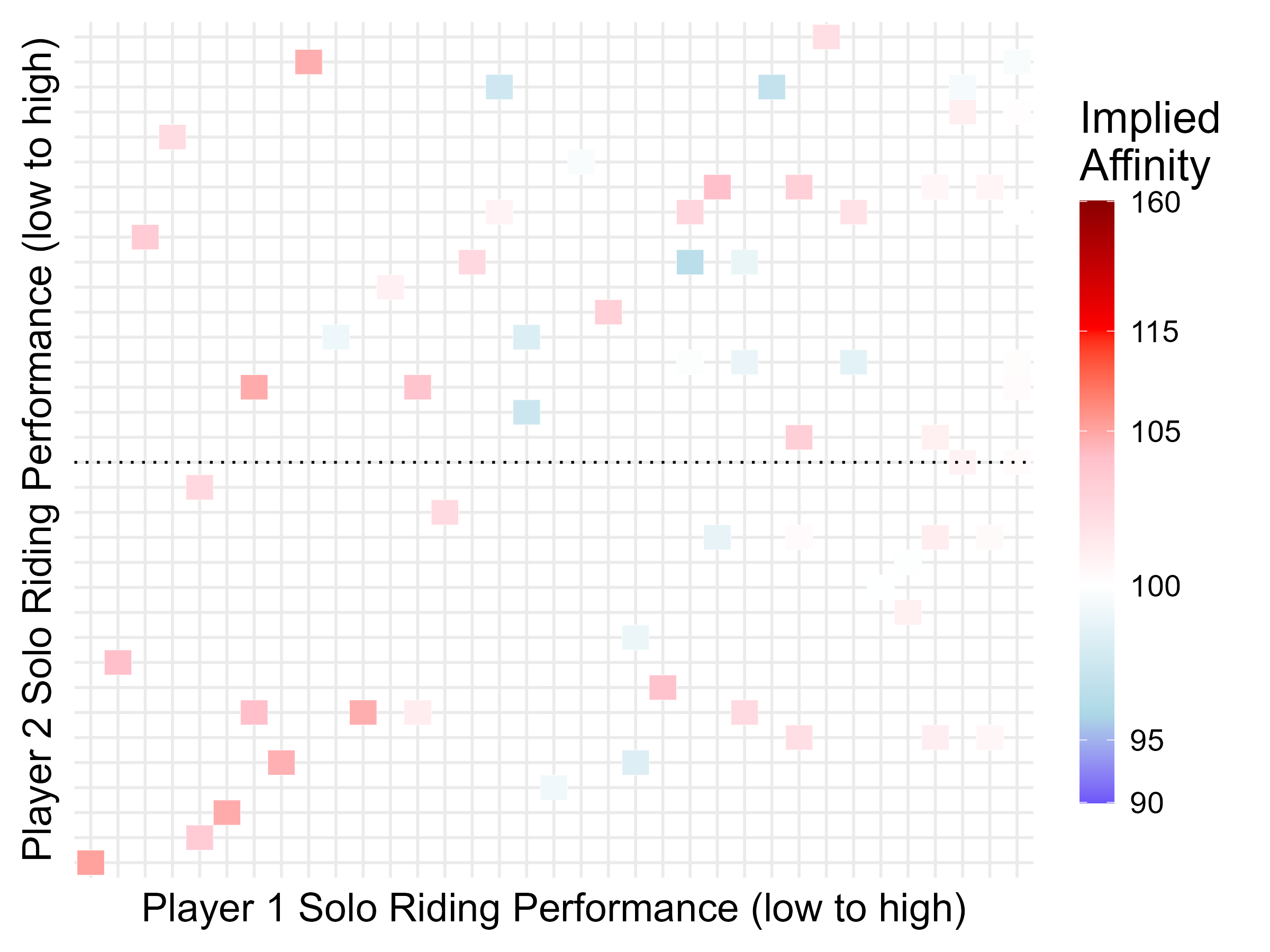}
      \caption{Riding phase, first attempt}
    \end{subfigure}
    \hfill
    \begin{subfigure}{0.47\textwidth}
      \centering
      \includegraphics[width=\textwidth]{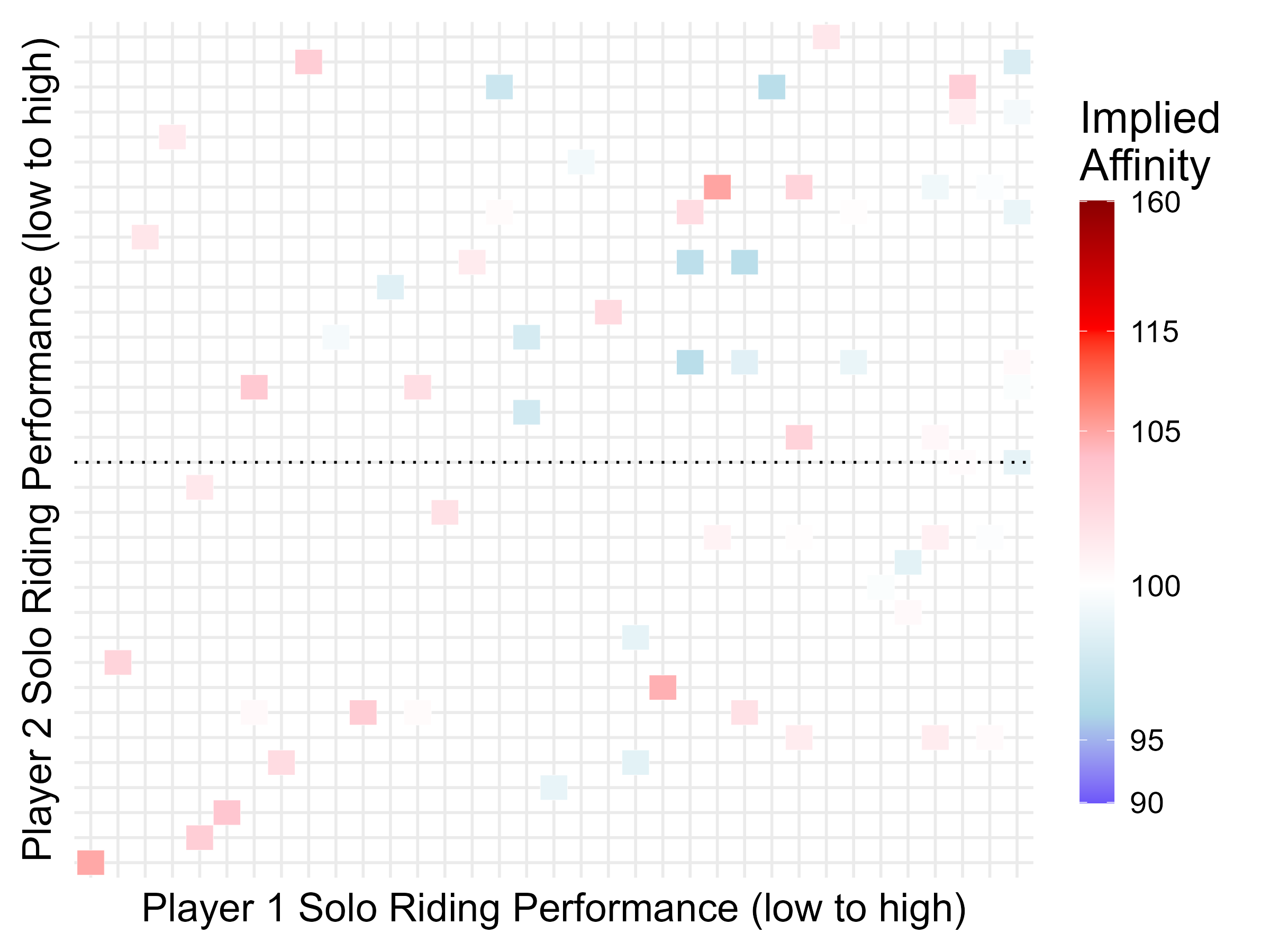}
      \caption{Riding phase, second attempt}
    \end{subfigure}
    \caption{Implied Affinity for Crew--Event Records with Exactly Two Attempts}
    \label{fg:efficiency_implied_riding_2_only_two_attempts}
  \end{center}
  {\footnotesize \textit{Note}: The restricted sample contains 237 crew--event participations. The dotted line marks the Player~2 skill of the normalization observation, whose reported affinity is set to 100 (white). Both axes run from lower to higher solo performance, based on fixed effects estimated with event and starting-order controls. Because normalization is sample-specific, raw affinity levels are not compared across samples.}
\end{figure}

\FloatBarrier

\subsection{Counterfactual Details and Additional Results}\label{app:counterfactual_details}\label{app:additional_figures}

This subsection reports implementation details and supplementary tables for the counterfactual assignment exercises in Section~\ref{sec:counterfactual}.

The exercises require assigning an affinity value to every candidate pair. The benchmark extends Equation~\eqref{eq:conditional_independence} with conditional exchangeability and common support between observed and never-observed candidate pairs. The counterfactual implementation assigns a never-observed pair a kernel estimate of conditional mean affinity given the driver's skill, consistent with the conditioning direction in Equation~\eqref{eq:conditional_independence}, while an observed pair receives its realized pair-mean affinity for each task and attempt. Team output is then predicted by evaluating the second-order projection at that assigned affinity. Because the projection contains a squared effective-input term, this plug-in uses $E[A\mid X]^2$ where an integrated expected-output calculation generally requires $E[A^2\mid X]$. The counterfactual values are therefore illustrative and would be sharpened by integrating over the conditional affinity distribution.

Gains are expressed in physical seconds by reversing Equation~\eqref{eq:positive_transform}. Conditional inference is described in the note to Table~\ref{tb:counterfactual_bootstrap_summary}.

\begin{table}[!htbp]
  \begin{center}
    \caption{Nation-Level Rematching Gains by Manager Information Set}
    \label{tb:counterfactual_nation_rematching}
    \begin{subtable}{\textwidth}
      \centering
      \caption{National pools}
      \begin{adjustbox}{max width=\textwidth}
        \input{figuretable/counterfactual_nation_pools}
      \end{adjustbox}
    \end{subtable}

    \vspace{0.75em}
    \begin{subtable}{\textwidth}
      \centering
      \caption{Fitted gains by manager (seconds)}
      \begin{adjustbox}{max width=\textwidth}
        \input{figuretable/counterfactual_nation_gains}
      \end{adjustbox}
    \end{subtable}
  \end{center}
  {\footnotesize \textit{Note}: Drivers, Brakemen, and Particip.\ are the numbers of included drivers, eligible brakemen, and weighted participations in each national pool. Start, Riding, and Skills abbreviate the Start-only, Riding-only, and Skills-only managers. Gains are fitted seconds per weighted two-attempt participation relative to historical pairings. SR uses separate task--attempt affinities; Start-only and Riding-only replace affinity in the other task by its conditional mean; Pooled uses the unweighted mean across all four panels; Skills-only uses conditional means throughout. All assignments are scored using SR's affinity inputs. National pools initially require at least two observed drivers and two observed brakemen. Drivers are retained in descending participation order subject to having enough distinct brakemen; no athlete can fill both roles or multiple crew positions. All managers face the same rosters. The pooled comparison uses the maintained panel normalizations.}
\end{table}

\ifsupp
\ifmsanon\theendnotes\fi
\bibliographystyle{ecca}
\bibliography{team_production}
\fi
\fi

\end{document}

%% file: figuretable/summary_statistics_player_level.tex
\begin{tabular}[t]{lrrrrr}
\toprule
  & N & mean & sd & min & max\\
\midrule
Residualized Solo Finish Outcome & 59 & 59.97 & 1.43 & 52.31 & 61.56\\
Residualized Solo Start Outcome & 59 & 6.11 & 0.31 & 4.61 & 6.69\\
Residualized Solo Riding Outcome & 59 & 53.86 & 1.29 & 46.30 & 55.27\\
\bottomrule
\end{tabular}

%% file: figuretable/summary_statistics_team_level.tex
\begin{tabular}[t]{lrrrrr}
\toprule
  & N & mean & sd & min & max\\
\midrule
Residualized Team Finish Outcome 1st attempt & 260 & 56.81 & 0.81 & 52.06 & 58.55\\
Residualized Team Finish Outcome 2nd attempt & 260 & 56.82 & 0.92 & 51.35 & 59.31\\
Residualized Team Start Outcome 1st attempt & 260 & 5.71 & 0.17 & 5.08 & 6.28\\
Residualized Team Start Outcome 2nd attempt & 260 & 5.71 & 0.17 & 4.88 & 6.19\\
Residualized Team Riding Outcome 1st attempt & 260 & 51.10 & 0.70 & 46.35 & 52.67\\
Residualized Team Riding Outcome 2nd attempt & 260 & 51.11 & 0.82 & 45.58 & 53.49\\
Starting No & 260 & 8.00 & 4.79 & 1.00 & 21.00\\
\bottomrule
\end{tabular}

%% file: figuretable/num_of_players_team_plays.tex
\begin{tabular}[t]{lrrrrrr}
\toprule
Nat & athletes & drivers & brakemen & pairs & participations & events\\
\midrule
AUS & 5 & 3 & 4 & 4 & 27 & 26\\
BEL & 3 & 2 & 2 & 3 & 5 & 5\\
CAN & 14 & 12 & 9 & 27 & 82 & 65\\
CHN & 2 & 1 & 1 & 1 & 1 & 1\\
GBR & 4 & 2 & 2 & 2 & 6 & 6\\
GER & 4 & 2 & 2 & 2 & 4 & 4\\
ITA & 3 & 1 & 2 & 2 & 22 & 22\\
JAM & 2 & 1 & 1 & 1 & 3 & 3\\
KOR & 4 & 2 & 2 & 2 & 3 & 3\\
POL & 4 & 2 & 2 & 3 & 11 & 10\\
ROU & 2 & 1 & 1 & 1 & 35 & 35\\
RUS & 2 & 1 & 1 & 1 & 1 & 1\\
SVK & 2 & 1 & 1 & 1 & 1 & 1\\
USA & 8 & 4 & 4 & 13 & 59 & 42\\
\bottomrule
\end{tabular}

%% file: figuretable/affinity_signal_correlation.tex
\begin{tabular}[t]{lrr}
\toprule
Statistic & Estimate & 95\% CI\\
\midrule
Start attempt persistence & 0.976 & {}[0.905, 0.997]\\
Riding attempt persistence & 0.902 & {}[0.725, 0.985]\\
Same-attempt cross-task correlation & 0.289 & {}[0.151, 0.605]\\
Cross-attempt cross-task correlation & 0.301 & {}[0.154, 0.609]\\
Persistent Start-Riding correlation & 0.321 & {}[0.166, 0.686]\\
\bottomrule
\end{tabular}

%% file: figuretable/counterfactual_reversal_share.tex
\begin{tabular}[t]{lrrr}
\toprule
Nat & blocks & reversals & share\\
\midrule
CAN & 6 & 2 & 0.333\\
USA & 15 & 2 & 0.133\\
Total & 21 & 4 & 0.190\\
\bottomrule
\end{tabular}

%% file: figuretable/counterfactual_event_swap_summary.tex
\begin{tabular}[t]{lr}
\toprule
Statistic & Value\\
\midrule
Event--nation cells & 32\\
Two-sled cells & 28\\
Three-sled cells & 4\\
Cells with improvement & 23\\
Cells where the observed assignment is optimal & 9\\
Share of cells with improvement & 0.719\\
Mean gain (seconds) & 0.744\\
Mean gain in improved cells (seconds) & 1.036\\
Maximum gain (seconds) & 4.581\\
\bottomrule
\end{tabular}

%% file: figuretable/counterfactual_bootstrap_summary.tex
\begin{tabular}[t]{lrrr}
\toprule
Statistic & Point & 95\% CI lower & 95\% CI upper\\
\midrule
Within-event: mean aggregate gain & 0.744 & 0.608 & 0.993\\
National: SR gain over historical assignment & 0.460 & 0.395 & 0.596\\
National: SR - Start-only & 0.073 & 0.049 & 0.246\\
National: SR - Riding-only & 0.017 & 0.000 & 0.046\\
National: SR - Pooled & 0.027 & 0.000 & 0.127\\
National: SR - Skills-only & 0.444 & 0.316 & 0.679\\
\bottomrule
\end{tabular}

%% file: figuretable/monobob_events_runs_matched_athletes_summary.tex
\begin{tabular}[t]{lrrrrrrrrrr}
\toprule
Sample & Obs. & Athletes & One event & Mean & SD & Min & Q1 & Median & Q3 & Max\\
\midrule
All & 1729 & 143 & 18 & 12.09 & 12.51 & 1 & 2 & 7 & 20 & 53\\
Matched & 937 & 59 & 5 & 15.88 & 14.41 & 1 & 3 & 10 & 24 & 53\\
\bottomrule
\end{tabular}

%% file: figuretable/start_riding_skill_correlation_summary.tex
\begin{tabular}[t]{lrrrrr}
\toprule
Sample & N athletes & Correlation & CI lower & CI upper & p-value\\
\midrule
All monobob athletes & 143 & 0.437 & 0.293 & 0.561 & 0.00\\
Matched two-woman athletes & 59 & 0.334 & 0.085 & 0.544 & 0.01\\
\bottomrule
\end{tabular}

%% file: figuretable/skill_pairing_summary_statistics_start.tex
\begin{tabular}[t]{cccc}
\toprule
\multicolumn{1}{c}{ } & \multicolumn{3}{c}{P2 Start-skill Bin} \\
\cmidrule(l{3pt}r{3pt}){2-4}
P1 Start-skill Bin & 1 & 2 & 3\\
\midrule
1 & 12 & 9 & 7\\
2 & 4 & 7 & 7\\
3 & 6 & 7 & 4\\
\bottomrule
\end{tabular}

%% file: figuretable/skill_pairing_summary_statistics_riding.tex
\begin{tabular}[t]{cccc}
\toprule
\multicolumn{1}{c}{ } & \multicolumn{3}{c}{P2 Riding-skill Bin} \\
\cmidrule(l{3pt}r{3pt}){2-4}
P1 Riding-skill Bin & 1 & 2 & 3\\
\midrule
1 & 15 & 4 & 4\\
2 & 4 & 8 & 13\\
3 & 3 & 6 & 6\\
\bottomrule
\end{tabular}

%% file: figuretable/skill_pairing_summary_statistics_p1start_p2riding.tex
\begin{tabular}[t]{cccc}
\toprule
\multicolumn{1}{c}{ } & \multicolumn{3}{c}{P2 Riding-skill Bin} \\
\cmidrule(l{3pt}r{3pt}){2-4}
P1 Start-skill Bin & 1 & 2 & 3\\
\midrule
1 & 14 & 6 & 8\\
2 & 3 & 6 & 9\\
3 & 5 & 6 & 6\\
\bottomrule
\end{tabular}

%% file: figuretable/skill_pairing_summary_statistics_p1riding_p2start.tex
\begin{tabular}[t]{cccc}
\toprule
\multicolumn{1}{c}{ } & \multicolumn{3}{c}{P2 Start-skill Bin} \\
\cmidrule(l{3pt}r{3pt}){2-4}
P1 Riding-skill Bin & 1 & 2 & 3\\
\midrule
1 & 17 & 5 & 1\\
2 & 2 & 12 & 11\\
3 & 3 & 6 & 6\\
\bottomrule
\end{tabular}

%% file: figuretable/skill_pairing_run_counts_start.tex
\begin{tabular}[t]{cccc}
\toprule
\multicolumn{1}{c}{ } & \multicolumn{3}{c}{P2 Start-skill Bin} \\
\cmidrule(l{3pt}r{3pt}){2-4}
P1 Start-skill Bin & 1 & 2 & 3\\
\midrule
1 & 42 & 29 & 16\\
2 & 17 & 18 & 53\\
3 & 18 & 46 & 21\\
\bottomrule
\end{tabular}

%% file: figuretable/skill_pairing_run_counts_riding.tex
\begin{tabular}[t]{cccc}
\toprule
\multicolumn{1}{c}{ } & \multicolumn{3}{c}{P2 Riding-skill Bin} \\
\cmidrule(l{3pt}r{3pt}){2-4}
P1 Riding-skill Bin & 1 & 2 & 3\\
\midrule
1 & 49 & 11 & 10\\
2 & 19 & 26 & 60\\
3 & 18 & 50 & 17\\
\bottomrule
\end{tabular}

%% file: figuretable/skill_pairing_run_counts_p1start_p2riding.tex
\begin{tabular}[t]{cccc}
\toprule
\multicolumn{1}{c}{ } & \multicolumn{3}{c}{P2 Riding-skill Bin} \\
\cmidrule(l{3pt}r{3pt}){2-4}
P1 Start-skill Bin & 1 & 2 & 3\\
\midrule
1 & 63 & 11 & 13\\
2 & 12 & 47 & 29\\
3 & 11 & 29 & 45\\
\bottomrule
\end{tabular}

%% file: figuretable/skill_pairing_run_counts_p1riding_p2start.tex
\begin{tabular}[t]{cccc}
\toprule
\multicolumn{1}{c}{ } & \multicolumn{3}{c}{P2 Start-skill Bin} \\
\cmidrule(l{3pt}r{3pt}){2-4}
P1 Riding-skill Bin & 1 & 2 & 3\\
\midrule
1 & 54 & 14 & 2\\
2 & 2 & 65 & 38\\
3 & 21 & 14 & 50\\
\bottomrule
\end{tabular}

%% file: figuretable/pairing_continuation_by_affinity.tex
\begin{tabular}[t]{lrrr}
\toprule
Task & Low affinity & Middle affinity & High affinity\\
\midrule
Start & 65.3\% (47/72) & 52.8\% (38/72) & 56.9\% (41/72)\\
Riding & 63.9\% (46/72) & 50.7\% (36/71) & 60.3\% (44/73)\\
\bottomrule
\end{tabular}
\par\smallskip\begin{minipage}{\linewidth}\footnotesize Identified within-season next-event transitions of a driver. Continuation keeps the same brakeman. Columns use previous-event affinity terciles (average attempt percentile ranks). Percentages show continuation counts/eligible transitions; terminal or ambiguous next events are not coded as breakups.\end{minipage}

%% file: figuretable/first_observed_affinity_comparison.tex
\begin{adjustbox}{max width=\linewidth}

\begin{tabular}[t]{lrrr}
\toprule
Window & First / previous & Start difference [95\% CI] & Riding difference [95\% CI]\\
\midrule
2018--19 through 2024--25 & 53 / 207 & $1.728$ [$0.503$, $3.008$] & $0.036$ [$-0.701$, $0.696$]\\
\bottomrule
\end{tabular}
\end{adjustbox}
\par\smallskip\begin{minipage}{\linewidth}\footnotesize First minus previously observed event-weighted affinity; conditional pair-cluster percentile intervals from 2,000 draws. Evaluation uses 2018--19 through 2024--25; first observed refers to the available archive from 2015--16.\end{minipage}

%% file: figuretable/monte_carlo_affinity_sensitivity.tex
\begin{tabular}[t]{lrrrrrrrrr}
\toprule
\multicolumn{4}{c}{ } & \multicolumn{2}{c}{NP} & \multicolumn{2}{c}{CD-M} & \multicolumn{2}{c}{CD-I} \\
\cmidrule(l{3pt}r{3pt}){5-6} \cmidrule(l{3pt}r{3pt}){7-8} \cmidrule(l{3pt}r{3pt}){9-10}
Technology & N & Rho & R & Bias & RMSE & Bias & RMSE & Bias & RMSE\\
\midrule
Cobb-Douglas & 50 & 0.0 & 2000/2000 & -1.10 (0.16) & 9.76 (0.10) & -96.69 (0.19) & 97.04 (0.19) & 8946.50 (4736.05) & 9014.15 (4735.99)\\
Cobb-Douglas & 50 & 0.1 & 2000/2000 & -1.06 (0.16) & 9.41 (0.09) & -97.28 (0.19) & 97.62 (0.19) & 64159.33 (50964.34) & 64250.16 (50964.28)\\
Cobb-Douglas & 50 & 0.2 & 2000/2000 & -1.19 (0.16) & 9.33 (0.09) & -97.93 (0.19) & 98.28 (0.19) & 795.76 (293.84) & 912.29 (293.67)\\
Cobb-Douglas & 50 & 0.5 & 2000/2000 & -1.48 (0.16) & 9.34 (0.10) & -99.46 (0.19) & 99.79 (0.19) & -78.73 (5.56) & 104.45 (5.35)\\
Cobb-Douglas & 100 & 0.0 & 2000/2000 & -0.66 (0.14) & 8.37 (0.08) & -96.76 (0.19) & 97.10 (0.19) & 636.69 (269.94) & 691.84 (269.87)\\
Cobb-Douglas & 100 & 0.1 & 2000/2000 & -0.55 (0.13) & 7.98 (0.08) & -97.11 (0.18) & 97.44 (0.19) & 149.65 (53.00) & 240.63 (52.83)\\
Cobb-Douglas & 100 & 0.2 & 2000/2000 & -0.68 (0.13) & 7.60 (0.08) & -98.16 (0.19) & 98.50 (0.19) & -28.10 (4.34) & 98.22 (3.80)\\
Cobb-Douglas & 100 & 0.5 & 2000/2000 & -1.05 (0.13) & 7.34 (0.08) & -99.43 (0.19) & 99.75 (0.19) & -95.43 (0.29) & 96.02 (0.25)\\
Cobb-Douglas & 200 & 0.0 & 2000/2000 & -0.56 (0.13) & 7.71 (0.07) & -96.45 (0.18) & 96.78 (0.19) & 88.83 (9.54) & 135.34 (9.26)\\
Cobb-Douglas & 200 & 0.1 & 2000/2000 & -0.35 (0.12) & 7.05 (0.07) & -97.43 (0.18) & 97.74 (0.18) & -13.87 (2.74) & 76.10 (2.17)\\
Cobb-Douglas & 200 & 0.2 & 2000/2000 & -0.28 (0.11) & 6.57 (0.07) & -97.95 (0.19) & 98.27 (0.19) & -60.64 (1.42) & 74.26 (1.06)\\
Cobb-Douglas & 200 & 0.5 & 2000/2000 & -0.63 (0.10) & 5.69 (0.07) & -99.51 (0.18) & 99.82 (0.18) & -97.17 (0.19) & 97.49 (0.19)\\
Perfect substitutes & 50 & 0.0 & 2000/2000 & -0.78 (0.14) & 7.90 (0.09) & -6.80 (5.74) & 106.90 (5.22) & 19163.96 (13246.56) & 19270.35 (13246.49)\\
Perfect substitutes & 50 & 0.1 & 2000/2000 & -1.03 (0.14) & 8.15 (0.10) & -60.74 (1.86) & 84.41 (1.33) & 503.37 (413.41) & 637.46 (413.32)\\
Perfect substitutes & 50 & 0.2 & 2000/2000 & -1.50 (0.15) & 8.46 (0.10) & -80.34 (1.20) & 88.96 (0.87) & 369.50 (227.66) & 526.79 (227.51)\\
Perfect substitutes & 50 & 0.5 & 2000/2000 & -2.47 (0.17) & 10.13 (0.11) & -98.27 (0.20) & 98.61 (0.20) & -96.84 (0.41) & 97.87 (0.32)\\
Perfect substitutes & 100 & 0.0 & 2000/2000 & -0.58 (0.12) & 6.22 (0.09) & -51.84 (1.48) & 70.68 (1.03) & 47.35 (52.91) & 161.37 (52.80)\\
Perfect substitutes & 100 & 0.1 & 2000/2000 & -0.59 (0.11) & 6.29 (0.08) & -74.90 (0.74) & 79.10 (0.51) & -51.67 (4.01) & 92.00 (3.64)\\
Perfect substitutes & 100 & 0.2 & 2000/2000 & -0.94 (0.12) & 6.89 (0.09) & -88.65 (0.36) & 89.18 (0.34) & -82.37 (0.88) & 88.13 (0.56)\\
Perfect substitutes & 100 & 0.5 & 2000/2000 & -1.88 (0.15) & 9.09 (0.09) & -99.40 (0.18) & 99.72 (0.18) & -99.20 (0.19) & 99.52 (0.19)\\
Perfect substitutes & 200 & 0.0 & 2000/2000 & -0.08 (0.07) & 4.48 (0.05) & -60.08 (0.72) & 64.62 (0.52) & -44.40 (1.81) & 70.82 (1.34)\\
Perfect substitutes & 200 & 0.1 & 2000/2000 & -0.29 (0.08) & 4.79 (0.07) & -81.17 (0.39) & 81.85 (0.36) & -75.48 (0.79) & 80.24 (0.54)\\
Perfect substitutes & 200 & 0.2 & 2000/2000 & -0.19 (0.09) & 5.29 (0.07) & -91.01 (0.23) & 91.35 (0.23) & -89.03 (0.34) & 89.65 (0.30)\\
Perfect substitutes & 200 & 0.5 & 2000/2000 & -0.73 (0.14) & 8.34 (0.08) & -99.75 (0.18) & 100.06 (0.18) & -99.70 (0.18) & 100.01 (0.18)\\
\bottomrule
\end{tabular}

%% file: figuretable/summary_statistics_implied_efficiency.tex
\begin{tabular}[t]{lrrrrr}
\toprule
Task-attempt & N pairs & Mean & SD & Min & Max\\
\midrule
Start, 1st attempt & 63 & 103.84 & 5.54 & 98.65 & 138.49\\
Start, 2nd attempt & 63 & 104.19 & 5.30 & 98.46 & 137.61\\
Riding, 1st attempt & 63 & 101.05 & 3.33 & 95.78 & 122.00\\
Riding, 2nd attempt & 63 & 100.62 & 3.12 & 94.91 & 118.88\\
\bottomrule
\end{tabular}

%% file: figuretable/implied_affinity_structure_dispersion.tex
\begin{tabular}[t]{llrrrrrr}
\toprule
Task & Attempt & N & Mean & SD & HH & LL & HH $-$ LL\\
\midrule
Start & 1st & 63 & 103.84 & 5.54 & 101.88 & 109.68 & -7.80\\
Start & 2nd & 63 & 104.19 & 5.30 & 102.42 & 109.81 & -7.39\\
Riding & 1st & 63 & 101.05 & 3.33 & 99.11 & 101.73 & -2.62\\
Riding & 2nd & 63 & 100.62 & 3.12 & 99.96 & 101.34 & -1.38\\
\bottomrule
\end{tabular}

%% file: figuretable/implied_affinity_structure_diagonal.tex
\begin{tabular}[t]{llrrr}
\toprule
Task & Attempt & Diag. & Off-diag. & Off $-$ Diag.\\
\midrule
Start & 1st & 106.09 & 102.54 & -3.55\\
Start & 2nd & 106.25 & 103.01 & -3.24\\
Riding & 1st & 100.80 & 101.26 & 0.46\\
Riding & 2nd & 100.61 & 100.64 & 0.03\\
\bottomrule
\end{tabular}

%% file: figuretable/counterfactual_nation_pools.tex
\begin{tabular}[t]{lrrr}
\toprule
Nation & Drivers & Brakemen & Particip.\\
\midrule
AUS & 2 & 3 & 22\\
BEL & 1 & 2 & 4\\
CAN & 6 & 7 & 69\\
GBR & 2 & 2 & 6\\
GER & 2 & 2 & 4\\
KOR & 2 & 2 & 3\\
POL & 2 & 2 & 11\\
USA & 4 & 4 & 59\\
\bottomrule
\end{tabular}

%% file: figuretable/counterfactual_nation_gains.tex
\begin{tabular}[t]{lrrrrrrr}
\toprule
Nation & SR & Start & Riding & Pooled & Skills & SR - Pooled & SR - Skills\\
\midrule
AUS & 0.895 & 0.895 & 0.895 & 0.895 & 0.362 & 0.000 & 0.533\\
BEL & 0.283 & 0.283 & 0.283 & 0.283 & 0.283 & 0.000 & 0.000\\
CAN & 0.554 & 0.456 & 0.509 & 0.554 & -0.098 & 0.000 & 0.652\\
GBR & 0.000 & 0.000 & 0.000 & 0.000 & 0.000 & 0.000 & 0.000\\
GER & 0.000 & 0.000 & 0.000 & 0.000 & 0.000 & 0.000 & 0.000\\
KOR & 0.000 & -1.167 & 0.000 & 0.000 & -1.167 & 0.000 & 1.167\\
POL & 0.237 & -0.009 & 0.237 & 0.237 & -0.009 & 0.000 & 0.246\\
USA & 0.342 & 0.342 & 0.342 & 0.260 & 0.068 & 0.081 & 0.273\\
\bottomrule
\end{tabular}